\documentclass[10pt]{article}

\usepackage[english]{babel}
\usepackage{amsthm}
\usepackage{amssymb}
\usepackage{mathtools}

\newtheorem{theorem}{Theorem}[section]
\newtheorem{lemma}[theorem]{Lemma}
\newtheorem{proposition}[theorem]{Proposition}
\newtheorem{corollary}[theorem]{Corollary}
\newtheorem{remark}[theorem]{Remark}
\newtheorem{definition}[theorem]{Definition}
\makeatletter

\@addtoreset{equation}{section}
\usepackage{color}
\usepackage{graphicx}

\makeatother
\usepackage{newlfont}
\usepackage{pdfpages}
\def \n {\noindent}
\begin{document}

\begin{center}
 
 {\Large {\color{red} {\bf Presentation of some elementary properties of Segal-Bargmann space and of  unitary Segal-Bargmann transform }}} \\
 \quad\\
  {\color{blue}{\Large{\bf with applications}}}\\
 \quad\\

{\Large {\bf Abdelkader. INTISSAR}}

``Le Prador''\\
129,  rue commandant Rolland\\
13008 Marseille France\\
{\color{blue}abdelkader.intissar@orange.fr}\\
 \quad\\
\end{center}

\begin{center}
{\color{red}{\bf Abstract}}\\
\end{center}

\n {\color{black}In this work, we present  some elementary properties of Segal-Bargmann space and some properties of unitary Segal Bargmann transform with applications to differential operators arising out of diffusion problem  or of reggeon field theory.}\\

\newpage
 \n  {\Large { \bf Content  }}\\
 
 \n  {\large { \bf  1 Introduction}}   \\
 
 \n  {\large { \bf  2  The standard creation and annihilation operators on $L_{2}(\mathbb{R})$ }}  \\
 
\n 2.1 Basic properties of creation and annihilation operators on $L_{2}(\mathbb{R})$............. {\color{red}5} \\ 
\n 2.2 Determination of eigenvalues and eigenfunctions of Hermite's operator... {\color{red}11}   \\
\n  2.3 Elementary properties of Bargmann space ............................................... {\color{red}14}  \\
\n 2.4 Examples of reproducing kernel spaces  .................................................... {\color{red}23}  \\
\n 2.5 Fundamental results on reproducing kernel spaces ................................. {\color{red}28}  \\

 \n  {\large { \bf  3  Unitarity of the Segal-Bargmann transform and some properties of its adjoint}}  \\
 
 \n 3.1 On gaussian measure and Bargmann space  ............................................. {\color{red}31} \\
 \n 3.2 Some properties of Segal-Bargmann transform   ....................................... {\color{red}32} \\
 \n 3.3 Some properties of the adjoint of Segal-Bargmann transform   ................. {\color{red}44} \\
  \n 3.4 The relationship between Bargmann transform, Gabor transform and FBI transform ...................................................................................................... {\color{red}52} \\
\n  3.5 The standard creation  and annihilation  operators on Bargmann space... {\color{red}54}  \\
   \n 3.6 Some applications to differential operators arising in diffusion problem or in reggeon field theory .......................................................................................  {\color{red}68}  \\

 \n  {\large { \bf  4 Domination of $\displaystyle{\sum_{i+ j < 2k}a_{i,j}A^{*^{i}}A^{j}}$ by $\displaystyle{A^{*^{k}}A^{k}; \, a_{i,j} \in \mathbb{C}}$  }}\\
 
  \n 4.1 On the domination of $\displaystyle{\sum_{i+ j < 2k}a_{i,j}A^{*^{i}}A^{j}}$ by $\displaystyle{A^{*^{k}}A^{k}}$...................................  {\color{red}84}  \\
  \n 4.2 Some spectral properties of the Hamiltonian  $\displaystyle{\mathbb{H} = A^{*^{k}}A^{k} +\sum_{i+ j < 2k}a_{i,j}A^{*^{i}}A^{j}}$ \\on Bargmann space .....................................................................................  {\color{red}91}  \\
  
  \n  {\large { \bf  5 On  regularized trace formula of  order one of the Hamiltonian $\displaystyle{\mathbb{H} = A^{*^{k}}A^{k} +\sum_{i+ j < 2k}a_{i,j}A^{*^{i}}A^{j}}$ on Bargmann space}}\\
  
   \n 5.1 Results of Sadovnichii and Podolski on regularized traces of abstract discrete operators .......................................................................................................  {\color{red}94}  \\
    \n 5.2 The regularized trace formula for the Hamiltonian  $\displaystyle{\mathbb{H} = A^{*^{k}}A^{k} +\sum_{i+ j < 2k}a_{i,j}A^{*^{i}}A^{j}}$ on Bargmann space ........................................................................................  {\color{red}96}  \\   
    
 \n {\bf {\Large References}}...........................................................................................  {\color{red}101}  \\

 \section{ {\color{red}Introduction}}
 
\n {\bf{\color{red}$\rhd$}} We define Segal-Bargmann space {\color{blue}[4]} by:\\
 \begin{equation}
 \displaystyle{\mathbb{B} = \{\varphi : \mathbb{C} \longrightarrow \mathbb{C} \quad entire ; \int_{\mathbb{C}} \vert \varphi(z) \vert^{2} e^{-\vert z \vert^{2}} dxdy < \infty. \}}
 \end{equation}
\n with the classical convention, where the inner product is linear in the first and  anti-linear in second argument :\\
 \begin{equation}
 \displaystyle{< \varphi,  \psi > = \frac{1}{\pi}\int_{\mathbb{C}}  \varphi(z) \overline{\psi(z)} e^{-\vert z \vert^{2}} dxdy ; z = x + iy , (x, y) \in \mathbb{R}^{2}}
 \end{equation}
 \n and the associated norm is denoted by $\mid\mid . \mid\mid_{ \mathbb{B}}$ or simply by $\mid\mid . \mid\mid $:\\
 
  \begin{equation}
 \displaystyle{\vert\vert \varphi \vert\vert = \sqrt{ \frac{1}{\pi}\int_{\mathbb{C}} \vert \varphi(z)\vert^{2} e^{-\vert z \vert^{2}} dxdy}}
 \end{equation}
\n and \\

\n {\bf{\color{red}$\rhd$}} We define $ \displaystyle{L_{2}(\mathbb{C}, d\mu(z))}$-space by:\\
 \begin{equation}
 \displaystyle{L_{2}(\mathbb{C}, d\mu(z)) = \{\varphi : \mathbb{C} \longrightarrow \mathbb{C} \quad mesearable ; \int_{\mathbb{C}} \vert \varphi(z) \vert^{2} e^{-\vert z \vert^{2}} dxdy < \infty. \}}
 \end{equation}
 
\n where   $\displaystyle{d\mu(z) = e^{-\vert z \vert^{2}} dxdy }$ and with scalar product : \\
 \begin{equation}
 \displaystyle{< \varphi,  \psi > = \frac{1}{\pi} \int_{\mathbb{C}}  \varphi(z) \overline{\psi(z)} e^{-\vert z \vert^{2}} dxdy ; z = x + iy , (x, y) \in \mathbb{R}^{2}}
 \end{equation}
 \n The associated norm of $L_{2}(\mathbb{C}, d\mu(z))$ is denoted by $\mid\mid . \mid\mid_{ L_{2}(\mathbb{C}, d\mu(z))}$ \\
 
  \n This space admits the following Hilbertian orthogonal decomposition:\\
  \begin{equation}
   \displaystyle{L_{2}(\mathbb{C}, d\mu(z)) = \underset{m \in \mathbb{Z}_{+}} {\bigoplus} E_{m}(\mathbb{C})}
 \end{equation}
 
  \n where the Hilbert subspaces $E_{m}(\mathbb{C}) , m \in \mathbb{Z}_{+}$ are defined by\\
  \begin{equation}
 \displaystyle{E_{m}(\mathbb{C}) = \{\varphi \in L_{2}(\mathbb{C}, d\mu(z)) ; \Lambda^{*}\Lambda\varphi = m\varphi \}}
  \end{equation}
  
  \n In (1.7) the operators $\Lambda^{*}$ and  $\Lambda$ are given respectively by $\displaystyle{\Lambda^{*} = -  \frac {\partial}{\partial z} +  \overline{z}} $ and $\displaystyle{\Lambda =  \frac{ \partial}{ \partial  \overline{z}}}$ so that  \\
 \begin{equation}
  \displaystyle{ \Lambda^{*}\Lambda = -  \frac{ \partial^{2}}{ \partial z  \partial  \overline{z}} +  \overline{z}  \frac{\partial}{ \partial z}}
  \end{equation}
  
   \n is a second order elliptic differential operator of Laplacian type and \\

    \begin{equation}
     \displaystyle{E_{0}(\mathbb{C}) = \mathbb{B}} 
     \end{equation}
  
   \n  i.e $ \mathbb{B}$ is the kernel of $ \Lambda^{*}\Lambda$.\\
   
  \n In the second section, We will show that \\
  
 \n {\color{red}$\bullet$} Segal-Bargmann space $\mathbb{B}$ is closed  in $ \displaystyle{L_{2}(\mathbb{C}, d\mu(z))}$, and becomes a Hilbert space.\\
 
 \n {\color{red}$\bullet$}  Segal-Bargmann space $\mathbb{B}$ is a reproducing kernel Hilbert space and it is determined by its reproducing kernel.\\
 
 \n and\\
 
 \n {\bf{\color{red}$\rhd$}} we give some  examples  of classical reproducing kernel Hilbert space with their explicit kernels respectively.\\
 
 \n {\bf{\color{red}$\rhd$}} we show also that two reproducing kernel Hilbert spaces with the same kernel are equal.\\
 
 \n In section 3, we define the Segal-Bargmann transform $S\mathcal{B}$ as transformation of $L_{2}(\mathbb{R})$  into  $ \displaystyle{L_{2}(\mathbb{C}, d\mu(z))}$.  \\
 
 \n {\bf{\color{red}$\rhd$}} if we denote the range of Segal-Bargmann transform by $SB(L_{2}(\mathbb{R}))$ then we define an inner product:\\
 \begin{equation}
 \displaystyle{ < \varphi , \psi >_{SB} = < f, g >_{L_{2}(\mathbb{R})}}
  \end{equation}
  \quad\\
 \n where $f$ and  $g$ are chosen such that $SB(f) = \varphi$ and $SB(g) = \psi$.\\
 
 \n We show that $SB(L_{2}(\mathbb{R}))$ is a reproducing kernel Hilbert space and that the Segal-Bargmann transform is injective, therfore it is an unitary mapping from  $L_{2}(\mathbb{R})$  to $SB(L_{2}(\mathbb{R}))$ .\\
 
 \n {\bf{\color{red}$\rhd$}} we show that $SB(L_{2}(\mathbb{R}))$ and $\mathbb{B}$ have same kernel. This implies that $SB$ is an unitary transform from $L_{2}(\mathbb{R})$  to Segal-Bargmann space $\mathbb{B}$.\\
 
 \n {\bf{\color{red}$\rhd$}} We use knowledge basics of holomorphic functions of one complex variable and the properties of Hilbert spaces. In this work the knowledge of spectral theory and the functional analysis of standard level is required\\ 
   
\section{{\color{red}The standard creation and annihilation operators on $L_{2}(\mathbb{R})$}}

\subsection{ {\color{blue}Basic properties of standard creation and annihilation operators on $L_{2}(\mathbb{R})$}}

\begin{definition}
\n Let $( \mathcal{H}_{1}, < , >_{1})$,   $( \mathcal{H}_{2}, < , >_{2})$, be  two Hilbert spaces, consider an operator $T$ from   $\mathcal{H}_{1}$ to $ \mathcal{H}_{2}$ and  an operator $S$ from  $ \mathcal{H}_{1}^{*} = \mathcal{H}_{1}$ to $ \mathcal{H}_{2}^{*} = \mathcal{H}_{2}$, $T$ of domain $D(T) = \{ u \in \mathcal{H}_{1} , Tu \in \mathcal{H}_{2}\}$ and $S$  of domain $D(S) = \{ v \in \mathcal{H}_{2}^{*} , Sv \in \mathcal{H}_{1}^{*}\}$ are said to be adjoint to each other if \\

\begin{equation}
\displaystyle{ < Tu , v > = < u , Sv > , u \in D(T) , v \in D(S)}
\end{equation} 
\end{definition}

\n For each operator $T$ from $\mathcal{H}_{1}$ to $ \mathcal{H}_{2}$, there are in general many operators  $ \mathcal{H}_{1}^{*} = \mathcal{H}_{1}$ to $ \mathcal{H}_{2}^{*} = \mathcal{H}_{2}$ that are adjoint to $T$.\\

\begin{definition}

If $T$ is densely defined (D(T) is dense in  $\mathcal{H}_{1}$, however, there is a unique maximal operator $T^{*}$ adjoint to $T$. This means that $T^{*}$ is adjoint to $T$ while any other $S$ adjoint to $T$ is a restriction of $T^{*}$.\\
$T^{*}$ is called the adjoint of $T$.\\
\end{definition}

\begin{remark}
{\color{red}$\bullet_{1}$} $T^{*}$ is constructed  in the following way . $D(T^{*})$ consists of all $v \in  \mathcal{H}_{2}^{*}$ such that there exists an $v^{*} \in  \mathcal{H}_{1}^{*}$ with the property \\
\begin{equation}
\displaystyle{< Tu, v > = < u, v^{*} > \,\, \mbox{for all u} \,\, \in D(T)}
\end{equation}

\n {\color{red}$\bullet_{2}$} the $v^{*}$ is determined uniquely by $v$, for $u , v^{*} > = < u, v^{^{'}} > $ for all $u \in D(T)$ implies $v^{*} = v^{^{'}}$ because $D(T)$ is dense in $\mathcal{H}_{1}$ by assumption.\\

\n {\color{red}$\bullet_{3}$} Therefore, an operator $T^{*}$ from $ \mathcal{H}_{2}^{*}$ to $ \mathcal{H}_{1}^{*}$ is defined by  setting $T^{*} v = v^{*}$ . Obviously $T^{*}$ is a linear  operator, and comparison of $(2. 1)$  with $(2.3)$ shows that $S \subset T^{*}$ holds for any $S$ adjoint to $T$ while $T^{*}$ itself is adjopint to $T.$\\

\n {\color{red}$\bullet_{3}$} The adjointness relation $(2.1)$ admits a simple  interpretation in form of the graphs ( the graph $\displaystyle{ \Gamma(T)}$ of an operator $T$ from  $\mathcal{H}_{1}$ to $\mathcal{H}_{2}$ is by definition the subset of $\mathcal{H}_{1} \times \mathcal{H}_{2}$ consisting of all elements  of the form $\{u, Tu\}$ with $u \in D(T)$).\\
The equation $(2.1)$ can be written $\displaystyle{<u,  - Sv > +  < Tu , v > = 0}$ which that $T$ and $S$ are adjoint to each other if and only if the graph of $T$ and the inverse graph of $-S$annihilate each other $\Gamma(T) = \Gamma^{'}(-S)$.\\

\n Similarly $(2.2)$ shows  that the inverse graph of $-T^{*}$ is the annihilator of the graph of $T$ :\\
  
\begin{equation}
  \Gamma^{'}(-T^{*}) = \Gamma(T)^{\perp}
 \end{equation} 
\end{remark}

\n Let $\mathcal{H}$ be  a Hilbert space equipped with scalar product $< , >$ and  let $T$ be densely defined linear operator on $\mathcal{H}$.\\

 \begin{proposition} 

\n  Let us define $\displaystyle{\mathbb{J} : \mathcal{H}\times \mathcal{H} \longrightarrow \mathcal{H}\times \mathcal{H} , (u, v ) \longrightarrow (-v , u)}$. We equip $\mathcal{H}\times \mathcal{H}$ with the natural scalar product. If $T$ is an operator with {\color{red}dense domain} $D(T)$, then \\

\n (i) $\displaystyle{\Gamma(T^{*}) = \mathbb{J} (\Gamma(T))^{\perp}}$ where $T^{*}$ denotes its adjoint and $\Gamma(T)$ is the  graph of $T$ and $\Gamma(T))^{\perp}$ is the orthogonal of this graph. \\

\n (ii) $\displaystyle{\overline{\Gamma(T)} = \mathbb{J} (\Gamma(T^{*}))^{\perp}}$, where $\overline{\Gamma(T)}$ denotes the closure of $\Gamma(T)$ \\

\n (iii) In particular, $T^{*}$ is closed.\\

\n (iv) $T$  is {\color{red}closable} if and only if $D(T^{*})$ is {\color{red}dense}. In this case, $(T^{*})^{*} = T^{**} = \overline{T}$, where $\overline{T}$ denotes the closure of $T$ .\\
\end{proposition}

\n {\color{red}{\bf Proof}}\\

\n  (i), (ii) and (iii):\\

 \n Since $D(T)$ is dense, we can assert that \\

\n  $\displaystyle{\Gamma(T^{*}) = \{(u, v) ; v - T^{*}u = 0\}}$\\

\n $\displaystyle{= \{(u, v) ;  < v - T^{*}u , w > = 0,  \, \forall \, w \in D(T)\}}$\\

\n $\displaystyle{= \{(u, v) ;  < v , w >  -  < u , Tw > = 0,  \, \forall \, w \in D(T)\}}$\\

\n $\displaystyle{= \{(u, v) ;  < (u, v) , \mathbb{J}(w, Tw) = 0,  \, \forall \, w \in D(T)\} = \mathbb{J} (\Gamma(T))^{\perp}}$\\

\n It follows that \\

\n $\displaystyle{\Gamma(T^{*})^{\perp} = (\mathbb{J} (\Gamma(T))^{\perp})^{\perp} = \overline{\mathbb{J} (\Gamma(T))} = \mathbb{J} (\overline{\Gamma(T)})}$.\\

\n Since $\mathbb{J}o\mathbb{J} = -\mathbb{I}$, we have \\

\n $\displaystyle{\overline{\Gamma(A)} = -\mathbb{J} (\Gamma(T^{*})^{\perp}) = \mathbb{J} (\Gamma(T^{*})^{\perp}) = \mathbb{J} (\Gamma(T^{*}))^{\perp}}$.\\

\n The set $\Gamma(T^{*})$ is closed as the orthogonal of the set $\mathbb{J} (\Gamma(T))$.\\

\n (iv) Assume that  $D(T^{*})$ is {\color{red}dense}. Then, we have \\

\n $\displaystyle{\Gamma((T^{*})^{*}) = \mathbb{J} (\Gamma(T)^{*})^{\perp} = \mathbb{J} ( \mathbb{J}(\Gamma(T)^{\perp})^{\perp} =  \overline{\Gamma(T)} =  \overline{\Gamma((T^{*})^{*})} }$.\\

\n In other words $\displaystyle{(T^{*})^{*} = T^{**}}$  is closed with graph $\overline{\Gamma (T)}$.  It follows that $T$ is closable, and more precisely $\overline{T} = T^{**}$.\\

\n Now, assume that $T$ is closable. Select $v \in D(T^{*})^{\perp}$ . We have \\

\n $\displaystyle{ \forall \, w \in D(T^{*}) , 0 = < v , w >  = < (0, v) , (-T^{*}w, w) >  }$  which means that \\

\n $\displaystyle{ (0, v) \in \mathbb{J} (\Gamma(T)^{*})^{\perp}  = \overline{\Gamma(T)} = \Gamma(\overline{T})}$  and therefore $v = T0 = 0$. By this way, we can recover $\displaystyle{(D(T^{*})^{\perp})^{\perp} = \{0\}^{\perp} = \mathcal{H} = \overline{D(T^{*})}}$ as expected.\hfill { } $\blacksquare$\\

\begin{definition}
({\color{red} Core of a linear operator})\\

\n Let $T$ be a closed operator in Hilbert space $\mathcal{H}$ with domain $D(T)$. A linear submanifold  {\bf D} of $D(T)$ is called a {\color{red} core} of $T$ if the set of elements 
$\{\varphi , T\varphi\}$ with $ \varphi \in$ {\bf D} is {\color{red}dense} in $\Gamma(T)$ (where $\Gamma(T)$ is the graph of $T$). For this it is necessary (but not sufficient in general) that {\bf D} is dense in $D(T)$.
\end{definition}

\n The following theorem is due to von Neumann :\\

\begin{theorem}
({\color{red} von Neumann})\\

\n  Let $\mathcal{H}_{1}$, $\mathcal{H}_{2}$ be Hilbert space. Let $\displaystyle{T: \mathcal{H}_{1} \longrightarrow \mathcal{H}_{2}, \varphi \longrightarrow T\varphi}$  be a closed and densely defined operator Then\\

\n (i)  $T^{*}T$ is a {\color{red}self-adjoint} operator in $\mathcal{H}_{1}$\\

\n and\\

\n (ii) $D(T^{*}T)$ is a {\color{red}core} of $T$.\\
\end{theorem}

\n {\bf {\color{red}Proof}}\\ (see {\color{blue}[22]}  p. 275)\\

\n From eq. $(2.3)$ and (i) of proposition $(2.4)$,  It is known that the graphs $\Gamma(T)$ and $\Gamma^{'}(-T^{*})$ are complementary subspaces of the Hilbert space $\displaystyle{\mathcal{H}_{1} \times \mathcal{H}_{2}}$. This implies that any vector  $\displaystyle{(u, u') \in \mathcal{H}_{1} \times \mathcal{H}_{2}}$, there exist a unique $v \in D(T)$ and $ v' \in D(T^{*})$ such that it can be expressed in the form \\

$\displaystyle{ (u, u') = (-Tv, v ) + (v', T^{*}v')}$ with $v \in D(T)$ and $ v' \in D(T^{*})$. \hfill { } {\color{blue}($\star_{1}$)}\\

\n That is\\

\n $\left \{ \begin{array} {c} u = v' - Tv\\
\quad\\
u' = v + T^{*}v'\\
\end{array} \right.$  \hfill { } {\color{blue}($\star_{2}$)}\\

\n Furthermore, taking the $\displaystyle{\mathcal{H}_{1} \times \mathcal{H}_{2} norm}$ in {\color{blue}($\star_{1}$)} gives \\

\n $\displaystyle{\vert\vert u \vert\vert^{2} + \vert\vert u' \vert\vert^{2}  = \vert\vert v \vert\vert^{2} + \vert\vert Tv \vert\vert^{2} + \vert\vert v' \vert\vert^{2} + \vert\vert T^{*}v' \vert\vert^{2}}$. \hfill { } {\color{blue}($\star_{3}$)}\\

\n Applying {\color{blue}($\star_{2}$)} to $u = 0$ shows that for every $u'$ there exists a $v \in D(T)$ and a $v' \in  D(T^{*}$ such that \\

\n $\left \{ \begin{array} {c}  v' = Tv \quad \quad\\
\quad\\
u' = v + T^{*}v'\\
\end{array} \right.$  \hfill { } {\color{blue}($\star_{4}$)}\\

\n This is to say that $v \in  D(T^{*}T)$ and $u' = (I + T^{*}T)v$\\

\n The relation ({\color{blue}($\star_{3}$)} gives  $\displaystyle{\vert\vert v \vert\vert \leq \vert\vert u' \vert\vert }$  in this case, because $\vert\vert u \vert\vert = 0$.\\

\n Hence the operator $I + T^{*}T$ is injective and  as  since $u' \in \mathcal{H}$ was arbitrary, it follows that  $\displaystyle{\mathcal{S} = I + T^{*}T}$ has  range $\mathcal{H}$ so$\displaystyle{\mathcal{S}^{-1} = (I + T^{*}T)^{-1}}$ is a bounded operator.\\

\n Now for all $u', u'' \in H_{1}$ we have, for some $u_{1} , u_{2} \in D(T^{*}T)$ \\

\n $\displaystyle{ <  (I + T^{*}T)^{-1}u' , u'' > =  < u_{1} + T^{*}Tu_{1} , u_{2} > =  <  u_{1}  ,  u_{2}  > + < T u_{1}  ,  Tu_{2} >}$  \hfill { } {\color{blue}($\star_{5}$)}\\

\n In particular $I + T^{*}T)^{-1}$ is self-adjoint, hence $ I + T^{*}T$ and $T^{*}T$ are self-adjoint.\\ 

\n (ii)  If $w \in D(T)$ is orthogonal to every $v \in D(T^{*}T)$ for the scalar product $<  , > _{T}$ then, for all $v \in D(T^{*}T)$ we have\\

\n $\displaystyle{ 0 = < v, w >_{T} := < v , w > + <Tv , Tw > = < v , w > + < T^{*}Tv , w > }$  \hfill { } {\color{blue}($\star_{5}$)}\\

\n That is, $w$ belongs to $\mathcal{I}m(I + T^{*}T)^{\perp}$ (where $\mathcal{I}m(I + T^{*}T)$ denotes the range of $I + T^{*}T$) , which is $\mathcal{H}_{1}^{\perp}$ because $I + T^{*}T$ is surjective as was noticed above. Hence $w = 0$ and the density is proved. \hfill { } $\blacksquare$\\

\n {\color{red}$\rhd$}  Now, we discuss here two important operators. Take $\mathcal{H} = L^{2}(\mathbb{R})$ (Here $L^{2}(\mathbb{R})$ is the Hilbert space of square-integrable functions on $\mathbb{R}$ ).\\

\n Let us introduce the following differential operators, acting on \\

\n $\displaystyle{D(A) = D(B) = \mathcal{S}(\mathbb{R}), A := \frac{1}{\sqrt{2}}(\frac{d}{dx} + x) \, and \,  B := \frac{1}{\sqrt{2}}(-\frac{d}{dx} + x)}$, where:. \\

\n $\displaystyle{ \mathcal{S}(\mathbb{R}) = \{f : \mathbb{R} \longrightarrow \mathbb{C};  f \in C_{0}^{k}(\mathbb{R}), \,\, \mbox{for all}  \,\ k  \,\, \mbox{and} \,\, l \in \mathbb{N} \}}$ such that \\

\n $\displaystyle{  < x >^{-l}C_{0}^{k}(\mathbb{R}) = \{f  \in  C_{0}^{k}(\mathbb{R}); f = < x >^{-l}g ; g \in C_{0}^{k}(\mathbb{R}) \,\, \mbox{for all}  \,\ k  \,\ \mbox{and}  \,\ l \in \mathbb{N} \}}$\\

\n with norm, $\displaystyle{\vert\vert f \vert\vert_{k,l} = \vert\vert g \vert\vert_{C^{k}}, g =  < x >^{l}f }$ and $\displaystyle{< x > = \sqrt{1 + \vert x \vert^{2}}}$ a function which is the size of $\vert x \vert$ for $\vert x \vert$ large, but has the virtue of being smooth.\\

\n The operators $B$ and $A$ are called {\color{red}Creation} and {\color{red}annihilation} operators.\\

\n We observe that the domains of their adjoints are \\

\n $\displaystyle{D(A^{*}) = \{ \psi \in L^{2}(\mathbb{R}) ; (-\frac{d}{dx} + x)\psi \in  L^{2}(\mathbb{R})\}}$ with $\displaystyle{\mathcal{S}(\mathbb{R}) \subset D(A^{*})}$ \\

\n and \\

\n $\displaystyle{D(B^{*}) = \{ \psi \in L^{2}(\mathbb{R}) ; (\frac{d}{dx} + x)\psi \in  L^{2}(\mathbb{R})\}}$ with  $\displaystyle{\mathcal{S}(\mathbb{R}) \subset D(B^{*})}$ .\\

\n  It follows that $D(A^{*})$ and $D(B^{*})$  are dense in $L^{2}(\mathbb{R})$.\\

\n By the above property (iv), the two operators $A$ and $B$ are {\color{red}closable} with $\overline{A} = A^{**}$ and $\overline{B} = B^{**}$. On the other hand \\

\n {\color{red}$\bullet$}  $\displaystyle{\forall \, \psi \in D(A^{*}) ,  A^{*}\psi = \frac{1}{\sqrt{2}}(- \frac{d}{dx} + x)\psi = B\psi}$.\\

\n {\color{red}$\bullet$}  $\displaystyle{\forall \, \psi \in D(B^{*}) ,  B^{*}\psi = \frac{1}{\sqrt{2}}( \frac{d}{dx} + x)\psi = A\psi}$.\\

\n In other words, we have $A \subset B^{*}$ and $B \subset A^{*}$ , whereas (by direct computations) we can check that $\Gamma(A^{*})$ and  $\Gamma(B^{*})$ are closed, so that $\overline{A} \subset B^{*}$ and $\overline{B} \subset A^{*}$.\\

\begin{proposition}

\n Let $\overline{A}$ and $\overline{B}$ be the colosure of $A$ and $B$ respectively, then we have \\

\n {\bf {\color{red}(i)}}  $\displaystyle{D(\overline{A}) = D(\overline{B}) = {\color{blue}\mathbb{D}} = \{ \psi \in H^{1}(\mathbb{R}) ; x \psi \in L^{2}(\mathbb{R})\}}$.\\

\n {\bf {\color{red}(ii)}}  $\overline{A}$ and $\overline{B}$ are {\color{red} adjoints of one another} and they share the same domain {\color{blue} $\mathbb{D}$}\\ 
\end{proposition}

\n {\color{red}{\bf Proof}}\\

\n For all $\varphi \in \mathcal{S}(\mathbb{R})$, we have \\

\n (i) $\displaystyle{2\mid\mid A \varphi \mid\mid^{2} = \mid\mid \varphi' \mid\mid^{2} + \mid\mid x \varphi \mid\mid^{2} - \mid\mid \varphi \mid\mid^{2}}$.\\

\n (ii) $\displaystyle{2\mid\mid B \varphi \mid\mid^{2} = \mid\mid \varphi' \mid\mid^{2} + \mid\mid x \varphi \mid\mid^{2} + \mid\mid \varphi \mid\mid^{2}}$.\\

\n Now, take $\varphi \in D(\overline{A})$. By definition, we have $\displaystyle{(\varphi , \overline{A}\varphi)  \in \overline{\Gamma(A)}}$. There exists $\varphi_{n} \in D(A)$ such that $(\varphi_{n})$ converges to $\varphi$ and $(A\varphi_{n})$ converges to $\overline{A}u$. We deduce that $(\varphi_{n}^{'})$ and $(x \varphi_{n})$ are Cauchy sequences in $L^{2}(\mathbb{R})$. By this way, we get that $(\varphi_{n}^{'}) \in  L^{2}(\mathbb{R})$ and  $(x \varphi_{n}) \in  L^{2}(\mathbb{R})$, and therefore 
$ D(\overline{A}) \subset {\color{blue}\mathbb{D}}$. We can proceed in the same way for $\overline{B}$.\\

\n We deal now with the reversed inclusion. Take $\varphi \in  {\color{blue}\mathbb{D}}$ . By using a classical result : \\

\n $\displaystyle{\mathcal{C}_{0}^{\infty}(\mathbb{R}) }$ is {\color{red} dense} in $({\color{blue}\mathbb{D}, \mid\mid . \mid\mid_{\mathbb{D}}})$,  there exists a sequence $(\varphi_{n})$ of {\color{red}smooth} functions with {\color{red}compact support} such that $\varphi_{n}$ converges to $\varphi$ in {\color{blue}$\mathbb{D}$}. In particular  $(A\varphi_{n})$ and  $(B\varphi_{n})$ converge in $L^{2}(\mathbb{R})$,  so that $(\varphi,  A\varphi) \in \overline{\Gamma(A)} = \Gamma(A)$ as well as   
$(\varphi,  B\varphi) \in \overline{\Gamma(B)} = \Gamma(B)$. This implies $\varphi \in D(\overline{A})$ and  $\varphi \in D(\overline{B})$.\\
\n Now to prove that the closures $\overline{A}$ and $\overline{B}$ of $A$ and $B$ are {\color{red} adjoints of one another} and they share the same domain {\color{blue} $\mathbb{D}$}, we consider :\\

\n  $\displaystyle{ \mathcal{V}_{\pm} = \{ \psi \in L^{2}(\mathbb{R}); (\pm \frac{d}{dx} + x )\psi \in  L^{2}(\mathbb{R})\} \subset  L^{2}(\mathbb{R})}$.\\

\n We let, for all  $\displaystyle{\varphi , \psi \in  \mathcal{V}_{\pm}}$,\\

\n $\displaystyle{Q_{\pm}(\varphi, \psi) = < \varphi, \psi >_{L^{2}(\mathbb{R})} + <  (\pm \frac{d}{dx} + x )\varphi ,  (\pm \frac{d}{dx} + x )\psi  >_{L^{2}(\mathbb{R})}}$.\\

\n Then we have \\

\n ($\alpha$) $\displaystyle{ (\mathcal{V}_{\pm} , Q_{\pm})}$ is a Hilbert space.\\

\n ($\beta$) For $\displaystyle{ f \in \mathcal{V}_{\pm}}$ the sequence $\displaystyle{ f_{n} = \chi_{n}(\rho_{n} \star f)}$ converges in $\mathcal{V}_{\pm}$.\\

\n where $(\rho_{n})$ is a sequence of smooth non-negative functions such that $\displaystyle{\int_{\mathbb{R}} \rho_{n}(x) dx = 1}$ with $supp \rho_{n} = B(0, \frac{1}{n})$ and $\chi_{n} (.) = \chi(\frac{1}{n})$ with $\chi$ is  a smooth function with compact support $0 \leq \chi \leq 1 $ equal to $1$ in a neighborhood of $ 0$.\\

\n For example, if $\psi \in D(B^{*})$,  we have $\psi \in L^{2}(\mathbb{R})$ and $\displaystyle{(\frac{d}{dx} + x )\psi \in L^{2}(\mathbb{R})}$. There exists $\psi_{n} \in \mathcal{S}(\mathbb{R})$ such that $\psi_{n}$ converges to $\psi$ and $\displaystyle{(\frac{d}{dx} + x )\psi_{n}}$ converges to $\displaystyle{(\frac{d}{dx} + x )\psi \in L^{2}(\mathbb{R})}$. Using (ii) we get that $(\psi_{n})$ and $(x\psi_{n})$ are Cauchy sequences in $ L^{2}(\mathbb{R})$ with limit $\psi'$ and $x\psi$. Thus $\psi \in {\color{blue}\mathbb{D}}$.\\

\n From the inclusion $\overline{A} \subset B^{*}$, we deduce that\\

\n  $\displaystyle{D(B^{*}) \subset {\color{blue}\mathbb{D}} = D(\overline{A} ) \subset D(B^{*})}$.\\

\n We can deal with $A$ in the same way to obtain \\

\n  $\displaystyle{D(B^{*}) = D(\overline{A}) = {\color{blue}\mathbb{D}} = D(\overline{B} ) \subset D(A^{*})}$.\\

\n We deduce that\\

\n {\color{red}$\bullet$ } $\displaystyle{\overline{A} = B^{*} , \overline{B} = A^{*}}$\\

\n {\color{red}$\bullet$}  $\displaystyle{\overline{A}^{*} = A^{*} = \overline{B},  \overline{B}^{*} = B^{*} = \overline{A}}$. \hfill { } $\blacksquare$\\

\n \subsection{{\color{blue} Determination of eigenvalues and eigenfunctions of Hermite's operator}}

\n Let $\mathcal{H} = L^{2}(\mathbb{R})$, $\displaystyle{\mathcal{V} = \{u \in H^{1}(\mathbb{R}); \sqrt{1 + x^{2}}u \in L^{2}(\mathbb{R})\}}$ \\

\n and \\

\n $\displaystyle{a(u, v) = \int_{-\infty}^{+\infty}u'(x)\overline{v'(x)}dx +  \int_{-\infty}^{+\infty}(1 + x^{2})u(x)\overline{v(x)}dx }$. \\

\n Then the operator associated to this form  called Hermite's operator is defined by:\\

\n  $Au = - u'' + (1 + x^{2})u $ with domain $\displaystyle{D(A) = \{ u \in H^{2}(\mathbb{R});  (1 + x^{2})u \in L^{2}(\mathbb{R})\}}$.\\

\n Note that one clearly has $\displaystyle{\mathcal{S}(\mathbb{R}) \subset D(A) \subset H^{2}(\mathbb{R})}$, whence $A$ is densely defined. It is also a closed operator. To see this, we have to show that \\

\n $\displaystyle{ (u_{j} \in D(A) \longrightarrow u  \, and \, Au_{j}  \longrightarrow v, as \, j \longrightarrow \infty)  \Longrightarrow u \in D(A) \, and \, Au = v}$\\

\n Since $\displaystyle{u_{j}   \longrightarrow u \, (\, in \, L^{2}(\mathbb{R}))\,  \Longrightarrow u_{j}   \longrightarrow u \, (\, in \, \mathcal{D}'(\mathbb{R})) \, as  \, j \longrightarrow \infty}$.\\

\n we have on the one hand  $\displaystyle{ Au_{j}   \longrightarrow Au = v \, (\, in \, \mathcal{D}'(\mathbb{R})) \, as  \, j \longrightarrow \infty}$.\\

\n Since  $v \in L^{2}(\mathbb{R})$ then $u \in  D(H)$ and $Au = v$, which is what we wanted to prove.\\

\n Now, we think of $A$ as an unbounded operator on $L^{2}(\mathbb{R})$ with dense domain $D(A)$. Then we have the following facts.\\

\n {\color{red}$\bullet$} The injection of $D(A)$ in $L^{2}(\mathbb{R})$ is compact .\\

\n In fact, it suffices to use the  theorems of compact embeddings of Sobolev spaces in $L^{2}(\mathbb{R})$.\\

\n {\color{red}$\bullet$} Self-adjointness: $A = A^{*}$.\\

\n {\color{red}$\bullet$} $\displaystyle{\sigma(A) = \{2(n+1); n \in \mathbb{N}\}}$ with multiplicity $1$.\\

\n {\color{red}$\bullet$}  Define $\displaystyle{h_{0} = e^{-\frac{x^{2}}{2}}}$ and  $\displaystyle{h_{n} = a_{+}^{n}(h_{0}), n \in \mathbb{N}}$  where  $\displaystyle{a_{+} = -\frac{d}{dx} + x}$. Then $\displaystyle{a_{-} (h_{0}) = 0}$ where $\displaystyle{a_{-} = \frac{d}{dx} + x}$, all the functions $h_{n} $ belong to $\mathcal{S}(\mathbb{R})$ and are of the form   $\displaystyle{h_{n} = H_{n} e^{-\frac{x^{2}}{2}}}$, where $\displaystyle{ H_{n} \in \mathbb{R}[X]}$ is a polynomial of degree $n$ (i.e., of the form $\displaystyle{a_{n}x^{n} + a_{n-1}x^{n-1} + .... + a_{1}x + a_{0}}$, with non-zero leading coefficient $a_{n}$). The polynomials  $H_{n}$  are called {\color{red}Hermite polynomials}. Moreover, the $h_{n}$ form a complete orthogonal system of $L^{2}(\mathbb{R})$.\\

\n Hence, upon defining $\displaystyle{ u_{n } =  \frac{h_{n}}{\mid\mid h_{n} \mid\mid}}$, the system $\displaystyle{\{u_{n}\}_{n \in \mathbb{N}} }$  is an orthonormal
basis of $L^{2}(\mathbb{R})$. The functions $h_{n}$  are called {\color{red}Hermite functions}.\\

\n We have, by a direct computation that $\displaystyle{a_{+}a_{-} = A  - 2I}$, $\displaystyle{a_{+}a_{-} = A} $ and $\displaystyle{[a_{+}, a_{-}] = - 2I}$\\

\n The operators $a_{\pm}$ are the celebrated {\color{red}creation} ($a_{+}$) and {\color{red}annihilation} ($a_{-}$) operators which are adjoint  of aech  other (see the above subsection)\\

\n Note, in the first place, that for any given $u_{1},u_{2} \in  \mathcal{S}(\mathbb{R})$ we have \\

\n $\displaystyle{< Au_{1}, u_{2} > = < u_{1}, Au_{2} >}$  (i.e. $A$ is symmetric).\\

\n and\\

\n $\displaystyle{< Au, u > = < ( a_{+}a_{-} + 2I)u , u > = \mid\mid a_{-}u \mid\mid^{2} + 2\mid\mid u \mid\mid^{2} , \, \forall \, u \in \mathcal{S}(\mathbb{R})}$.\\

\n On the other hand, since $a_{-}h_{0} = 0$ and $h_{0} \in  \mathcal{S}(\mathbb{R})$, we indeed have that\\

\n $\displaystyle{min_{u \in \mathcal{S}(\mathbb{R})-\{0\}} \frac{< Au, u > }{\mid\mid u \mid\mid^{2}}}$.\\

\n Next, it is clear that the functions $\displaystyle{h_{n} = a_{+}^{n}h_{0} \in \mathcal{S}(\mathbb{R})\, \forall \, n \in \mathbb{N}}$. Hence we may
compute by induction :\\

\n $\displaystyle{Ah_{0} = 2h_{0}}$\\

\n $\displaystyle{Ah_{n} = ( a_{+}a_{-} + 2I)u_{n} = ( a_{+}a_{-} + 2I) a_{+}h_{n-1} = a_{+} Ah_{n-1}+ 2a_{+}h_{n-1}}$\\

\n $\displaystyle{ = (\lambda_{n-1} + 2)h_{n} = \lambda_{n} h_{n} }$.\\

\n This implies $\displaystyle{\lambda_{n} = \lambda_{n-1} + 2 }$  with $\lambda_{0} = 2$,  it follows that $\displaystyle{Ah_{n} = 2(n + 1)h_{n}}$.\\

\n Of course, we have to make sure that $h_{n} \neq 0$ for all $n$. Hence, we compute by induction:\\

\n $\displaystyle{\mid\mid h_{n} \mid\mid^{2} = < h_{n}, h_{n} > =  < a_{+}^{n}h_{0}, a_{+}^{n}h_{0} >  =  < a_{-}^{n}a_{+}^{n}h_{0}, h_{0} > }$\\

\n $\displaystyle{ = < a_{-}^{n-1}[ a_{-}a_{+}]a_{+}^{n-1}h_{0}, h_{0} >  }$ \\

\n $\displaystyle{ = 2(n-1)< a_{-}^{n-1}a_{+}^{n-1}h_{0}, h_{0} > = 2(n-1)< a_{+}^{n-1}h_{0}, a_{+}^{n-1}h_{0} > }$\\

\n $\displaystyle{= 2(n-1)\mid\mid h_{n-1} \mid\mid^{2} = 2^{n}(n-1)!\mid\mid h_{0} \mid\mid^{2}}$.\\ 

\n whence $h_{n} \neq 0$ for all $n \in \mathbb{N}$.\\

\n Now, since $a_{-}h_{0} = 0$  a similar computation also shows that \\

\n  $\displaystyle{ < h_{n} , h_{m} > = 0}$,\\

\n that yields the orthogonality of the system $\{h_{n}\}_{n \in \mathbb{N}}$. To prove its completeness, we note in the first place that it is clear that $h_{n} = H_{n}h_{0}$, for a real polynomial $H_{n}$ of degree $n$ (i.e., $H_{n}(x) = a_{n}x^{n} + a_{n-1}x^{n-1}+...+ a_{0}$, with the $a_{j } \in \mathbb{R}$  and non-zero leading coefficient $a_{n}$). \\ 

\n Hence, let $g \in L^{2}(\mathbb{R})$ be such that $\displaystyle{ \int_{ \infty}^{+\infty}g(x)h_{n}(x)dx = 0 \, \forall \, n \in \mathbb{N}}$.\\ 

\n It then follows that, since any {polynomial } $P \in  \mathcal{R}[x] $ can be written as a linear combination of the $H_{n}$ (by virtue of the fact that their leading coefficients are not zero), we have \\

\n $\displaystyle{ \int_{- \infty}^{+\infty} g(x)P(x)e^{-\frac{ x ^{2}}{2}}dx = 0 \quad \forall \, P \in \mathbb{R}[X]}$. \hfill { } {\bf {\color{blue}(*)}}\\ 

\n Using  $\displaystyle{e^{-ix\xi} = \sum_{j \geq 0}^{} \frac{(-ix\xi)^{j} }{j!}}$ and  {\bf {\color{blue}(*)}} to get \\

 \n $\displaystyle{ \int_{ -\infty}^{+\infty} e^{-ix\xi}g(x)e^{-\frac{ x ^{2}}{2}}(x)dx = \mathfrak{F}_{x \longrightarrow \xi}(ge^{-\frac{ x ^{2}}{2}})(\xi) = 0 , \, \forall \, \xi \in \mathbb{R}}$.\\
 
\n But an $L^{2}(\mathbb{R})$-function which has zero Fourier-transform must be zero, that is\\

\n  $\displaystyle{ ge^{-\frac{ x ^{2}}{2}} = 0 \Longrightarrow g = 0 \, in \, L^{2}(\mathbb{R})}$.\\

\n Hence $\{h_{n}\}_{n \in \mathbb{N}}$ is an orthogonal basis of $L^{2}(\mathbb{R})$ and $\{u_{n}\}_{n \in \mathbb{N}}$, where $\displaystyle{u_{n} := \frac{h_{n}}{\mid\mid h_{n} \mid\mid}}$, is an orthonormal basis of $L^{2}(\mathbb{R})$.\\

\n Using this, one then sees that $\displaystyle{\mathcal{R}(H \pm iI) = L^{2}(\mathbb{R})}$.\\

\n whence, by general arguments see, for instance, {\bf {\color{blue}[11]}} or  {\bf {\color{blue}[22]}} \\
 
 \n  $A = A^{*}$. \\
 
 \n This therefore shows that $\displaystyle{\sigma(A) = \{ 2(n+1) , n \in \mathbb{N}}$ with multiplicity $1$, and concludes the determination of eigenvalues and eigenfunctions of Hermite's operator. \hfill { } $\blacksquare$\\

\subsection{{\color{blue} Elementary properties of Bargmann space}}

 We denote $\displaystyle{L_{2}(\mathbb{C}, d\mu(z))}$  the Hilbert space of complex measurable functions $f(z, \overline{z})$ on the complex plane $\mathbb{C}$, not necessarily analytical, which are square integrable with respect to the Gaussian measure \\
  \begin{equation}
 \displaystyle{d\mu(z) = \frac{1}{\pi}e^{-\vert z \vert^{2}}dzd\overline{z}; z = x +i y, i^{2} = - 1 , (x, y) \in \mathbb{R}^{2}}.
 \end{equation}
 [being $dzd\overline{z}$ the Lebesgue measure on $\mathbb{C}$ and $\displaystyle{\overline{z} = x - i y) , (x, y) \in \mathbb{R}^{2}}$].\\

We recall that the Bargmann space {\color{blue}[4]}, denoted by \\
\begin{equation}
\displaystyle{\mathbb{B} = \{ \varphi : \mathbb{C} \longrightarrow \mathbb{C}\,\, \mbox{ analytic}\,\, ; \int_{\mathbb{C}} \vert \varphi(z) \vert^{2} e^{-\vert z \vert^{2}}dxdy < \infty \}}
\end{equation}

\n is the subspace of all entire functions in $\displaystyle{L_{2}(\mathbb{C}, d\mu(z))}$ with the inner product (1.2) \\

\begin{lemma}
 The Bargmann space $\mathbb{B} $ is injected continuously into the space of distributions   $\mathfrak{D}'(\mathbb{R}^{2})$
 \end{lemma}
 
 \n {\bf {\color{red} Proof}}\\
 
 \n  Let $\varphi(x, y) \in \mathfrak{D}(\mathbb{R}^{2})$ and $\psi(x, y) \in \mathfrak{D}(\mathbb{R}^{2})$ where $\mathfrak{D}(\mathbb{R}^{2})$ is the space of indefinitely differentiable functions of compact support.\\

\n Then we have:\\

\n $\displaystyle{\int_{\mathbb{R}}\int_{\mathbb{R}}\phi(x, y)\bar{\psi}(x, y)dxdy} = \displaystyle{\int_{\mathbb{R}}\int_{\mathbb{R}}e^{\frac{-(x^{2}+y^{2})}{2}}\phi(x, y)e^{\frac{x^{2}+y^{2}}{2}}\bar{\psi}(x, y)dxdy }.$\\

\n and apply Schwarz's inequality to obtain the following inequality :\\

\n $\displaystyle{\int_{\mathbb{R}}\int_{\mathbb{R}}\phi(x, y)\bar{\psi}(x, y)dxdy} \leq \mid\mid \phi \mid\mid_{E}\mid\mid e^{\frac{x^{2}+y^{2}}{2}}\psi \mid\mid_{L^{2}(I\!\!R^{2})}$.\\

\n From this inequality we deduce that $\mathbb{B} $ is injected continuously into the space of distributions   $\mathfrak{D}'(\mathbb{R}^{2})$.\hfill { } $\blacksquare$\\

\begin{lemma}
(i) Any Cauchy sequence in $\mathbb{B}$ is also Cauchy in the space of continuous functions $\displaystyle{\mathcal{C}(\mathbb{C})}$ endowed with the topology of compact
convergence.\\

\n (ii) The Bargmann space is a Hilbert space.
\end{lemma}

\n {\bf{\color{red}Proof}}\\

(i) In this topology a sequence of functions $\{\varphi_{n}\}_{n=1}^{\infty}$ converges if and only if for every compact set $K \subset \mathbb{C}$ the sequence converges uniformly on $K$.\\
The goal is to find for any $K$ as above an estimate of the form \\

\begin{equation}
\displaystyle{ \underset{z \in K}{sup}  \vert \varphi_{n} - \varphi_{m} \vert \leq C_{K} \vert\vert \varphi_{n} - \varphi_{m} \vert\vert_{B}}
\end{equation}

\n where $C_{K}$ is a constant depending on the compact set $K$. Once we have this, we find for each $K$ a uniform bound and hence the sequence f $\{\varphi_{n}\}_{n=1}^{\infty}$ is Cauchy in $\displaystyle{\mathcal{C}(\mathbb{C})}$ if it is Cauchy in $\mathbb{B}$.\\

\n Let $K \subset \mathbb{C}$  be compact. To simplify some computations choose $r > 0 $ such that $K$ is contained in $B_{r}$, the ball with radius $r$ around the origin. Then pick any smooth function with compact support $\displaystyle{\chi \in \mathcal{C}_{c}^{\infty}(\mathbb{C}}$ such that $\displaystyle{\chi(z)  = 1}$ for all $\displaystyle{z \in B_{r+1}}$. Here and in the following, $B_{r}$ denotes the sphere of radius  $r$. \\

\n Set $\displaystyle{U = supp\,\, \chi - B_{r+1}}$. This is arranged so that $U$ is a compact set with distance one to $B_{r}$ and $\chi$ vanishes on all parts of the boundary of $U$ except on $\partial B_{r+1}$, where it is one. Since $K\subset B_{r}$  we have \\
\begin{equation}
\displaystyle{ \underset{z \in K}{sup}  \vert \varphi_{n} - \varphi_{m} \vert \leq    \underset{z \in B_{r}}{sup}  \vert \varphi_{n} - \varphi_{m} \vert}
\end{equation}

\n By the Bochner-Martinelli generalization of Cauchy's integral theorem and Stokes? theorem, we find that the norm in Lebesgue measure $dz$
is equivalent to the norm in Gauss measure $\displaystyle{d\mu(z) = e^{-\vert z \vert^{2}}dz}$. Thus\\
\begin{equation}
\displaystyle{ \vert \vert \varphi_{n} - \varphi_{m} \vert\vert_{L_{2}(U)} \leq c \vert \vert \varphi_{n} - \varphi_{m} \vert\vert_{L_{2}(U, d\mu)} \leq c \vert \vert \varphi_{n} - \varphi_{m} \vert\vert_{\mathbb{B}}}
\end{equation}
\n for some constant c depending on $U$.\\
\n Putting all this together we arrive at the desired estimate (2.6). Notice that the constant depends on $r$ and on $U$, but it is readily seen that $r$ and $U$ depend on $K$ only.\\

\n (ii)  Let $\{\varphi_{n}\}_{n=1}^{\infty}$ be a Cauchy sequence in $\mathbb{B}$. Then by the above result it is a Cauchy sequence in $\displaystyle{\mathcal{C}(\mathbb{C})}$ with the topology of compact convergence. Since this space is complete, there exists a limit function $\displaystyle{\varphi \in \mathcal{C}(\mathbb{C})}$. By a theorem of Weierstrass (see for example {\color{blue}[24], Ch.V]}) this function is holomorphic.\\

The sequence $\{\varphi_{n}\}_{n=1}^{\infty}$  is Cauchy in $\mathbb{B}$, so that the $\vert\vert \varphi_{n}\vert\vert_{\mathbb{B}}$  are bounded by a constant $C$. From the uniform convergence on compacta it follows that for any compact set $K \subset \mathbb{C}$\\

\begin{equation}
\displaystyle{\int_{K} \vert \varphi(z) \vert^{2} d\mu(z) = \lim\limits_{n \longrightarrow +\infty} \int_{K} \vert \varphi_{n}(z) \vert^{2} d\mu(z) \leq \vert\vert \varphi_{n} \vert\vert_{\mathbb{B}}^{2} \leq C^{2}}
\end{equation}
and we see that $\varphi$ is square integrable on every compact subset. By monotone convergence we find that $\vert\vert \varphi \vert\vert \leq C$, so $\varphi \in \mathbb{B}$.. Using the triangle inequality we see that $\displaystyle{\vert\vert \varphi - \varphi_{n} \vert\vert_{\mathbb{B}} \leq 2C \,\, \mbox{for} n= 1, 2, ...} $. Thus we may apply Fubini theorem to obtain\\
\begin{equation}
\begin{array} {c}\displaystyle{\lim\limits_{n \longrightarrow + \infty} \vert\vert \varphi - \varphi_{n} \vert\vert_{\mathbb{B}} =  \lim\limits_{n \longrightarrow + \infty} \underset{K}{sup}\int_{K} \vert \varphi(z) - \varphi_{n}(z) \vert^{2} d\mu(z)} \\
\quad\\
\displaystyle{=  \underset{K}{sup}\int_{K} \lim\limits_{n \longrightarrow + \infty} \vert \varphi(z) - \varphi_{n}(z) \vert^{2} d\mu(z)}
\end{array}
\end{equation}
which is zero because the $\varphi_{n}$ converge uniformly to $\varphi$ on each compact set $K$. We conclude that \\

\begin{equation}
\displaystyle{\vert\vert \varphi - \varphi_{n} \vert\vert_{\mathbb{B}} \longrightarrow 0 \,\, \mbox{as} \,\, n \longrightarrow + \infty, \,\, \mbox{so} \,\, \varphi_{n} \longrightarrow \varphi \,\, \mbox{in} \,\, \mathbb{B} }
\end{equation}
This completes the proof.\hfill { } $\blacksquare$\\

\begin{definition}
Recall that a linear operator $T$ on a Hilbert space $\mathcal{H}$ with domain $\displaystyle{D(T) \subset \mathcal{H}}$ is said to be well defined
if its domain is dense in this space. 
 \end{definition}

 \n  As a holomorphic function $\varphi$ can be expressed as a power series:\\
 
 \begin{equation}
 \displaystyle{\varphi(z) = \sum_{n=0}^{\infty}a_{n}z^{n}}
 \end{equation}
 
 \n We will show that it is possible to rewrite the inner product on $\mathbb{B}$  in terms of the expansion coefficients $a_{n}$.\\
 
 \n To do this we will calculate (1.2) for $\psi = \varphi$ using polar coordinates $z = re^{i\theta}$;  $r \in [0, +\infty[$ et $\theta \in [0, 2\pi]$ and the power series expression of $\varphi$. We then get \\
 
  \begin{lemma}
  \begin{equation}
 \displaystyle{\vert\vert \varphi\vert\vert_{\mathbb{B}}^{2} = \pi \sum_{n=0}^{\infty} n! \vert a_{n}\vert^{2}}
 \end{equation}
 \end{lemma}

 \n {\bf{\color{red}Proof}}\\
 
\n  Let $\varphi \in \mathbb{B}$ and if we put  $z = re^{i\theta}$;  $r \in [0, +\infty[$ and $\theta \in [0, 2\pi]$ then \\
$\displaystyle{\varphi(z) = \sum_{n=0}^{\infty}a_{n}z^{n} = \sum_{n=0}^{\infty}a_{n}r^{n}e^{in\theta}}$ and we can write: \\

\n $\displaystyle{\vert\vert \varphi \vert\vert_{\mathbb{B}}^{2} = \int_{\mathbb{C}}e^{-\mid z\mid^{2}}\mid\varphi(z)\mid^{2}dzd\bar{z}} =\displaystyle{\lim\limits_{\sigma \longrightarrow \infty} \int_{0}^{\sigma}\int_{0}^{2\pi}e^{-r^{2}}\mid \phi(re^{i\theta})\mid^{2}rdrd\theta}$\\

\n and\\

\n $\mid \varphi(re^{i\theta})\mid^{2} = \displaystyle{\sum_{n=0}^{+\infty}\displaystyle{\sum_{p+q =n}a_{p}\bar{a}_{q}r^{p+q}e^{i(p-q)\theta}}}$\\

\n By Beppo-Levi and Fubini theorems, we obtain that:\\

\n $\displaystyle{\int_{\mathbb{C}}e^{-\mid z\mid^{2}}\mid\varphi(z)\mid^{2}dzd\bar{z}} = \displaystyle{\lim\limits_{\sigma \longrightarrow \infty} \int_{0}^{\sigma}e^{-r^{2}}}\displaystyle{\sum_{n=0}^{+\infty}}\displaystyle{\sum_{p+q =n}a_{p}\bar{a}_{q}r^{p+q+1} \int_{0}^{2\pi}e^{i(p-q)\theta}drd\theta}$\\

\n As $\displaystyle{\int_{0}^{2\pi}e^{i(p-q)\theta}d\theta} = 2\pi\delta_{pq}$ where $\delta_{pq}$ is Kronoecker's symbol then we get:\\

\n $\displaystyle{\int_{\mathbb{C}}e^{-\mid z\mid^{2}}\mid\varphi(z)\mid^{2}dzd\bar{z}} = \displaystyle{\lim\limits_{\sigma \longrightarrow \infty} \int_{0}^{\sigma}e^{-r^{2}}} \displaystyle{\sum_{n=0}^{+\infty}} \displaystyle{\sum_{2p =n}a_{p}\bar{a}_{p}r^{2p+1}   2\pi dr}$\\

\n $= 2\pi\displaystyle{\lim\limits_{\sigma \longrightarrow \infty} \int_{0}^{\sigma}e^{-r^{2}}}\displaystyle{\sum_{k=0}^{+\infty}a_{k}\bar{a}_{k}r^{2k+1}dr}$\\

\n Again by applying Beppo-Levi's theorem, we obtain:\\

\n $\displaystyle{\int_{\mathbb{C}}e^{-\mid z\mid^{2}}\mid\varphi(z)\mid^{2}dzd\bar{z}} = \displaystyle{\sum_{k=0}^{+\infty}\mid a_{k}\mid^{2}}\displaystyle{\lim\limits_{\sigma \longrightarrow \infty} \int_{0}^{\sigma}e^{-r^{2}}r^{2k+1}2\pi dr}$\\

\n But \\

\n $\displaystyle{\lim\limits_{\sigma \longrightarrow \infty} \int_{0}^{\sigma}e^{-r^{2}}r^{2k+1}2\pi dr = \pi k!}$\\

\n Therefore:\\

\n $\displaystyle{\int_{\mathbb{C}}e^{-\mid z\mid^{2}}\mid\phi(z)\mid^{2}dzd\bar{z}} = \pi\displaystyle{\sum_{k=0}^{+\infty}k! \mid a_{k}\mid^{2} < +\infty}$\\

\n from which we conclude that the condition for a function $\varphi$ to be an element of $\mathbb{B}$ is given by \\

\begin{equation}
\displaystyle{\sum_{n=0}^{\infty}n! \vert a_{n} \vert^{2} < \infty}
\end{equation}

\n Using the linearity of the inner product:\\

\n $\displaystyle{ < \varphi + \psi , \varphi + \psi >_{\mathbb{B}}=  < \varphi , \varphi >_{\mathbb{B}} +  < \varphi , \psi >_{\mathbb{B}} +  < \psi , \varphi >_{\mathbb{B}} +  < \psi , \psi >_{\mathbb{B}}}$\\

\n For $\displaystyle{\varphi(z) = \sum_{n=0}^{\infty}a_{n}z^{n}}$ and   $\displaystyle{\psi(z) = \sum_{n=0}^{\infty}b_{n}z^{n}}$, we also find\\
\begin{equation}
\n \displaystyle{ < \varphi , \psi >_{\mathbb{B}} = \pi \sum_{n=0}^{\infty} n! a_{n}\overline{b_{n}}}
\end{equation}
\par \hfill { } $\blacksquare$\\
\n We also find\\

\n \begin{lemma} 

{\color{red}(Complete orthonormal system in $\mathbb{B}$)}

\n Let $\displaystyle{e_{n}(z) = \frac{z^{n}}{\sqrt{n!}}, n \in \mathbb{N}}$ then \\

\n The system $\displaystyle{\{ e_{n}(z) ; n \in \mathbb{N} \}}$ forms a complete orthonormal system in $\mathbb{B}$
\end{lemma}

\n {\bf{\color{red}Proof}}\\

\n {\color{red}$\bullet$} $\displaystyle{< z^{n} , z^{m} > = \frac{1}{\pi}\int_{\mathbb{C}} z^{n}\overline{z}^{m} e^{-\vert z \vert^{2}}dxdy}$\\
$\displaystyle{ = \frac{1}{2\pi} \int_{0}^{2\pi} e^{i(n - m)\theta}d\theta \int_{0}^{\infty} r^{n + m}e^{-r^{2}} 2r dr}$\\
$\displaystyle{= \delta_{n,m}\int_{0}^{\infty}t^{n}e^{-t}dt}$\\
$\displaystyle{= n! \delta_{n,m}}$.\\

\n In particular\\

\n {\color{red}$\bullet$} $\displaystyle{ < \frac{z^{n}}{\sqrt{n!}} , \frac{z^{m}}{\sqrt{m!}} > =  \delta_{n,m} }$\\

\n This gives that the system  $\displaystyle{\{ e_{n}(z) ; n \in \mathbb{N} \}}$ is orthonormal with the norm of each $e_{n}$ given by $\displaystyle{\vert\vert e_{n} \vert\vert_{\mathbb{B}} = 1}$.\\

\n In this computation the substitutions $z = re^{i\theta}$ with $\theta \in [0, 2\pi]$, $r \in [0, \infty[$ and $t = r^{2}$ have been made.\\

\n {\color{red}$\bullet$} Let $\varphi \in \mathbb{B}$. Then $\varphi$ is holomorphic and we can write it as a power series $\displaystyle{\varphi(z) = \sum_{n=0}^{\infty}a_{n}z^{n}}$, converging uniformly on compact subsets of $\mathbb{C}$.\\

\n {\color{blue}$\bullet$}  Suppose that $\displaystyle{< z^{m}, \varphi > = 0}$ for all $m \in \mathbb{N}$\\

\n Using first dominated convergence and then the uniform convergence on compact sets, we find that\\

\n $\displaystyle{< z^{m} , \varphi > = \frac{1}{\pi}\int_{\mathbb{C}} z^{m}  \overline{ \varphi(z)} e^{-\vert z \vert^{2}} dxdy}$

\n $\displaystyle{ = \frac{1}{\pi}\int_{\mathbb{C}} z^{m}  \sum_{n=0}^{\infty} \overline{a}_{n} \, \overline{z}^{n} dxdy}$

\n $\displaystyle{ = \frac{1}{2\pi} \lim\limits_{r \longrightarrow + \infty} \int_{0}^{r}\int_{0}^{2\pi}\sum_{n=0}^{\infty}\overline{a}_{n}e^{i(m - n)\theta} r^{m+ n}e^{-r^{2}}2 rdr}$\\

\n $\displaystyle{ =  \lim\limits_{r \longrightarrow + \infty} \sum_{n=0}^{\infty}\overline{a}_{n}\delta_{n,m}\int_{0}^{\infty} t^{n}e^{-t}dt}$\\

\n $\displaystyle{ = m! \overline{a}_{m}}$.\\

\n and hence $a_{m} = 0$ for all $m \in \mathbb{N}$, giving $\varphi = 0$. This shows the completeness of the system $\displaystyle{\{ e_{n}(z) ; n \in \mathbb{N} \}}$.\hfill { } $\blacksquare$\\

\begin{corollary}

\n The set of polynomials $\mathcal{P}$ is dense in $\mathbb{B}$.
\end{corollary}

\n Now we define :\\

\begin{definition}
\begin{equation}
\displaystyle{\mathbb{B}_{S} = \{ (a_{n})_{n=0}^{\infty} \in \mathbb{C} ; \displaystyle{\sum_{n=0}^{\infty}n! \vert a_{n} \vert^{2} < \infty \}}}
\end{equation}
equipped with inner product\\
\begin{equation}
\displaystyle{ < (a_{n})_{n=0}^{\infty}  ,  (b_{n})_{n=0}^{\infty} > _{\mathbb{B}_{S}} = \pi \sum_{n=0}^{\infty} n! a_{n}\overline{b_{n}}}
\end{equation}
\end{definition}
\begin{proposition}
\n $\mathbb{B}$ is related to $\mathbb{B}_{S}$ by an unitary transform of $\mathbb{B}$ onto $\mathbb{B}_{S}$,  i.e. an
isomorphism between $\mathbb{B}$ and $\mathbb{B}_{S}$ that preserves the inner product, given by the following transform {\bf I}:\\
\begin{equation}
\displaystyle{ I = \mathbb{B} \longrightarrow \mathbb{B}_{S}, \varphi \longrightarrow I(\varphi) = (\frac{1}{n!}\varphi^{(n)}(0))_{n=0}^{\infty} = (a_{n})_{n=0}^{\infty}}
\end{equation}
\end{proposition}

\n{\bf{\color{red}Proof}}\\

\n It follows from (2.9) that \\
\begin{equation}
 \mid\mid \phi \mid\mid_{\mathbb{B}} = \mid\mid (\frac{1}{n!}\varphi^{(n)}(0))_{n\in \mathbb{N}}\mid\mid_{\mathbb{B}_{S}}\\
\end{equation}
and from (2.14), we deduce that $I$ is continuous and  if $I(\varphi) = 0 $ then  $\varphi  = 0$, i.e. $I$ is injective.\\

\n  Now, as $I$ is continuous and $\mathcal{R}^{\perp}(I) = \{0\}$ where $\mathcal{R}(I)$ denotes the range of $I$ and $\mathcal{R}^{\perp}(I)$ is its orthogonal, then $I$ is a surjective map between  $\mathbb{B}$  and $\mathbb{B}_{S}$, it follows that $I$ is an isomorphism.\hfill { } $\blacksquare$\\

\begin{proposition}

\n (1)  For all $\varphi \in \mathbb{B}$, we have the following inequalities:\\

\n $(i) \mid \varphi(z) \mid \leq e^{\frac{\mid z\mid^{2}}{2}}\mid\mid \varphi \mid\mid_{ \mathbb{B}},$ for all $z \in \mathbb{C}$ \\

\n  $(ii) \mid \varphi^{'}(z) \mid \leq (1 + \mid z\mid^{2})^{\frac{1}{2}}e^{\frac{\mid z\mid^{2}}{2}}\mid\mid \varphi \mid\mid_{ \mathbb{B}}$,  for all $z \in \mathbb{C}$ \\

\n $(iii) \mid \varphi^{''}(z) \mid \leq (\alpha_{0} + \alpha_{1}\mid z\mid^{2} +  \alpha_{2}\mid z\mid^{4})^{\frac{1}{2}}e^{\frac{\mid z\mid^{2}}{2}}\mid\mid \varphi \mid\mid_{\mathbb{B}}$,  for all $z \in \mathbb{C}$ and $\alpha_{i} > 0 , i = 0, 1, 2$. \\

\n (2) $\mathbb{B}$ is a closed subspace of $L_{2}(\mathbb{C}, d\mu(z))$.\\
\end{proposition}

\n {\bf {\color{red}Proof}}\\

{\color{blue}$\bullet$} (1) \\

\n (i) Let $\varphi \in \mathbb{B}$ then by using Cauchy-Shwarz inequality we deduce that:\\

\n $\displaystyle{\mid \varphi(z) \mid \leq \displaystyle{\sum_{k=0}^{+\infty}\mid a_{k}z^{k}\mid}}$ $\displaystyle{= \sum_{k=0}^{+\infty}\mid \sqrt{k!}\,\,a_{k}\frac{z^{k}}{\sqrt{k!}}\mid}$\\

$\leq (\displaystyle{\sum_{k=0}^{+\infty}k!\mid a_{k}\mid^{2})^{\frac{1}{2}}}$ $(\displaystyle{\sum_{k=0}^{+\infty}\frac{\mid z\mid^{2k}}{k!})^{\frac{1}{2}}}$ $\leq \displaystyle{e^{\frac{\mid z\mid^{2}}{2}}\mid\mid \varphi \mid\mid_{\mathbb{B}}}.$\\

 \n {\color{red}$\bullet$} This inequality (i) says that pointwise evaluation is continuous, i.e.  for each  $z  \in \mathbb{C}$, the map that takes a function $\varphi \in \mathbb{B}$ to the number $\varphi(z)$ is a continuous linear functional on $\mathbb{B}$

\n (ii) Let $\varphi \in \mathbb{B}$ then we have:\\

\n $\varphi'(z) = \displaystyle{\sum_{k=0}^{+\infty}(k+1) a_{k+1}z^{k} =} \displaystyle{\sum_{k=0}^{+\infty}\sqrt{k!}\sqrt{k+1} a_{k+1} \sqrt{k + 1}\frac{z^{k}}{\sqrt{k!}}}$ \\

\n  $\displaystyle{=\sum_{k=0}^{+\infty}\sqrt{(k+1)!} a_{k+1} \sqrt{k + 1}\frac{z^{k}}{\sqrt{k!}}}$\\

\n Then we get the following inequalities :\\

\n $\mid \varphi'(z) \mid^{2} \leq (\displaystyle{\sum_{k=0}^{+\infty}\mid \sqrt{(k+1)!} a_{k+1} \sqrt{k + 1}\frac{z^{k}}{\sqrt{k!}}\mid)^{2}}$\\

 \n By applying the Cauchy-Shwarz's inequality, we obtain:\\
 
\n $\leq (\displaystyle{\sum_{k=0}^{+\infty}(k+1)! \vert a_{k+1}\vert^{2} \sum_{k=0}^{+\infty} (k+1)\frac{\vert z\vert^{2k}}{k!}}$\\

\n $\leq (\displaystyle{\vert\vert \varphi \vert\vert_{\mathbb{B}}^{2}[ \sum_{k=0}^{+\infty}k\frac{\vert z\vert^{2k}}{k! } + e^{\vert z\vert^{2}}]}$\\

\n $\leq (\displaystyle{\vert\vert \varphi \vert\vert_{\mathbb{B}}^{2}[\vert z \vert^{2} e^{\vert z\vert^{2}}  +  e^{\vert z\vert^{2}}]}$\\

i.e.\\

\n $\displaystyle{ \mid \varphi'(z) \mid^{2} \leq (1 + \mid z \mid^{2})e^{\mid z\mid^{2}}\mid\mid \varphi \mid\mid_{\mathbb{B}}^{2}.}$\\

\n then \\

\n $\displaystyle{\mid \varphi'(z) \mid \leq \sqrt{1 + \mid z \mid^{2}}e^{\frac{\mid z\mid^{2}}{2}}\mid\mid \varphi \mid\mid_{\mathbb{B}}}.$\\

\n (iii) Let $\varphi \in \mathbb{B}$, then we have :\\

\n $\varphi(z) = \displaystyle{\sum_{k=0}^{+\infty}a_{k}z^{k}}$\\

\n $\varphi'(z) = \displaystyle{\sum_{k=0}^{+\infty}(k+1)a_{k+1}z^{k}}$\\

\n $\varphi''(z) = \displaystyle{\sum_{k=0}^{+\infty}(k+1)(k+2)a_{k+2}z^{k}}$\\

\n In similar, we obtain :\\

\n $\mid \varphi''(z)\mid^{2} \leq (\displaystyle{\sum_{k=0}^{+\infty}\mid \sqrt{k!}\sqrt{k+1}\sqrt{k+2}\vert a_{k+2}\vert \sqrt{k+1}\sqrt{k+2}\frac{\vert z\vert^{k}}{\sqrt{k!}}})^{2}$\\

\n $\displaystyle{\mid \varphi''(z)\mid^{2} \leq}$ $\displaystyle{\sum_{k=0}^{+\infty} (k+2)!\mid a_{k+2}\mid^{2}}$ $\displaystyle{\sum_{k=0}^{+\infty}(k^{2} + 3k + 2)\frac{\mid z \mid^{2k}}{k!}}$\\

\n $\displaystyle{\leq (\alpha_{0} + \alpha_{1}\mid z \mid^{2} + \alpha_{2}\mid z \mid^{4}) e^{\mid z\mid^{2}}\mid\mid \varphi \mid\mid_{\mathbb{B}}^{2}}$ with $\alpha_{i} > 0 , i =0, 1, 2$\\

\n Then:\\

\n $\mid \varphi''(z)\mid \leq \sqrt{(\alpha_{0} + \alpha_{1}\mid z \mid^{2} + \alpha_{2}\mid z \mid^{4})}e^{\frac{\mid z\mid^{2}}{2}}\mid\mid \varphi \mid\mid_{\mathbb{B}}$\\

\n {\color{blue}$\bullet$} (2) \\

\n For $z \in \mathbb{C}$, we can find a neighborhood $U$ of $z$ and a constant $C_{z}$ such that \\
\begin{equation}
\displaystyle{ \vert \varphi(\xi) \vert^{2} \leq C_{z} \vert\vert \varphi \vert\vert_{ \mathbb{B}}^{2} , \,\, \mbox{for all}\,\, \xi \in U \,\, \mbox{and all}\,\, \varphi \in \mathbb{B}}
\end{equation}

\n Now, suppose that we have a sequence $\varphi_{n} \in \mathbb{B}$ and $\varphi \in L_{2}(\mathbb{C} , d\mu(z))$ such that $\displaystyle{ \varphi_{n} \longrightarrow \varphi  \,\, in \,\, L_{2}(\mathbb{C} , d\mu(z)), \,\, \mbox{as} \,\, n \longrightarrow + \infty}$.Then $\varphi_{n} $ is certainly a Cauchy sequence in $L_{2}(\mathbb{C} , d\mu(z))$;  But then \\

\begin{equation}
\displaystyle{\underset{ \xi \in U}{sup} \vert \varphi_{n}  - \varphi_{m} \vert \sqrt{C_{z}} \vert\vert \varphi_{n}  - \varphi_{m} \vert\vert_{L_{2}(\mathbb{C} , d\mu(z))} \,\, \mbox{as} \,\, n, m \longrightarrow \infty} 
\end{equation} 
\n This shows that the sequence $\varphi_{n}$ converges locally uniformly to some limit function, which must be $\varphi$. A standard theorem shows that a locally uniform limit of holomorphic functions is always holomorphic. So the limit function $\varphi$ is actually in $\mathbb{B}$. It follows that  $\mathbb{B}$ is closed.\hfill { } $\blacksquare$\\ 

\n { \color{red}$ \rhd$} Since $\mathbb{B}$ is a closed subspace of  $\displaystyle{ L_{2}(\mathbb{C} , d\mu(z)) = \{\psi  \,\, \text{measurable} \,\, ; \int_{\mathbb{C}} \vert \psi(z)\vert^{2} d\mu(z) < \infty \}}$ we have the following orthogonal decomposition:\\
$$\displaystyle{ L_{2}(\mathbb{C} , d\mu(z)) = \mathbb{B} \oplus \mathbb{B}^{\perp}}$$

\n { \color{red}$ \rhd$} The projection $\displaystyle{ P:  L_{2}(\mathbb{C} , d\mu(z)) \longrightarrow \mathbb{B}}$ is the following operator :\\
$$ \displaystyle{ (P\psi)(z) = \frac{1}{\pi}\sum_{n=0}^{\infty} < \psi , \frac{z^{n}}{\sqrt{n!}} >  \frac{z^{n}}{\sqrt{n!}}  \quad\quad ( \psi \in  L_{2}(\mathbb{C} , d\mu(z)) )}$$
\n or\\
$$ \displaystyle{ (P\psi)(z) = \frac{1}{2i\pi}\int_{\mathbb{C}}e^{z\overline{w}} \psi(w) e^{-\vert w \vert^{2}} dw \wedge d\overline{w} \quad ( \psi \in  L_{2}(\mathbb{C} , d\mu(z)) )}$$

\n where $\displaystyle{e^{z\overline{w}}}$ is Bergman (reproducing) kernel with respect to Gaussian measure $\displaystyle{\frac{1}{2i\pi} e^{-|w|^{2}} dw \wedge d\overline{w}}$

 \begin{definition}
  (Hilbert function spaces)  \\

 \n A Hilbert { \color{red}function} space on a set $E$ is a Hilbert space $ \mathcal{H}$ which is a subspace of the vector space of all functions $E \rightarrow \mathbb{C}$, with one additional and crucial property: \\

 \n { \color{red}$ \bullet$} For every $z \in E$ the functional of { \color{red}point evaluation} at $z$, given by $ \varphi \mapsto  \varphi(z)$, is a { \color{red}bounded functional} on $ \mathcal{H}$ \\
 \end{definition}
 
  \n { \color{red}$ \rhd$} Let $ \mathcal{H}$ be a Hilbert function space on a set $E$. As point evaluation is a bounded functional,  then by the Riesz representation theorem there is, for every $z  \in E$, an element $ k_z \in  \mathcal{H}$ such that $\varphi(z) = \langle \varphi, k_z \rangle$ for all $\varphi \in  \mathcal{H}$. \\
  
  \begin{lemma}
  The  Bargmann space $\mathbb{B}$ is a Hilbert function space on $\mathbb{C}$.
  \end{lemma}
  
  \n {\bf {\color{red} Proof}}\\
  
\n To show that Bargmann space $\mathbb{B}$ is a Hilbert function space on $\mathbb{C}$, we need to show that for every $z \in \mathbb{C}$, the functional  $ \varphi \mapsto  \varphi(z)$ is bounded. From (i) of above proposition, this functional is bounded, then $\mathbb{B}$ is a Hilbert function space on $\mathbb{C}$ and by the Riesz representation theorem there exists some $k_{z} \in \mathbb{B}$ such that $\displaystyle{\varphi(z) = \langle \varphi, k_z \rangle = \int_{\mathbb{C}} \varphi(w)\overline{k_{z}(w)}d\mu(w)}$ for each $\varphi \in \mathbb{B}$. \hfill { } $\blacksquare$\\

  \begin{definition}
    (Reproducing kernels)  \\
  
 \n  The function $k_z$ is called the kernel function at $z$. \\ We define a function  $\displaystyle{ K: E \times E \longrightarrow \mathbb{C}}$ by $K(z, w) = k_w(z)$, the function $K$ is called the reproducing kernel  of $ \mathcal{H}$. \\
 
\n The family of functions $k_w = K(\cdot, w)$ enable one to reproduce the values of any $\varphi \in \mathbb{H}$ via the relationship  $\varphi(w) = \langle \varphi, k_w \rangle$. Note that $K(z, w) = \langle k_w, k_z \rangle$\\

\n The reproducing kernel $K(z,w)$ is positive definite if for all choices of $N  \in \mathbb{N}$, $z_{1}, . . . , z_{N} \in E$ and $c_{1}, . . . , c_{N} \in \mathbb{C}$, we have:\\

\n $\displaystyle{\vert\vert \sum_{j=1}^{N}k(., z) c_{j}\vert\vert_{\mathcal{H}}^{2} = \sum_{i,j=1}^{N} \overline{c_{i}}k(z_{j}, z_{j}) c_{j}  \geq 0 }$ \\
 \end{definition}
 
 \begin{definition}
 {\color{red} (Reproducing kernel Hilbert space[RKHS])}\\
  Given a set $E$, we will say that $\mathcal{H}$ is a reproducing kernel Hilbert space(RKHS) on $E$ over  $\mathbb{K}$ where $\mathbb{K} = \mathbb{C}$ or $\mathbb{K} = \mathbb{R}$ , provided that:\\
  
\n (i) $\mathcal{H}$ is a vector vector space,\\
  
\n (ii) $\mathcal{H}$ is endowed with an inner product, $\langle , \rangle$, making it into a Hilbert space,\\
  
\n (iii) for every $z \in E$, the linear evaluation functional, $\displaystyle{ev_{z} : \mathcal{H}  \longrightarrow ,\mathbb{K}}$ defined by $\displaystyle{\varphi \longrightarrow ev(\varphi) = \varphi(z)}$, is bounded.\\
 \end{definition} 
\begin{remark}
\n $\displaystyle{\vert\vert ev_{z} \vert\vert^{2} =  \vert\vert k_{z} \vert\vert^{2}  = \langle k_{z} , k_{z} \rangle = K(z, z)}$.\\
\end{remark}

 \subsection{ {\color{blue} Examples of reproducing kernel spaces}}
 
 \n {\color{red} $\rhd_{1}$} $\mathbb{C}^{n}$ can be  considered as a Hilbert function space on $\displaystyle{ E = \{ 1; 2, .., n\}}$. The vector $\displaystyle{z = (z_{1}, z_{2}, ....., z_{n})}$ corresponds to the function $\displaystyle{\varphi :  E \longrightarrow \mathbb{C}}$ defined by $\displaystyle{\varphi(j) = z_{j}}$.\\
 \n As $\mathbb{C}^{n}$ is finite dimensional then all linear functional is continuous, and therefore $\mathbb{C}^{n}$ is a Hilbert function space. We have for each $z \in \mathbb{C}^{n}$, and for $1 j  n$ $z_{j} = < z, e_{j} >$ where $e_{j}$ is the vector containing a $1$ in position $j$ and $0$ elsewhere.\\
 
 \n Then $k_{j} = e_{j}$ and \\
 \begin{equation}
 K(i, j) = \left \{\begin{array} {c} 1 \quad \mbox{if} \,\, i = j \\
 \quad\\
 0 \quad \mbox{if} \,\, i \neq j \\
 \end{array} \right.
\end{equation}
\par \hfill { } $\blacksquare$\\

 \n {\color{red} $\rhd_{2}$} Let $\displaystyle{\ell^{2}(E) = \{ \varphi : E  \longrightarrow \mathbb{C} \,\, \mbox{such that} \,\, \sum_{z \in E} \vert \varphi (z) \vert^{2} < \infty\}}$\\
 
 \n For each $z \in E$ and for each $ \displaystyle{\varphi \in \ell^{2}(E)}$, we have\\
 
 \begin{equation}
 \vert \varphi(z) \vert = \sqrt{ \vert \varphi(z) \vert^{2}} \leq \sqrt{\sum_{w \in E} \vert \varphi (w) \vert^{2}} = \vert\vert \varphi \vert\vert _{\ell^{2}(E)}
 \end{equation}
  
\n and therefore the evaluation functional at $z$ is bounded. That $\displaystyle{\ell^{2}(E)}$  is a Hilbert function space and for each $\varphi \in \ell^{2}(E)$ and for each  $z \in E$, we have\\
 
 \n $\displaystyle{\varphi(z) = < \varphi , e_{z} > }$ where $\displaystyle{ e_{z} : E \longrightarrow \mathbb{C} }$ is the function defined by\\
 
  \begin{equation}
 e_{z} (w) = \left \{\begin{array} {c} 1 \quad \mbox{if} \,\, w = z \\
 \quad\\
 0 \quad \mbox{if} \,\, w \neq z \\
 \end{array} \right.
\end{equation}

\n It follows that $k_{z} = e_{z}$ and 

 \begin{equation}
 K(z, w) = \left \{\begin{array} {c} 1 \quad \mbox{if} \,\, z = w \\
 \quad\\
 0 \quad \mbox{if} \,\, z \neq w \\
 \end{array} \right.
\end{equation}
\par \hfill { } $\blacksquare$\\
  \n {\color{red} $\rhd_{3}$} The Hardy space $\mathcal{H}^{2}$ is the class of holomorphic functions $\varphi$ on the open unit disk $\mathbb{D}$ satisfy:\\
  \begin{equation}
  \n \displaystyle{\varphi(z) = \sum_{n=0}^{\infty} a_{n}z^{n} \,\, ,   \vert \vert \varphi \vert \vert_{\mathcal{H}^{2}}^{2} = \sum_{n=0}^{\infty} \vert a_{n} \vert^{2} < \infty}
  \end{equation}
  
  \n Let $ w  \in  \mathbb{D}$. If $\displaystyle{\varphi(z) = \sum_{n=0}^{\infty} a_{n}z^{n}}$ then by using Cauchy-Schwarz inequality we get \\
  
    \begin{equation}
  \n \displaystyle{\vert \varphi(w) \vert \leq \sum_{n=0}^{\infty} \vert a_{n}w^{n}\vert \leq \sqrt{\sum_{n=0}^{\infty} \vert a_{n} \vert^{2}} \sqrt{\sum_{n=0}^{\infty} \vert  w \vert^{2n}} = \sqrt{\frac{1}{1 - \vert w \vert^{2}}} \,\,\vert\vert \varphi \vert\vert_{\mathcal{H}^{2}}}
  \end{equation}
  
 \n Hence, the evaluation functional at $w$ is bounded. That $\mathcal{H}^{2}$  is a Hilbert function space and for each $\varphi \in \mathcal{H}^{2}$ and for each  $z \in \mathbb{D}$, we have\\
 
 \begin{equation}
   \begin{array}{c}\displaystyle{< \varphi, k_{w} > =  \varphi(w)  = \sum_{n=0}^{\infty} a_{n}w^{n} = < \sum_{n=0}^{\infty} a_{n}z^{n}  ,  \sum_{n=0}^{\infty} \overline{w}^{n}z^{n}> }\\
    \displaystyle{= < \varphi, \frac{1}{1 - z\overline{w}} >} \quad
    \end{array}
  \end{equation}
  
 \n We have used the Taylor series :
  
   \begin{equation}
   \displaystyle{\frac{1}{1 - u} = \sum_{n=0}^{\infty}u^{n}}
   \end{equation}
  
  \n Then \\
  
\n  $ \displaystyle{k_{w}(z) = \frac{1}{1 - z\overline{w}}}$ and  $ \displaystyle{K(z, w) = k_{w}(z) = \frac{1}{1 - z\overline{w}}}$.\hfill { } $\blacksquare$ \\
  
   \n {\color{red} $\rhd_{4}$}  The Bergman space $\mathcal{A}^{2}$  is the class of holomorphic functions $\varphi$ on the open unit disk $\mathbb{D}$ satisfy:\\
  \begin{equation}
  \n \displaystyle{\varphi(z) = \sum_{n=0}^{\infty} a_{n}z^{n} \,\, ,   \vert \vert \varphi \vert \vert_{\mathcal{A}^{2}}^{2} = \frac{1}{\pi}\int_{\mathbb{D}} \vert \varphi(z) \vert^{2} dxdy < \infty}
  \end{equation}
By Parseval's formula and the monotone  convergence theorem, we deduce that

 \n $\displaystyle{\frac{1}{\pi}\int_{\mathbb{D}} \vert \varphi(z) \vert^{2} dxdy = \frac{1}{\pi}\int_{0}^{1}\int_{0}^{2 \pi}  \vert  \sum_{n=0}^{ \infty} a_{n} r^{n} e^{in \theta} \vert^{2} rdr d \theta = 2  \int_{0}^{1}  \sum_{n=0}^{ \infty}  \vert a_{n}  \vert^{2} r^{2n + 1}dr }$\\
 
\n $\displaystyle{ = \sum_{n=0}^{ \infty}  \frac{ \vert a_{n}  \vert^{2}}{n + 1}}$ \\

\n Hence

\n $\displaystyle{\varphi  \in \mathcal{A}^{2} \iff   \sum_{n=0}^{ \infty}  \frac{ \vert a_{n}  \vert^{2}}{n + 1} < \infty}$ \\

\n and\\

\n $\displaystyle{< \varphi ,  \psi >_{ \mathcal{A}^{2}}  =   \sum_{n=0}^{ \infty}  \frac{ a_{n}  \overline{b_{n}}}{n + 1} }$ \\

\n  where  $ \displaystyle{\varphi(z) = \sum_{n=0}^{\infty} a_{n}z^{n} }$ and  $\displaystyle{\psi(z) = \sum_{n=0}^{\infty} b_{n}z^{n}}$ \\

\n $\displaystyle{(\mathcal{A}^{2} , \, \,<  . >_{\mathcal{A}^{2}})}$ is a Hilbert space \\

 \n Let $ w  \in  \mathbb{D}$. If $\displaystyle{\varphi(z) = \sum_{n=0}^{\infty} a_{n}z^{n}}$ then by using Cauchy-Schwaerz inequality we get \\
  
    \begin{equation}
  \n \displaystyle{\vert \varphi(w) \vert \leq \sum_{n=0}^{\infty} \vert a_{n}w^{n}\vert \leq \sqrt{\sum_{n=0}^{\infty} \frac{\vert a_{n} \vert^{2}}{n + 1}} \sqrt{\sum_{n=0}^{\infty}(n + 1) \vert  w \vert^{2n}} = \sqrt{\frac{1}{(1 - \vert w \vert^{2})^{2}}} \,\,\vert\vert \varphi \vert\vert_{\mathcal{A}^{2}}}
  \end{equation}
  
   \n  We have used the Taylor series :\\
  
   \begin{equation}
   \displaystyle{\frac{1}{(1 - u)^{2}} = \sum_{n=1}^{\infty}n u^{n-1}}
   \end{equation}
   
 \n Hence, the evaluation functional at $w$ is bounded. That $\mathcal{A}^{2}$  is a Hilbert function space and for each $\varphi \in \mathcal{A}^{2}$ and for each  $z \in \mathbb{D}$, we have\\
 
 \n  $ \displaystyle{ < \varphi , k_{w} >_{\mathcal{A}^{2}}  =  \varphi(w) = \sum_{n=0}^{\infty}  a_{n}w^{n} = \sum_{n=0}^{\infty} \frac{a_{n} \overline{(n+1)\overline{w}^{n}}}{n + 1} = < \varphi ,  \sum_{n=0}^{\infty} (n + 1)(z\overline{w})^{n} >}$
 
 \n Then\\
    
   \n  $ \displaystyle{k_{w}(z) = \frac{1}{(1 - z\overline{w})^{2}}}$ and  $ \displaystyle{K(z, w) = k_{w}(z) = \frac{1}{(1 - z\overline{w})^{2}}}$. \hfill { } $\blacksquare$ \\
   
 \begin{lemma}
 The Bargmann space $\mathbb{B}$ is a reproducing kernel Hilbert space with reproducing kernel $\displaystyle{K(z, w) = e^{z\overline{w}}}$; i.e., for every $\varphi  \in \mathbb{B}$ we have\\
 \begin{equation}
 \displaystyle{\varphi (z) = < \varphi  , K(z , .) >_{\mathbb{B}}}
 \end{equation}
 \end{lemma}
  \n {\bf {\color{red}Proof}} \\
  
 \n The Bergmann space $\mathbb{B}$  is the class of entire functions $\varphi$  on $\mathbb{C}$ satisfy:\\
  \begin{equation}
  \n \displaystyle{\varphi(z) = \sum_{n=0}^{\infty} a_{n}z^{n} \,\, ,   \vert \vert \varphi \vert \vert_{\mathbb{B}}^{2} = \frac{1}{\pi}\int_{\mathbb{C}} \vert \varphi(z) \vert^{2} e^{-\vert z \vert^{2}} dxdy < \infty}
  \end{equation}
By Fubini theorem  and the monotone  convergence theorem, we deduce that

 \n $\displaystyle{\frac{1}{\pi}\int_{\mathbb{C}} \vert \varphi(z) \vert^{2} e^{-\vert z \vert^{2}}dxdy = \frac{1}{\pi}\int_{0}^{\infty}\int_{0}^{2 \pi}  \vert  \sum_{n=0}^{ \infty} a_{n} r^{n} e^{in \theta} \vert^{2} rdr d \theta }$  = $\displaystyle{ = \sum_{n=0}^{ \infty}  n! \vert a_{n}  \vert^{2}}$ \\

\n Hence

\n $\displaystyle{\varphi  \in \mathbb{B} \iff   \sum_{n=0}^{ \infty}  n! \vert a_{n}  \vert^{2}  < \infty}$ \\

\n and\\

\n $\displaystyle{< \varphi ,  \psi >_{ \mathbb{B}}  =   \sum_{n=0}^{ \infty} n! a_{n}  \overline{b_{n}} }$ \\

\n  where  $ \displaystyle{\varphi(z) = \sum_{n=0}^{\infty} a_{n}z^{n} }$ and  $\displaystyle{\psi(z) = \sum_{n=0}^{\infty} b_{n}z^{n}}$ \\

\n $\displaystyle{(\mathbb{B} , \, \,<  . >_{\mathbb{B}})}$ is a Hilbert space \\

 \n Let $ w  \in  \mathbb{C}$. If $\displaystyle{\varphi(z) = \sum_{n=0}^{\infty} a_{n}z^{n}}$ then by using Cauchy-Schwarz inequality we get \\
  
    \begin{equation}
  \n \displaystyle{\vert \varphi(w) \vert \leq \sum_{n=0}^{\infty} \vert a_{n}w^{n}\vert \leq \sqrt{\sum_{n=0}^{\infty} \frac{\vert w \vert^{2n}}{n!}} \sqrt{\sum_{n=0}^{\infty} n! \vert  a_{n} \vert^{2}} = \sqrt{e^{\vert w \vert^{2}}}\,\,\vert\vert \varphi \vert\vert_{\mathbb{B}}}
  \end{equation}
   
 \n Hence, the evaluation functional at $w$ is bounded,  which implies  \\
 
   \begin{equation}
   \displaystyle{\vert \varphi(w) - \psi(w) \vert \leq \sqrt{e^{\vert w \vert^{2}}}\,\,\vert\vert \varphi  - \psi \vert\vert_{\mathbb{B}} \,\, \mbox{for all }\,\, \varphi, \psi \in \mathbb{B}}
  \end{equation}
  
  \n and\\
  
  \n For all $w \in \mathbb{C}$ and for all $\psi \in \mathbb{B}$, it holds that: \\
  
   \begin{equation}
  \n \displaystyle{\forall \,\,  \varphi  \in  \mathbb{B},  \,  \, \forall \,\,   \epsilon > 0  \, \,  \exists  \, \,  \delta > 0  \, \,  \mbox{;} \,  \, (\vert\vert \varphi  - \psi \vert\vert_{\mathbb{B}} < \delta \Longrightarrow \vert \varphi(w) - \psi(w) \vert < \epsilon)}
  \end{equation}
\n i.e. all the point evaluation maps are continuous.\\
 
\n That $\mathbb{B}$  is a Hilbert function space and for each $\varphi \in \mathbb{B}$ and for each  $z \in \mathbb{C}$, we have\\
 
 \n  $ \displaystyle{ < \varphi , k_{w} >_{\mathbb{B}}  =  \varphi(w) = \sum_{n=0}^{\infty}  a_{n}w^{n} = \sum_{n=0}^{\infty} n! a_{n} \overline{\frac{\overline{w}^{n}}{n!}} = < \varphi ,  \sum_{n=0}^{\infty} \frac{ (z\overline{w})^{n} }{n!}>}$
 
 \n Then\\
    
   \n  $ \displaystyle{k_{w}(z) = e^{z\overline{w}}}$ and  $ \displaystyle{K(z, w) = k_{w}(z) =  e^{z\overline{w}}}$. \hfill { } $\blacksquare$ \\

   \begin{proposition}
   
   \n Let $\displaystyle{(z, w) \in \mathbb{C}^{2}, u \in \mathbb{R}  \,\,\mbox{ and} \,\, \mathcal{A}(z , u) = e^{-\frac{u^{2}}{2} + \sqrt{2}uz -\frac{z^{2}}{2}}= \sum_{n=0}^{ \infty}\frac{z^{n}}{ \sqrt{n!}}h_{n}(u)}$\\
   
 \n where $ (h_{n}(u))_{n \geq 0}$  denote the normalized Hermite functions. \\
   
  \n  Then \\
  \begin{equation}
  \displaystyle{\int_{\mathbb{R}}\mathcal{A}(z , u) \overline{\mathcal{A}(w , u) } du = \sqrt{\pi} e^{z\overline{w}}} 
  \end{equation}
  \end{proposition}
  \n {\bf{\color{red}Proof}}\\
  
  \n  $\displaystyle{\int_{\mathbb{R}}\mathcal{A}(z , u) \overline{\mathcal{A}(w , u) } du = \int_{\mathbb{R}}e^{-\frac{u^{2}}{2} + \sqrt{2}uz -\frac{z^{2}}{2}}e^{-\frac{u^{2}}{2} + \sqrt{2}u\overline{w} -\frac{\overline{w}^{2}}{2}}du}$\\
  
 \n  $\displaystyle{= e^{-\frac{1}{2}(z^{2} + \overline{w}^{2}} e^{\frac{(z + \overline{w})^{2}}{2}} \int_{\mathbb{R}} e^{-(u - \frac{z + \overline{w}}{\sqrt{2}})^{2} }du}$\\
 
 \n  $\displaystyle{= e^{z\overline{w}} \int_{\mathbb{R}}e^{-t^{2}}dt}$\\
 
  \n  $\displaystyle{= \sqrt{\pi}e^{z\overline{w}}}$\\
  
  \n where  $\displaystyle{t = u - \frac{z + \overline{w}}{\sqrt{2}}}$. \hfill { } $\blacksquare$\\
  
 \n   It is advisable to consult at this topic {\color{blue}[3]}.\\

 \begin{remark}
 1. Polynomials and entire functions of exponential type belong to Bargmann space. For example, sinc function  $\displaystyle{\frac{sin z}{z} }$ is entire function of exponential type (see {\color{blue}[42]} or {\color{blue}[46]}). Hence it belongs to Bargmann space.\\
 
 \n 2. $\displaystyle{\sigma(z) = z \prod(1-\frac{z}{\lambda_{m,n}})\exp(\frac{z}{\lambda_{m,n}}+\frac{z^{2}}{2\lambda_{m,n}^{2}})}$, is Weierstrass $\sigma$- function and $\lambda_{m,n}$ are lattice points in $\mathbb{C}$.\\
Under suitable conditions on lattice points, $\displaystyle{ \frac{\sigma(z)}{z}}$ belongs to Bargmann space see {\color{blue}[8]}.\\
 
\end{remark}
 
 \subsection{{\color{blue} Fundamental results on reproducing kernel spaces}}
  
  \begin{proposition}
 
 \n  Let $\mathcal{H}$ be a reproducing kernel Hilbert space on a set $E$ and $K(z, w)$ is its reproducing kernel. Let $\mathcal{H}_{0}$ be denote the space of finite linear combinations
 \begin{equation}
 \displaystyle{\sum_{j} \alpha_{j} K_{z_{j}}, \alpha_{j} \in \mathbb{C} \,\, \mbox{and} \, \, z_{j} \in E} \,\\, i.e. \,\, \mathcal{H}_{0} = span\{k_{z} , z \in E \}
 \end{equation} 
 
\n then \\
 
\n $\mathcal{H}_{0}$ is dense in $\mathcal{H}$ and  it follows that $\mathcal{H}$ is determined by the reproducing kernel.\\
\end{proposition}

\n {\bf {\color{red}Proof}}\\

\n Let $\displaystyle{\varphi \in  span\{k_{z} , z \in E \}^{\perp}}$. Then for all $z \in E$, we have $\varphi(z) = < \varphi , k_{z} > = 0$ this implies that $\varphi = 0$ and $\displaystyle{span\{k_{z} , z \in E \}^{\perp} = \{0\}}$\,\, i.e. \,\, $\mathcal{H}_{0}$ is dense in $\mathcal{H}$.\hfill { } $\blacksquare$\\

 \begin{theorem}
\n  Let  $\mathcal{H}_{1}$ and $\mathcal{H}_{2}$ be two reproducing kernel Hilbert spaces on $E$ with a {\color{red}same} kernel then they are {\color{red}equal}.
  \end{theorem}
\n {\bf {\color{red}Proof}}\\

\n As $\mathcal{H}_{1}$ and $\mathcal{H}_{2}$ are  two reproducing kernel Hilbert spaces on $E$ with  {\color{red}same} kernel then we  have $\displaystyle{\vert\vert \varphi \vert\vert_{\mathcal{H}_{1}} = \vert\vert \varphi \vert\vert_{\mathcal{H}_{1}}}$ for all $\displaystyle{\varphi \in  span\{k_{z} , z \in E \}}$. In fact : \\

\n If  $\displaystyle{\varphi = \sum_{j=1}^{n} \alpha_{j} k_{z_{j}}}$ then we have\\

 \n $\displaystyle{\vert\vert \varphi \vert\vert_{\mathcal{H}_{1}}^{2} = < \sum_{j=1}^{n} \alpha_{j} k_{z_{j}} , \sum_{i=1}^{n} \alpha_{i} k_{z_{i}} > =  \sum_{i,j}^{n} \alpha_{j} \overline{\alpha_{i}}< k_{z_{j}} , k_{z_{j}} >_{\mathcal{H}_{1}} =  \sum_{i,j}^{n} \alpha_{j} \overline{\alpha_{i}} K(z_{i}, z_{j})}$.  \\
 
  \n Similar, we get  \\
  
   \n $\displaystyle{\vert\vert \varphi \vert\vert_{\mathcal{H}_{2}}^{2} = \sum_{i,j}^{n} \alpha_{j} \overline{\alpha_{i}} K(z_{i}, z_{j})}$.  \\
   
 \n Now let $\psi \in  \mathcal{H}_{1}$ then from above lemma, we deduce there exists an sequence $\displaystyle{\psi_{n} \in  span\{k_{z} , z \in E \}}$ such that $\displaystyle{\psi_{n} \longrightarrow  \varphi  \, \,  \mbox{in} \mathcal{H}_{1}}$. As $(\psi_{n})$ is convergent then in particular, it is a Cauchy sequence : \\
 
  \n $\displaystyle{ \forall \,\, \epsilon > 0, \, \exists \,\, N > 0 ; \,\   \forall \,\ n > N , \,\   \forall \,\ m > N  \Longrightarrow \vert\vert \psi_{n} - \psi_{m} \vert\vert_{\mathcal{H}_{1}} < \epsilon}$ \\
  
   \n But  $\displaystyle{\vert\vert \psi_{n} - \psi_{m} \vert\vert_{\mathcal{H}_{1}} =  \vert\vert \psi_{n} - \psi_{m} \vert\vert_{\mathcal{H}_{2}}}$ thus $(\psi_{n})$ is  Cauchy sequence in  $\mathcal{H}_{2}$, so converges to some $ \varphi  \in \mathcal{H}_{2}$  because $\mathcal{H}_{2}$ is complete.  \\
   
 \n But the convergence in an reproducing kernel Hilbert space (RKHS) implies the ``simple convergence.'' So $ \psi_{n} \longrightarrow \psi $ as $n \longrightarrow + \infty$ and $ \psi_{n} \longrightarrow \varphi $ as $n \longrightarrow + \infty$. So $\psi = \varphi$ and $\psi \in \mathcal{H}_{2}$ . By the same reasoning, any function in $\mathcal{H}_{2}$ is in $\mathcal{H}_{1}$. \\
 
 \n Finally, we have \\
 
 $\displaystyle{\vert\vert \psi \vert\vert_{\mathcal{H}_{1}}  = \lim\limits_{n \longrightarrow + \infty} \vert\vert \psi_{n} \vert\vert_{\mathcal{H}_{1}} = \lim\limits_{n \longrightarrow + \infty} \vert\vert \psi_{n} \vert\vert_{\mathcal{H}_{2}} = \vert\vert \psi \vert\vert_{\mathcal{H}_{2}} }$. \hfill { } $\blacksquare$\\   
 
 \begin{theorem}
 ({\color{red}Moore})\\
 \n Let $E$ be a set and $K: E \times E \longrightarrow \mathbb{C}$ a semi-definite function positive. Then there is a unique RKHS on $E$ whose kernel is $K$.
 \end{theorem}
 
 \n {\bf{\color{red}Proof}}\\
 
 \n  Uniqueness follows immediately from above theorem. Let us pose, for $z , w \in E, k_{w}(z) = K(z, w)$ and $\displaystyle{\mathcal{H}_{0} = span\{k_{z} , z \in E \}}$ \\
 
 \n If  $\displaystyle{\varphi(z) = \sum_{i=1}^{n} \alpha_{i} k_{z_{i}},}$ and $\displaystyle{\psi(z)= \sum_{i=1}^{m} \beta_{j} k_{w_{j}},}$ then we define \\
 
 \begin{equation}
 \displaystyle{< \varphi , \psi > = \ \sum_{i, j}\alpha_{i} \overline{\beta_{j}} K(w_{j} , z_{i})}
 \end{equation}

\n We must first verify that $< , >$ is well defined. suppose that\\

 \n $\displaystyle{\varphi(z) = \sum_{i=1}^{n} \alpha_{i} k_{z_{i}},}$ $\displaystyle{= \sum_{i=1}^{n} \alpha_{i}^{'} k_{z_{i}^{'}},}$ \\

  \n $\displaystyle{\psi(z)= \sum_{j=1}^{m} \beta_{j} k_{w_{j}},}$ $\displaystyle{= \sum_{j=1}^{m} \beta_{j}^{'} k_{w_{j}^{'}},}$  \\

\n are two different representations for $\varphi$ and $\psi$ (we can assume that there are the same number of terms in representations by adding null coefficients if necessary). So for each $z \in E$, we have\\

 \n $\displaystyle{ \sum_{i=1}^{n} \alpha_{i} k_{z_{i}},}$ $\displaystyle{= \sum_{i=1}^{n} \alpha_{i}^{'} k_{z_{i}^{'}},}$  \\

 \n $\displaystyle{ \sum_{j=1}^{m} \beta_{j} k_{w_{j}},}$  $\displaystyle{= \sum_{i=1}^{m} \beta_{j}^{'} k_{w_{j}^{'}},}$  \\

\n and so \\

\n $\displaystyle{ \sum_{i, j}\alpha_{i} \overline{\beta_{j}} K(w_{j} , z_{i}) = \sum_{j=1}^{m}. \overline{\beta_{j}} ( \sum_{i=1}^{n}  \alpha_{i}k_{z_{i}}(w_{j}) = \sum_{j=1}^{m}. \overline{\beta_{j}} ( \sum_{i=1}^{n}  \alpha_{i}^{'}k_{z_{i}^{'}}(w_{j})}$

\n $\displaystyle{ =\sum_{i, j}\alpha_{i}^{'} \overline{\beta_{j}} \overline{K(z_{i}^{'}, w_{j})} = \sum_{i=1}^{n} \alpha_{i}^{'} (\overline{\sum_{j=1}^{m}\beta_{j}k_{w_{j}}(z_{i}^{'}}})$  \\

\n $\displaystyle{ =\sum_{i=1}^{n}\alpha_{i}^{'} (\overline{\sum_{j=1}^{m} \beta_{j}^{'}k_{w_{j}}(z_{i}^{'} })= \sum_{i,j} \alpha_{i}^{'} \beta_{j}^{'} K(w_{j} , z_{ji}^{'})}$  \\
 
\n  Thus, $< , >$ is well defined. We verify that $< , > $ is left linear and ``anti-symmetric''. Moreover, as $K \geq 0$, we have $< \varphi , \varphi  > \geq 0$ for all $\varphi \in \mathcal{H}_{0}$. Note that for all$\varphi \in \mathcal{H}_{0}$, we have $< \varphi , k_{z} > = \varphi(z)$.\\ 
 
\n  Now suppose that $< \varphi , \varphi  > =  0$. By the Cauchy-Schwarz inequality (which is valid for positive semidefinite anti-symmetric bilinear forms),\\
 
 $$ \vert < \varphi , k_{z} > \vert \leq < \varphi , \varphi  > < k_{z} , k_{z} > = 0 $$
 
\n So, for all $z \in E$\\

$$ \varphi(z) = < \varphi , k_{z} > = 0$$

\n So $\varphi = 0$. So $< , > $ is a inner product. \hfill { } $\blacksquare$\\

\n We define $\mathcal{H}$ as being the completion of $\mathcal{H}_{0}$ with respect to the scalar product. We can see $\mathcal{H}$ as a space of functions on $E$. Indeed, if $ \varphi \in \mathcal{H}$, then we associate to it the function $\tilde{\varphi}$ defined by\\

\begin{equation}
\tilde{ \varphi}(z) := < \varphi , k_{z} > 
\end{equation}

\n This uniquely defines $\tilde{\varphi}$ because $\mathcal{H}_{0}$ is dense in $\mathcal{H}$ . The set $\displaystyle{\{\tilde{\varphi} ; \varphi \in \mathcal{H}\}}$ is a subspace of the vector space $\displaystyle{\mathfrak{F}(E , \mathbb{C})}$ of functions on $E$ with values in $\mathbb{C}$. This is isomorphic (as a vector space) to $\mathcal{H}$. Hence $\mathcal{H}$ is an reproducing kernel Hilbert space (RKHS) with kernel $K$.\\

\section{ {\color{red} Unitarity of the Segal-Bargmann transform and some properties of its adjoint}}
  
\subsection{ {\color{blue}  On gaussian measure and Bargmann space}}

\n Let be $O(\mathbb{C})$ the set of entire functions, $\mathfrak{M}$ is the operator of multiplication by $z = x + iy ; (x, y) \in \mathbb{R}^{2}$ and $i^{2} = -1$ (i.e $\mathfrak{M}\varphi = z\varphi, \varphi \in O(\mathbb{C})$) and  $\mathfrak{D}$ is the operator of derivative in order $z$ (i.e   $\displaystyle{\mathfrak{D}\varphi = \frac{\partial}{\partial z}\varphi, \varphi \in O(\mathbb{C})}$). we observe that on $O(\mathbb{C})$ the operators $\mathfrak{D}$ and $\mathfrak{M}$ satisfy : \\

\begin{equation}
\displaystyle{[\mathfrak{D}, \mathfrak{M}] = I} 
\end{equation}

\begin{lemma}
({\color{red}Gaussian measure})\\ 

\n If we request that $\mathfrak{M}$ is adjoint to $\mathfrak{D}$ in $O(\mathbb{C})$ with inner product :\\
 $\displaystyle{  < \varphi, \psi > = \int_{\mathbb{C}} \varphi(z) \overline{\psi(z)}\rho(z, \bar{z})dxdy}$ where  $\rho(z, \bar{z})$ is a  measure.\\
 Then $\displaystyle{ \rho(z, \bar{z}) = e^{-\mid z\mid^{2}}}$ (Gaussian measure).\\
 \end{lemma}
 
\n {\bf {\color{red}Proof}}\\

\n The requirement that $\mathfrak{M} = \mathfrak{D}^{*}$ gives :\\

\n $\displaystyle{\int_{\mathbb{C}}\frac{\partial}{\partial z }[\varphi(z)]\overline{\psi(z)}\rho((z, \bar{z})dxdy =  \int_{\mathbb{C}}\varphi(z)\overline{z\psi(z)}\rho((z, \bar{z})dxdy}$\\

\n As \\

\n $\displaystyle{\frac{\partial}{\partial z }[\varphi(z)]\overline{\psi(z)}\rho((z, \bar{z}) = \frac{\partial}{\partial z }[\varphi(z)\overline{\psi(z)}\rho((z, \bar{z})] - \varphi(z)\frac{\partial}{\partial z }[\overline{\psi(z)}]\rho((z, \bar{z}) - \varphi(z)\overline{\psi(z)}\frac{\partial}{\partial z }[\rho((z, \bar{z})]}$\\

\n then\\

\n $\displaystyle{\int_{\mathbb{C}}\frac{\partial}{\partial z }[\varphi(z)]\overline{\psi(z)}\rho((z, \bar{z})dxdy = \int_{\mathbb{C}}\frac{\partial}{\partial z }[\varphi(z)\overline{\psi(z)}\rho((z, \bar{z})]dxdy - \int_{\mathbb{C}}\varphi(z)\frac{\partial}{\partial z }[\overline{\psi(z)}]\rho((z, \bar{z})dxdy }$\\

\n $\displaystyle{- \int_{\mathbb{C}}\varphi(z)\overline{\psi(z)}\frac{\partial}{\partial z }[\rho((z, \bar{z})]dxdy}$\\

\n In the right hand side, the first term of the integrand vanishes if we assume that the inner product between $\varphi$ and $\psi$  is finite, so that $\varphi \overline{\psi}\rho \longrightarrow 0$  sufficiently fast  as $\mid z \mid \longrightarrow \infty$. The second term also vanishes,because  $\psi$ is holomorphic, so that  $\overline{\psi}$ is anti-holomorphic and hence $\displaystyle{\frac{\partial}{\partial z}\overline{\psi} = 0}$.This gives:\\
\begin{equation}
\displaystyle{ \int_{\mathbb{C}}\varphi(z)\overline{z\psi(z)}\rho((z, \bar{z})dxdy + \int_{\mathbb{C}}\varphi(z)\frac{\partial}{\partial z }[\overline{\psi(z)}]\rho((z, \bar{z})dxdy = 0}\\
\end{equation}

\n which is solved for arbitrary $\varphi$ and $\psi$ if \\
\begin{equation}
\displaystyle{\bar{z}\rho((z, \bar{z}) + \frac{\partial}{\partial z }[\rho((z, \bar{z})] = 0}\\
\end{equation}
\n giving \\
\begin{equation}
\displaystyle{ \rho((z, \bar{z}) = Ce^{-\mid z\mid^{2}}}\\
\end{equation}
\n The constant $C$ is chosen to be $\displaystyle{\frac{1}{\pi}}$, so that the norm of the constant function $\varphi(z) \equiv 1$ is one. This explains why the space of holomorphic functions equipped with a Gaussian measure gives Bargmann space $\mathbb{B}$. \hfill { } $\blacksquare$\\

\n {\color{red}$\bullet$} It is well known that the classical Bargmann space is the only space of entire functions on which the creation and annihilation operators are adjoints of
each others and satisfy the classical commutation rules. Of course, this is not true anymore on $L_{2}(\mathbb{C}, d\mu)$, see Proposition 7.2 in {\color{blue}[37]}. \\

\subsection{{\color{blue}Some properties of Segal-Bargmann transform}}

\begin{definition}
({\color{red}Segal-Bargmann transform})\\

\n The Segal-Bargmann transform $\displaystyle{SB : L_{2}(\mathbb{R}) \longrightarrow \mathbb{B} : f  \longrightarrow \varphi}$ defined by\\
\begin{equation}
\displaystyle{SB[f](z) = \varphi(z) =  \int_{\mathbb{R}}\mathcal{A}(z, u)f(u)du, \quad \quad {\color{blue}{\Large(\star)}}}
\end{equation}
\n with \\
\begin{equation}
\displaystyle{\mathcal{A}(z, u) = {\color{red}c} e^{-\frac{u^{2}}{2} + \sqrt{2}uz -\frac{z^{2}}{2}}; \,\,{\color{red}c > 0}\quad  \quad \quad \quad \quad \quad \quad {\color{blue}{\Large(\star\star)}}}
\end{equation}
\end{definition}

\begin{remark}

\n  (i) The above transformation was introduced by Bargmann in ({\color{blue}[4], p. 198}) with $\displaystyle{ c = \frac{1}{\pi^{\frac{1}{4}}}}$\\

\n (ii) Let $u = \sqrt{2}q$ then $\displaystyle{du = \sqrt{2}dq}$ and  \\

\begin{equation}
\displaystyle{SB[f](z) = c' \int_{\mathbb{R}} e^{- q^{2} + 2qz -\frac{z^{2}}{2}} f(q)dq; \,\,{\color{red}c' = c\sqrt{2}}}
\end{equation}

\n (iii) Different forms of this transformation were used in the literature in particular, a parametrized form of the Bargmann transform given by K. Zhu {\color{blue}[48]} as follows:\\

\n $\displaystyle{{\mathcal{B}_{\alpha} : L_{2}(\mathbb{R}) \longrightarrow \mathbb{B}_{\alpha} \subset  L_{2}(\mathbb{C} , d\mu_{\alpha}(z) : f  \longrightarrow \mathcal{B}_{\alpha}[f](z)}}$ \\ 

\n where $\mathbb{B}_{\alpha}$ is the subspace of all entire functions in $ L_{2}(\mathbb{C} , d\mu_{\alpha}(z))$ with respect to the weight \\ 

\begin{equation}
\displaystyle{\mu_{\alpha}(z) = \frac{\alpha}{\pi}e^{- \alpha \vert z \vert^{2}} ; \alpha > 0}
\end{equation}

\n as follows\\

\begin{equation}
\displaystyle{\mathcal{B}_{\alpha}[f](z) = (\frac{2\alpha}{\pi})^{\frac{1}{4}} e^{\frac{\alpha}{2}z^{2}} \int_{\mathbb{R}} e^{-\alpha(z - u)^{2}} f(u)du, \,\, z \in \mathbb{C}}
\end{equation}
\n provided that the integral is finite.\\ 

\n (iv)  {\color{red}$\bullet_{1}$}  Monograph on Fock spaces {\color{blue}[48]} covers many properties of the Bargmann transform in its general normalization with $\alpha > 0$.\\ 

\n {\color{red}$\bullet_{2}$} Analysis of the Bose-Einstein condensation in the semi-classical limit in {\color{blue}[1]} uses $\displaystyle{ \alpha = \frac{1}{\sqrt{h}}}$ with the small parameter $h > 0$.\\

\n {\color{red}$\bullet_{3}$} Introduction of the Bargmann transform in harmonic analysis in ({\color{blue}[13]}, Section 1.6] uses $ \alpha = \frac{1}{\sqrt{\pi}}$\\

\n {\color{red}$\bullet_{4}$} The work on the lowest Landau level in {\color{blue}[15]} uses $\alpha = 1$\\

\n   {\color{red}$\bullet_{5}$} It is advisable to consult at this topic the recent paper {\color{blue}[36]}.\\
 
 \n   {\color{red}$\bullet_{6}$} Let $\displaystyle{ \mathcal{N}(z, u) = e^{ \frac{\alpha}{2}z^{2} -\alpha(z - u)^{2} = e^{- \frac{\alpha}{2}z^{2}  + 2\alpha uz - \alpha u^{2}}} = e^{- \frac{ (\sqrt{\alpha} z)^{2}}{2}  + 2\sqrt{\alpha} u\sqrt{\alpha}z - \alpha u^{2}}}$, \\
 
 \n $\displaystyle{[\mathbb{I}_{\alpha}f](z) = \int_{\mathbb{R}}  \mathcal{N}(z, u) f(u)du}$ and $\xi = \sqrt{\alpha}z$. Then \\
 
 \n $\displaystyle{ \tilde{\mathcal{N}}(\xi, u) = \mathcal{N}(\sqrt{\alpha}z , u) = e^{- \frac{\xi^{2}}{2}  + 2\sqrt{\alpha} u\xi - \alpha u^{2}}}$,\\
 
  \n $\displaystyle{e^{-\alpha \vert z \vert^{2}} dz \Lambda d\overline{z} = \frac{1}{\alpha} e^{- \vert \xi \vert^{2}} d\xi \Lambda d\overline{\xi}}$
 
 \n  and\\
 
 \n $\displaystyle{[\mathbb{I}_{\alpha}f](\xi) = \int_{\mathbb{R}} \tilde{ \mathcal{N}}(\xi, u) f(u)du}$.\\
 
 \n If we use $\alpha = \frac{1}{2}$, we obtain the kernel of classical Bargmann transform :\\
 
 \n   $\displaystyle{ \tilde{\mathcal{N}}(\xi, u) =  e^{- \frac{\xi^{2}}{2}  + \sqrt{2} u\xi - \frac{u^{2}}{2}}}$ , $\xi \in \mathbb{C} $ and $u \in \mathbb{R}$.\\
 
\end{remark}

\n {\color{red}$\rhd$} It is usefull to observe that the following linear transform \\

\n $\displaystyle{ \mathcal{T} : L_{2}(\mathbb{R}) \longrightarrow L_{2}(\mathbb{R}, e^{-u^{2}}du)}$, $\displaystyle{f  \longrightarrow \mathcal{T}(f) = e^{\frac{u^{2}}{2}}f}$\\

\n is a unitary isomorphism.\\

\n {\color{red}$\rhd$}  We recall that in $L_{2}(\mathbb{R})$ representation, the operators $\displaystyle{a = \frac{1}{\sqrt{2}}( \frac{\partial}{\partial u} + u)}$  and  $\displaystyle{a^{*} =  \frac{1}{\sqrt{2}}( -\frac{\partial}{\partial u} + u)}$ are adjoint with respect to the $L_{2}(\mathbb{R})$-inner product and satisfy $[a, a^{*}] = I$.\\

\n The Segal-Bargmann transform was introduced (depently) by I. Segal and V. Bargmann near 1960 and also by F.A Berezin in the same time give a generalization of this transformation). It  has to satisfy the following two diagrams:\\

\n $\displaystyle{\begin{array}{c} L_{2}(\mathbb{R}) \quad \longrightarrow ^{\underset {\longrightarrow}{SB}} \longrightarrow  \quad  \mathbb{B}\\
\quad \quad \downarrow a \quad \quad  \quad \quad \quad \quad \quad \downarrow \mathfrak{D}\\
L_{2}(\mathbb{R}) \quad \longrightarrow   ^{\underset {\longrightarrow}{SB}} \longrightarrow \quad  \mathbb{B}\\
\end{array}}$ \,\,and \,\, $\displaystyle{\begin{array}{c} L_{2}(\mathbb{R}) \quad \longrightarrow ^{\underset {\longrightarrow}{SB}} \longrightarrow  \quad  \mathbb{B}\\
\quad \quad \downarrow a^{*} \quad \quad  \quad \quad \quad \quad \quad  \downarrow \mathfrak{M}\\
L_{2}(\mathbb{R}) \quad \longrightarrow   ^{\underset {\longrightarrow}{SB}} \longrightarrow \quad  \mathbb{B}\\
\end{array}}$\\

\begin{lemma}
\n $ (i) \,\,  \displaystyle{\mathfrak{D} A(z, u) = a^{*}A(z, u)}$\\

\n $ (ii) \,\, \displaystyle{\mathfrak{M} A(z, u) = a A(z,u)}$\\

\n $ (iii) \,\, \displaystyle{\int_{\mathbb{R}} A(z, u)\overline{A(\xi,u)} du = c e^{z\overline{\xi}}, c > 0}$\\
\end{lemma}
\n{\bf{\color{red}Proof}}\\

\n (i) As $ \displaystyle{\mathfrak{D} A(z, u) = \frac{\partial}{\partial z} A(z, u) = (-z + \sqrt{2}u)A(z, u)}$\\
\n  and \\
\n $\displaystyle{a^{*} A(z, u) = \frac{1}{\sqrt{2}}(- \frac{\partial}{\partial u}  + u) A(z, u) = (-z + \sqrt{2}u)A(z, u)}$ \\

\n Then $\displaystyle{\mathfrak{D} A(z, u) = a^{*}A(z, u)}$\\

\n (ii) As  $\displaystyle{\frac{\partial}{\partial u} A(z, u) = (-u + \sqrt{2}z)A(z, u)}$ then $ \displaystyle{ \frac{1}{\sqrt{2}}(\frac{\partial}{\partial u}  + u) A(z, u) = zA(z, u)}$.\\
Hence $ \displaystyle{\mathfrak{M} A(z, u) = a A(z,u)}$\\

\n (iii) $\displaystyle{< \mathcal{A}(z, u),   \mathcal{A}(\xi, u) >_{L_{2}(\mathbb{R})} = \int_{\mathbb{R}} A(z, u)\overline{A(\xi,u)} du }$ \\
$\displaystyle{ = \int_{\mathbb{R}}  e^{-\frac{u^{2}}{2} + \sqrt{2}uz -\frac{z^{2}}{2}}\overline{ e^{-\frac{u^{2}}{2} + \sqrt{2}u\xi -\frac{\xi^{2}}{2}}}du}$  \\

\n $\displaystyle{ = \int_{\mathbb{R}}  e^{-\frac{u^{2}}{2} + \sqrt{2}uz -\frac{z^{2}}{2}} e^{-\frac{u^{2}}{2} + \sqrt{2}u\overline{\xi} -\frac{\overline{\xi}^{2}}{2}}du}$v

\n $\displaystyle{ = e^{ -\frac{z^{2}}{2}}e^{ -\frac{\overline{\xi}^{2}}{2}} \int_{\mathbb{R}}  e^{-\frac{u^{2}}{2} + \sqrt{2}uz} e^{-\frac{u^{2}}{2} + \sqrt{2}u\overline{\xi}}du}$\\

\n $\displaystyle{ = e^{ -\frac{1}{2}(z^{2} + \overline{\xi}^{2})} \int_{\mathbb{R}}  e^{- u^{2} + \sqrt{2}u(z + \overline{\xi})}du}$ \\

\n $\displaystyle{ = e^{ -\frac{1}{2}(z^{2} + \overline{\xi}^{2})} e^{\frac{(z + \overline{\xi})^{2}}{2}}\int_{\mathbb{R}}  e^{- (u - \frac{z + \overline{\xi}}{\sqrt{2}})^{2}}du}$ \\

\n $\displaystyle{ = e^{z\overline{\xi}}\int_{\mathbb{R}}e^{-t^{2}} dt}$\\

\n $\displaystyle{ = \sqrt{\pi}e^{z\overline{\xi}}}$ \\

\n where $\displaystyle{t = u - \frac{z + \overline{\xi}}{\sqrt{2}}}$. \hfill { } $\blacksquare$ \\

\begin{corollary}
\n (i) $ \displaystyle{ \mathcal{A}(z, u)}$ is in  $L_{2}(\mathbb{R})$ for all $z \in \mathbb{C}$\\

\n (ii) $ \displaystyle{ \mathcal{A}(z, u) f(u)}$ is integrable for $f \in  L_{2}(\mathbb{R})$ \\

\n (iii) $ \displaystyle{ \vert SB(f)(z)  \vert  \leq \sqrt{\pi} e^{z\overline{z}}}$\\

\end{corollary}

\n{\bf{\color{red}Proof}}\\

\n (i) From (iii) of above lemma we get that, for every $z$ in $\mathbb{C}$, we have 

\n $\displaystyle{< A(z, u) , A(z, u) >_{L_{2}(\mathbb{R})} = \int_{\mathbb{R}} A(z, u)\overline{A(z,u)} du = \sqrt{\pi} e^{z\overline{z}}}$\\
 
\n Therefore $\displaystyle{ A(z, u)}$ is in  $L_{2}(\mathbb{R})$ for all $z \in \mathbb{C}$\\

\n (ii) Combining (i) with Cauchy-Schwarz's inequality and the fact that $f$ in $L_{2}(\mathbb{R})$ implies $\overline{f}$ in $L_{2}(\mathbb{R})$, we find that $ \displaystyle{ \mathcal{A}(z, u) f(u)}$ is integrable for $f \in  L_{2}(\mathbb{R})$ i.e. $\displaystyle{SB(f)(z)}$ is well defined for all $f$ in $L_{2}(\mathbb{R})$  \\

\n (iii) From Schwarz' inequality it also follows that $ \displaystyle{ \vert SB(f)(z)  \vert  \leq \sqrt{\pi} e^{z\overline{z}}}$ since $\displaystyle{\vert\vert A(z, u) \vert\vert_{L_{2}(\mathbb{R})} = \sqrt{\pi} e^{z\overline{z}}}$.\hfill { } $\blacksquare$\\
 
\n {\bf Theorem 6.2.3}\\

\n The Segal-Bargmann transform $\displaystyle{SB : L_{2}(\mathbb{R}) \longrightarrow \mathbb{B} : f  \longrightarrow \varphi}$ defined by \\

\n $\displaystyle{SB[f](z) = \varphi(z) = \int_{\mathbb{R}}\mathcal{A}(z, u)f(u)du}$ with $\displaystyle{\mathcal{A}(z, u) = e^{-\frac{u^{2}}{2} + \sqrt{2}uz -\frac{z^{2}}{2}}}$ is a holomorphic function and it is injective.\\

\n{\color{red} {\bf Proof}}\\

\n {\color{red} $\rhd$} Showing that for any $f \in L_{2}(\mathbb{R})$ the image $SBf$ is a holomorphic function. We will use the following classical theorem:\\

\begin{theorem}
({\color{red} Classical theorem})\\

\n Let $h :\mathbb{R}\times \mathbb{C} \longrightarrow \mathbb{C}$ such that:\\

\n (1) For all $z \in \mathbb{C}$ the function $ u \longrightarrow h(u, z)$  is {\color{red}integrable}.\\

\n (2) For all $u \in \mathbb{R}$ the function $ z \longrightarrow h(u, z)$  is {\color{red}holomorphic}.\\

\n (3) For each $z_{0} \in \mathbb{C}$ there exists an open neighbourhood $U$ of $z_{0}$ and a {\color{red}non-negative} function $g \in L_{1}(\mathbb{R})$ (the space of integrable functions) such that for all $ z \in U$ $\displaystyle{\vert h(u, z) \vert \leq g(u)}$.\\

\n Let $\displaystyle{H(z) = \int_{\mathbb{R}} h(u, z)du}$. Then $H$ is holomorphic on $\mathbb{C}$ and we have :\\

\begin{equation}
\displaystyle{\frac{\partial H}{\partial z}(z_{0}) = \int_{\mathbb{R}} \frac{\partial}{\partial z} h(u, z_{0}) du}
\end{equation}
\end{theorem}

\n {\color{red} $\rhd$} Let $\displaystyle{ = h(u, z) = \mathcal{A}(z, u)f(u) = e^{-\frac{u^{2}}{2} + \sqrt{2}uz -\frac{z^{2}}{2}}f(u)}$. We have already proven that $\mathcal{A}(z; u) f(u)$ is integrable for all $z$ in $\mathbb{C}$. Because $\mathcal{A}(z, u) \in L_{2}(\mathbb{R})$ and $f(u) \in L_{2}(\mathbb{R})$ \\

\n {\color{red} $\rhd$} The second condition follows from the fact that the exponential function is holomorphic.  \\

\n {\color{red} $\rhd$} By applying the following inequalities 

$$\displaystyle{(\epsilon \vert \sqrt{2}z\vert  - \vert u \vert)^{2} \geq 0 \,\, \forall \,\, \epsilon > 0}$$

\n and \\

 $$ \displaystyle{\vert \sqrt{2}zu \vert = \vert \sqrt{2}z\vert \vert u \vert  \leq  \epsilon^{2}\vert z \vert^{2} + \frac{1}{2\epsilon^{2}} \vert u \vert^{2}\,\, \forall \,\, \epsilon > 0}$$

\n we deduce that for $\displaystyle{\epsilon = \sqrt{2}}$ we get \\

$$ \displaystyle{\vert \sqrt{2}zu \vert \leq 2 \vert z \vert^{2} + \frac{1}{4} \vert u \vert^{2}}$$

\n hence the following inequality \\

\n $\displaystyle{ \vert \mathcal{A}(z, u)f(u)  \vert =  \vert e^{-\frac{u^{2}}{2} + \sqrt{2}uz -\frac{z^{2}}{2}}f(u) \vert =  \vert e^{-\frac{u^{2}}{2}} \vert .\vert e^{ \sqrt{2}uz }\vert .\vert e^{-\frac{z^{2}}{2}} \vert . \vert f(u) \vert}$\\

\n $\displaystyle{  \leq   \vert e^{-\frac{u^{2}}{2}} \vert .\vert e^{  2 \vert z \vert^{2} + \frac{1}{4} \vert u \vert^{2} }\vert .\vert e^{-\frac{z^{2}}{2}} \vert . \vert f(u) \vert}$ $\displaystyle{  =   \vert e^{-\frac{u^{2}}{2} + \frac{1}{4} \vert u \vert^{2}} \vert .\vert e^{  2 \vert z \vert^{2} }\vert .\vert e^{-\frac{z^{2}}{2}} \vert . \vert f(u) \vert}$\\ 

\n $\displaystyle{ \leq   \vert e^{- \frac{1}{4} \vert u \vert^{2}} \vert .\vert e^{  2 \vert z \vert^{2} + \frac{1}{2}\vert  z \vert^{2} }\vert . \vert f(u) \vert}$ $\displaystyle{ =   \vert e^{- \frac{1}{4} \vert u \vert^{2}} \vert .\vert e^{\frac{5}{2}\vert  z \vert^{2} }\vert  . \vert f(u) \vert}$ $\displaystyle{ =   \vert e^{\frac{5}{2}\vert  z \vert^{2} - \frac{1}{4} \vert u \vert^{2}} \vert . \vert f(u) \vert}$\\ 
\n Then we observe that the function $\displaystyle{g(u) = e^{\frac{5}{2}\vert  z \vert^{2} - \frac{1}{4} \vert u \vert^{2}}  \vert f(u) \vert}$  is non-negative and integrable (because  $\displaystyle{ e^{\frac{5}{2}\vert  z \vert^{2} - \frac{1}{4} \vert u \vert^{2}} \in L_{2}(\mathbb{R}) \,\,\mbox{and}\,\, f(u) \in L_{2}(\mathbb{R})}$). \\

\n Therefore the third condition of above theorem is also satisfied and we conclude that  \\

\n $\displaystyle{SB[f](z) = \varphi(z) = \int_{\mathbb{R}}\mathcal{A}(z, u)f(u)du}$ is a holomorphic function.\\

\n {\color{red} $\rhd$} Now, we will show that the Segal-Bargmann transform is injective, i.e. prove that $SB( f ) = 0 $ implies $ f = 0.$ \\

\n Let $\displaystyle{z = \frac{iy}{\sqrt{2}}}$ then we have \\

\n $\displaystyle{SB[f](i\frac{y}{\sqrt{2}}) = e^{\frac{y^{2}}{4}} \int_{\mathbb{R}} e^{\frac{-u^{2}}{2}}f(u)e^{iyu}du}$  $\displaystyle{ = e^{\frac{y^{2}}{4}} \mathfrak{F}( e^{\frac{-u^{2}}{2}}f(u))(y)}$  where $\mathfrak{F}$ is the Fourier transform on $ L_{1}(\mathbb{R})$\\

\n The function $\displaystyle{ e^{\frac{-u^{2}}{2}}f(u)}$ is an element of $ \displaystyle{ L_{1}(\mathbb{R}) \cap L_{2}(\mathbb{R})}$ and the Fourier transform is a
bijective operator on $L_{2}(\mathbb{R})$. Hence $\displaystyle{SB[f](z) = 0}$ implies $\displaystyle{ e^{\frac{-u^{2}}{2}}f(u) = 0}$ Consequently, we also have $f(u) = 0$, which proves injectivity. \hfill { } $\blacksquare$\\

\n { \color{red}$\rhd$} Frequently, we denote the Segal-Bargmann $SB$ by $\mathcal{B}$ called Bargmann transform.We denote its range by $SB(L_{2}(\mathbb{R}))$, on this space we define  the following inner product:\\
 \begin{equation}
 \displaystyle{ < \varphi , \psi >_{SB} = < f, g >_{L_{2}(\mathbb{R})}}
  \end{equation}
  \quad\\
 \n where\\
 
 \n $f$ and  $g$ are chosen such that $SB(f) = \varphi$ and $SB(g) = \psi$.\\
 
 \n Then we have \\
 
 \begin{theorem}
 ({\color{red}Unitarity of Segal-Bargmann transform})\\
 
\n (i) $SB$ is an unitary transform from $L_{2}(\mathbb{R})$  to $SB(L_{2}(\mathbb{R}))$ i.e.  an isomorphism between $L_{2}(\mathbb{R})$ and  $SB(L_{2}(\mathbb{R}))$) that preserves the inner product..\\
 
 \n  (ii) $(SB(L_{2}(\mathbb{R})), < , >_{SB})$ is a reproducing kernel Hilbert space.\\
 
\n (iii) $SB(L_{2}(\mathbb{R}))$ and  Bargmann space $\mathbb{B}$ have same kernel.\\
 
\n (iv)   $SB(L_{2}(\mathbb{R})) = \mathbb{B}$.\\
\end{theorem}

\n {\bf{\color{red}Proof}}\\

\n (i) The range $SB(L_{2}(\mathbb{R}))$ of the Segal-Bargmann transform is a vector space. \\

\n As the Segal-Bargmann transform is injective, then the inner product defined by \\
 $$\displaystyle{ < \varphi , \psi >_{SB} = < f, g >_{L_{2}(\mathbb{R})}}$$
  \n where $f$ and  $g$ are chosen such that $SB(f) = \varphi$ and $SB(g) = \psi$.\\

\n is well defined and it is clear that the Segal-Bargmann transform is an unitary mapping from $L_{2}(\mathbb{R})$  to $SB(L_{2}(\mathbb{R}))$.\\

\n (ii) From (1.13) we deduce that, for $\varphi$ in $SB(L_{2}(\mathbb{R}))$ and $z \in \mathbb{C}$, we have\\

$$ \vert \varphi(z) \vert \leq \pi^{\frac{1}{4}} e^{\frac{z\overline{z}}{2}} \vert\vert \varphi \vert\vert_{SB}$$

\n This implies that all evaluation maps are continuous. Hence  $SB(L_{2}(\mathbb{R}))$ is a reproducing kernel Hilbert space.\\

\n (iii) By the Riesz-representation , this implies that for each evaluation map there exists a $k_{z}$ in $SB(L_{2}(\mathbb{R}))$ such that, for all $\varphi$ in $SB(L_{2}(\mathbb{R}))$,\\

$$\displaystyle{\varphi(z) = < \varphi , k_{z} >_{SB}}$$

and also by definition of the Segal-Bargmann transform,\\
$$\displaystyle{SB[f](z) = \varphi(z) = \int_{\mathbb{R}} \mathcal{A}(z, u) f(u)du}$$

$$\displaystyle{ =  < f , \overline{\mathcal{A}(z, .)} >_{L_{2}(\mathbb{R})}}$$

$$\displaystyle{ =  < SB f , SB\overline{\mathcal{A}(z, .)} >_{SB}}$$

$$\displaystyle{ =  < \varphi , SB\overline{\mathcal{A}(z, .)} >_{SB}}$$ for all $\varphi \in SB(L_{2}(\mathbb{R}))$

\n Thus we find that \\

$$\displaystyle{K(z, w) = k_{w}(z) = SB(\overline{\mathcal{A}(w, .)}(z)}$$

$$\displaystyle{ \quad \quad \quad = \int_{\mathbb{R}} \mathcal{A}(z, u) \overline{\mathcal{A}(w, u)}du}$$

$$\displaystyle{ = e^{z\overline{w}} \quad \quad \quad \quad }$$

\n where we have used (iii) of lemma 3.4.\\

\n (iii) The above kernel is equal to the kernel we found for $\mathbb{B}$ in lemma 2.22 page 25.\\

\n Therefore, \\

\n (iv) We have by using theorem 2.25 page 27 that $SB(L_{2}(\mathbb{R})$ is equal to Segal-Bargmann space $\mathbb{B}$.\hfill { } $\blacksquare$ \\

\begin{proposition}

One has\\
\begin{equation}
\displaystyle{(z + \frac{\partial}{\partial z})\mathcal{B}[f](z) = \sqrt{2}\mathcal{B}[uf(u)](z)}
\end{equation}
\begin{equation}
\displaystyle{\sqrt{2}z\mathcal{B}[f](z) = \mathcal{B}[(u - \frac{\partial}{\partial u})f(u)](z)}
\end{equation}
\end{proposition}
\n {\bf { \color{red}Proof}}\\

\n One has\\

\n (i) $\displaystyle{\frac{\partial}{\partial z}\mathcal{B}[f](z) =  \int_{\mathbb{R}}\frac{\partial}{\partial z}A(z, u)f(u) du = \int_{\mathbb{R}}(- z +  \sqrt{2}u)A(z, u)f(u) du}$\\

 \n $\displaystyle{ = - z  \int_{\mathbb{R}}A(z, u)f(u) du +  \sqrt{2}\int_{\mathbb{R}}A(z, u)uf(u) du }$ \\
 
\n and\\

\n $\displaystyle{z\mathcal{B}[f](z) =  z\int_{\mathbb{R}}A(z, u)f(u) du}$ \\

\n Then \\

\n $\displaystyle{(z + \frac{\partial}{\partial z})\mathcal{B}[f](z) = \sqrt{2}\mathcal{B}[uf(u)](z)}$ \\

 \n(ii) Consider the rapidly decreasing function $\mathcal{A}(z, u)$ of $u$ with a parameter $z$ arising in the definition of the Bargmann transform then we integrate by parts to get\\

 \n $\displaystyle{\int_{\mathbb{R}}A(z, u)[-\frac{\partial}{\partial u }f(u) ] du = {\color{red}[- A(z, u)f(u)]_{_{-\infty}}^{^{}+ \infty}} + \int_{\mathbb{R}}\frac{\partial}{\partial u }[A(z, u)] f(u) du}$\\
 
 \n $\displaystyle{ = {\color{red}0 } + \int_{\mathbb{R}}(- u +  \sqrt{2}z)A(z, u)f(u)du}$  $\displaystyle{ = - \int_{\mathbb{R}}(A(z, u)uf(u)du}$ +  $\displaystyle{  \sqrt{2}z\int_{\mathbb{R}}A(z, u)f(u)du}$\\
 
 \n This implies \\
 
 \n $\displaystyle{\sqrt{2}z\mathcal{B}[f](z) = \mathcal{B}[(u - \frac{\partial}{\partial u})f(u)](z)}$.\hfill { } $\blacksquare$\\

\begin{corollary}

\n One has\\

 \n (i) $\displaystyle{z\mathcal{B}[f](z) = \mathcal{B}[ \frac{1}{\sqrt{2}}(u - \frac{\partial}{\partial u})f(u)](z)}$ \\
 
  \n (ii) $\displaystyle{\frac{\partial}{\partial z}\mathcal{B}[f](z) = \mathcal{B}[ \frac{1}{\sqrt{2}}(u + \frac{\partial}{\partial u})f(u)](z)}$ \\
\end{corollary}
\n {\bf { \color{red}Proof}}\\

\n (i) From (3.12), we deduce that \\

 \n $\displaystyle{z\mathcal{B}[f](z) = \mathcal{B}[ \frac{1}{\sqrt{2}}(u - \frac{\partial}{\partial u})f(u)](z)}$
 
 \n and \\
 
\n $\displaystyle{z\mathcal{B}[f](z)   = -   \mathcal{B}[ \frac{1}{\sqrt{2}} \frac{\partial}{\partial u})f(u)](z) +  \mathcal{B}[ \frac{1}{\sqrt{2}}uf(u)](z)}$ \hfill { } {\color{red}($\star$)}\\

\n (ii) From (3.11) and {\color{red}($\star$)}, we deduce that\\

\n $\displaystyle{\frac{\partial}{\partial z}\mathcal{B}[f](z) =  - z \mathcal{B}[f](z) + \sqrt{2}\mathcal{B}[uf(u)](z)}$\\

\n and\\

\n $\displaystyle{\frac{\partial}{\partial z}\mathcal{B}[f](z) =    \mathcal{B}[ \frac{1}{\sqrt{2}} \frac{\partial}{\partial u})f(u)](z) -  \mathcal{B}[ \frac{1}{\sqrt{2}}uf(u)](z) +  \sqrt{2}\mathcal{B}[uf(u)](z)}$

\n $\displaystyle{= \mathcal{B}[ \frac{1}{\sqrt{2}} \frac{\partial}{\partial u})f(u)](z)  + \mathcal{B}[ (\sqrt{2} - \frac{1}{\sqrt{2}})uf(u)](z) }$\\

\n $\displaystyle{= \mathcal{B}[ \frac{1}{\sqrt{2}} \frac{\partial}{\partial u})f(u)](z)  + \mathcal{B}[ \frac{1}{\sqrt{2}}uf(u)](z) }$\\

\n $\displaystyle{= \mathcal{B}[ \frac{1}{\sqrt{2}} (\frac{\partial}{\partial u} + u)f(u)](z) }$.\hfill { } $\blacksquare$\\

\begin{remark}

\n {\color{red}$\bullet_{1}$} Therefore the Segal-Bargmann transform $\mathcal{B}$  maps the dimensionless creation and annihilation operators \\

$$ \displaystyle{a^{*} = \frac{1}{\sqrt{2}}(- \frac{\partial}{\partial u} + u) \,\,\, \mbox{and} \,\,\,a = \frac{1}{\sqrt{2}}( \frac{\partial}{\partial u} + u)}$$

\n on $L_{2}(\mathbb{R})$, onto the respective creation and annihilation operators :\\
$$ \displaystyle{A^{*} = z \,\,\, \mbox{and} \,\,\ A = \frac{\partial}{\partial z}}$$
\n on Segal-Bargmann space.\\

\n {\color{red}$\bullet_{2}$} The Segal-Bargmann transform of the normalised Hermite functions\\

\begin{equation}
\displaystyle{ h_{n}(u) = \frac{1}{2^{n}.n! \sqrt{\pi}} (u - \frac{\partial}{\partial u})^{n} e^{-\frac{u^{2}}{2}}}
\end{equation}
\n is given by\\
\begin{equation}
\displaystyle{\mathcal{B}[\frac{1}{2^{n}.n! \sqrt{\pi}} (u - \frac{\partial}{\partial u})^{n} e^{-\frac{u^{2}}{2}}](z) = \frac{z^{n}}{\sqrt{n! \sqrt{\pi}}} \mathcal{B}[e^{-\frac{u^{2}}{2}}] = \frac{z^{n}}{\sqrt{n!}}}
\end{equation}
\n Because  $\displaystyle{ \mathcal{B}[e^{-\frac{u^{2}}{2}}] = \pi^{\frac{1}{4}}}$\\

\n {\color{red}$\bullet_{3}$} These are exactly the normalised basis functions $e_{n}$ of the Segal-Bargmann space.\\

\n {\color{red}$\bullet_{4}$} Given that the Segal-Bargmann transform maps the basis of $L_{2}(\mathbb{R})$ on the basis of $\mathbb{B}$, it is a natural question to ask whether also an inverse transform exists.\\
\end{remark}

 \begin{definition} 
  ({\color{blue} [20]})\\
  
 \n The inverse Segal-Bargmann transform $\displaystyle{\mathcal{B}^{-1} : \mathbb{B} \longrightarrow L_{2}(\mathbb{R})}$ of a function, $\varphi$, is given by:\\

\begin{equation}
\displaystyle{(\mathcal{B}^{-1}\varphi) (z) = c \int_{\mathbb{C}} \varphi(z) e^{-\frac{1}{2}(u^{2} + \overline{z}^{2}) + \sqrt{2} \overline{z}u}e^{- \vert z \vert^{2}}dxdy}
\end{equation}
\end{definition}

\n {\color{red}$\rhd$} We end this subsection on the Segal-Bargmann transform $\mathcal{B}_{\alpha}$ by addressing its classical properties.\\ 

\n We recall that the Segal-Bargmann transform $\mathcal{B}_{\alpha}$ of $ \displaystyle{ f(u) : \mathbb{R} \longrightarrow \mathbb{R}}$ given in (3.9) is:\\ 
$$\displaystyle{\mathcal{B}_{\alpha}[f](z) = c_{\alpha} e^{\frac{\alpha}{2}z^{2}} \int_{\mathbb{R}} e^{-\alpha(z - u)^{2}} f(u)du =  c_{\alpha} \int_{\mathbb{R}} e^{-\alpha u^{2} + 2\alpha u z - \frac{\alpha}{2}z^{2}} f(u)du,  z \in \mathbb{C},  u \in \mathbb{R}}$$
\n or\\

\n $\displaystyle{ \mathcal{B}_{\alpha}[f](z) = c_{\alpha} \int_{\mathbb{R}}\mathcal{N}(z, u)  f(u)du}$ \\

\n where\\

\n  $\displaystyle{ \mathcal{N}(z, u)   = e^{-\alpha u^{2} + 2\alpha u z - \frac{\alpha}{2}z^{2}}}$ \hfill { } {\color{blue} ($\mathcal{N}$)}\\

\n and \\

\n $\displaystyle{c_{\alpha} = (\frac{2\alpha}{\pi})^{\frac{1}{4}}}$ is a constant.\\

\n This transform is associated to Segal-Bargmann space :\\ 
$$\displaystyle{\mathbb{B}_{\alpha} = \{\varphi \in L_{2}(\mathbb{C}, \mu_{\alpha}) ; \varphi(z) \,\, \mbox{ is entire in} \,\, z \in \mathbb{C} \}}$$
\n where\\ 
$$\displaystyle{\mu_{\alpha}(z) = \frac{\alpha}{\pi}e^{-\alpha \vert z \vert^{2}}}$$

$$\displaystyle{ \vert\vert \varphi \vert\vert_{L_{2}(\mathbb{C}, d\mu_{\alpha})}^{2}  = \frac{\alpha}{\pi}\int_{\mathbb{C}}\vert \varphi(z) \vert^{2}e^{-\alpha \vert z \vert^{2}}dxdy}$$

\n  Since $\mathcal{B}_{\alpha}[f](z)$ does not depend on $\overline{z}$ for every $z \in \mathbb{C}$, then  $\mathcal{B}_{\alpha}[f](z)$ is entire in $\mathbb{C}$. \\

\begin{lemma}
 $$\displaystyle{\vert\vert [\mathcal{B}_{\alpha}f] \vert\vert_{L_{2}(\mathbb{C}, d\mu_{\alpha})} = \vert\vert f \vert\vert_{L_{2}(\mathbb{R})} \,\, \forall \,\, f \in  L_{2}(\mathbb{R})}$$
 \end{lemma}
 
 \n {\bf{\color{red} Proof}}\\
 
 \n As $\displaystyle{\mathcal{B}_{\alpha}[f](z) = (\frac{2\alpha}{\pi})^{\frac{1}{4}} e^{\frac{\alpha}{2}z^{2}} \int_{\mathbb{R}} e^{-\alpha(z - u)^{2}} f(u)du }$ \\
 \n and \\
 \n $\displaystyle{\overline{\mathcal{B}_{\alpha}[f](z)} = (\frac{2\alpha}{\pi})^{\frac{1}{4}} e^{\frac{\alpha}{2}\overline{z}^{2}} \int_{\mathbb{R}} e^{-\alpha(\overline{z} - u')^{2}} \overline{f(u')}du' }$ \\
 
 \n  Then we deduce that \\
 
\n  $\displaystyle{\vert \mathcal{B}_{\alpha}[f](z)\vert^{2} = (\frac{2\alpha}{\pi})^{^{\frac{1}{2}}} e^{\frac{\alpha}{2}(z^{2} + \overline{z}^{2})} \int_{\mathbb{R}^{2}} e^{-\alpha[(z - u)^{2} + (\overline{z} - u')^{2}] } f(u)\overline{f(u')}dudu' }$ \\

\n  $\displaystyle{ = (\frac{2\alpha}{\pi})^{^{\frac{1}{2}}} e^{\frac{\alpha}{2}(z^{2} + \overline{z}^{2})} \int_{\mathbb{R}^{2}} e^{-\alpha[z^{2} -2zu + u^{2} + \overline{z}^{2} - 2\overline{z}u' + u'^{2}] }f(u)\overline{f(u')}dudu'}$\\

\n  $\displaystyle{ = (\frac{2\alpha}{\pi})^{^{\frac{1}{2}}} e^{\frac{\alpha}{2}(z^{2} + \overline{z}^{2})} \int_{\mathbb{R}^{2}} e^{-2\alpha[x^{2} -y^{2} - x(u +u') + iy(u -u') + \frac{u^{2}}{2} +  \frac{u'^{2}}{2}] }f(u)\overline{f(u')}dudu'}$\\

\n  $\displaystyle{ = (\frac{2\alpha}{\pi})^{^{\frac{1}{2}}} e^{\frac{\alpha}{2}(z^{2} + \overline{z}^{2})} \int_{\mathbb{R}^{2}} e^{2\alpha y^{2}} e^{-2\alpha [x^{2} - x(u +u') +\frac{u^{2}}{2} +  \frac{u'^{2}}{2}]} e^{-2i\alpha y(u -u')}f(u)\overline{f(u')}dudu'}$\\

\n  $\displaystyle{ = (\frac{2\alpha}{\pi})^{^{\frac{1}{2}}} e^{\frac{\alpha}{2}(z^{2}+ \overline{z}^{2})} \int_{\mathbb{R}^{2}} e^{2\alpha y^{2}} e^{-2\alpha [x - \frac{1}{2}(u +u')]^{2}} e^{-\frac{\alpha}{2}(u - u')^{2}} e^{-2i\alpha y(u -u')}f(u)\overline{f(u')}dudu'}$\\

\n so \\

\n $\displaystyle{\vert\vert [\mathcal{B}_{\alpha}f] \vert\vert_{L_{2}(\mathbb{C}, d\mu_{\alpha})}^{2} = \frac{\alpha}{\pi}\int_{\mathbb{C}} \vert [\mathcal{B}_{\alpha}f](z) \vert^{2} e^{-\alpha \vert z \vert^{2}} dxdy}$\\

\n $\displaystyle{ = \sqrt{2}(\frac{\alpha}{\pi})^{\frac{3}{2}}\int_{\mathbb{\mathbb{R}^{2}}} e^{\alpha(x^{2} - y^{2})}\int_{\mathbb{R}^{2}} e^{2\alpha y^{2}} e^{-2\alpha [x - \frac{1}{2}(u +u')]^{2}} e^{-\frac{\alpha}{2}(u - u')^{2}} e^{-2i\alpha y(u -u')} }$\\

\n $\displaystyle{e^{-\alpha (x^{2} + y^{2})}  f(u) \overline{f(u')}du du' dxdy}$\\

\n $\displaystyle{ = \sqrt{2}(\frac{\alpha}{\pi})^{\frac{3}{2}}\int_{\mathbb{\mathbb{R}^{2}}} \int_{\mathbb{R}^{2}}e^{-2\alpha [x - \frac{1}{2}(u +u')]^{2}} e^{-\frac{\alpha}{2}(u - u')^{2}} e^{-2i\alpha y(u -u')}  f(u) \overline{f(u')}du du' dxdy}$\\

\n $\displaystyle{ = \sqrt{2}(\frac{\alpha}{\pi})^{\frac{3}{2}}\int_{\mathbb{\mathbb{R}^{2}}} \int_{\mathbb{R}}e^{-2\alpha [x - \frac{1}{2}(u +u')]^{2}} e^{-\frac{\alpha}{2}(u - u')^{2}} f(u) \overline{f(u')}du du' dx\int_{\mathbb{R}} e^{-2i\alpha y(u -u')} dy}$\\

\n Now, as $\displaystyle{ \int_{\mathbb{R}} e^{-2i\alpha y(u -u')} dy = \delta (u - u')}$ where $\delta$ is Dirac-distribution, it follows that:\\

\n $\displaystyle{\vert\vert [\mathcal{B}_{\alpha}f] \vert\vert_{L_{2}(\mathbb{\mathbb{C}}, d\mu_{\alpha})}^{2} = \sqrt{2}(\frac{\alpha}{\pi})^{\frac{3}{2}}\int_{\mathbb{\mathbb{R}^{2}}} \int_{\mathbb{R}}e^{-2\alpha [x - \frac{1}{2}(u +u')]^{2}} e^{-\frac{\alpha}{2}(u - u')^{2}} f(u) \overline{f(u')}\delta (u - u')du du' dx}$\\

\n $\displaystyle{ =  \sqrt{2}(\frac{\alpha}{\pi})^{\frac{1}{2}}\int_{\mathbb{R}} \int_{\mathbb{R}}e^{-2\alpha (x - u)^{2}}\vert f(u)\vert^{2} du dx}$\\

\n $\displaystyle{ =  \sqrt{2}(\frac{\alpha}{\pi})^{\frac{1}{2}}\int_{\mathbb{R}} \vert f(u)\vert^{2} (\int_{\mathbb{R}}e^{-2\alpha (x - u)^{2}}dx) du}$\\

\n $\displaystyle{ = \int_{\mathbb{R}} \vert f(u)\vert^{2}du}$\\

\n $\displaystyle{ = \vert\vert f \vert\vert_{L_{2}(\mathbb{R})}^{2}}$. \hfill { } $\blacksquare$\\

\n From {\color{blue}[49]} or {\color{blue}[36]}, we can address the first classical properties of \\ $\mathcal{B}_{\alpha}$-transform : \\ 

\n{\color{red}$\bullet_{1}$} $\mathcal{B}_{\alpha}$ is an isometry from $L_{2}(\mathbb{R})$ to $\mathbb{B}_{\alpha}$\\

\n{\color{red}$\bullet_{2}$} $\displaystyle{\mathcal{B}_{\alpha}^{-1}[\varphi](u) = c_{\alpha} \int_{\mathbb{C}} e^{-\alpha u^{2} + 2\alpha u \overline{z} - \frac{\alpha}{2}\overline{z}^{2}} \varphi(z)d\mu(z)}$\\

\n{\color{red}$\bullet_{3}$} $\displaystyle{\mathcal{B}_{\frac{\alpha}{2}}[uf](z) = (\frac{1}{\alpha} \frac{\partial}{\partial z} + \frac{z}{2}) [\mathcal{B}_{\frac{\alpha}{2}}f](z) }$ \\

\n{\color{red}$\bullet_{4}$} $\displaystyle{\mathcal{B}_{\frac{\alpha}{2}}[\frac{\partial}{\partial u}f](z) = ( \frac{\partial}{\partial z} - \frac{\alpha}{2}z) [\mathcal{B}_{\frac{\alpha}{2}}f](z) }$ \\

\n{\color{red}$\bullet_{5}$} $\displaystyle{\mathcal{B}_{\frac{\alpha}{2}}[(\frac{\partial}{\partial u} -  \alpha u)f](z) = - \alpha z [\mathcal{B}_{\frac{\alpha}{2}}f](z) }$ \\

\n {\color{red}$\bullet_{6}$} $\displaystyle{\mathcal{B}_{\frac{\alpha}{2}}[(\frac{\partial}{\partial u} +  \alpha u)f](z) = 2\frac{\partial}{\partial z} [\mathcal{B}_{\frac{\alpha}{2}}f](z) }$ \\

\n In following subsection, we introduce the adjoint Bargmann transform of  \\

\n $\displaystyle{\mathcal{B}_{\alpha}: L_{2}(\mathbb{R})  \longrightarrow \mathbb{B}_{\alpha} \subset L_{2} (\mathbb{C} , d\mu_{\alpha})}$ defined by $\displaystyle{\mathcal{B}_{\alpha}^{*}:  L_{2} (\mathbb{C} , d\mu_{\alpha}) \longrightarrow  L_{2}(\mathbb{R})}$  :\\

 \begin{equation}
 \displaystyle{(\mathcal{B}_{\alpha}^{*}\varphi)(u) = \frac{2^{\frac{1}{4}}\alpha^{\frac{5}{4}}}{\pi^{\frac{5}{4}}} \int_{\mathbb{C}}e^{\frac{\alpha}{2}\overline{z}^{2}  - \alpha(u - \overline{z})^{2}}\varphi(z)e^{ - \alpha \vert z \vert^{2}} dxdy}
 \end{equation}
 \n or\\
 
 \n  $\displaystyle{(\mathcal{B}_{\alpha}^{*}\varphi)(u) = \frac{2^{\frac{1}{4}}\alpha^{\frac{5}{4}}}{\pi^{\frac{5}{4}}} \int_{\mathbb{C}}\mathcal{N}(\overline{z}, u)\varphi(z)e^{ - \alpha \vert z \vert^{2}} dxdy}$ \\
 
 \n where \\
 
\n $\displaystyle{\mathcal{N}(\overline{z}, u) = e^{\frac{\alpha}{2}\overline{z}^{2}  - \alpha(u - \overline{z})^{2}} = \overline{\mathcal{N}(z, u)}}$ \hfill { } {\color{blue}($\overline{\mathcal{N}}$)}\\
 
\subsection{{\color{blue}Some properties of the adjoint of Segal-Bargmann transform}}

\begin{proposition}
({\color{red} Classical properties of  $\mathcal{B}_{\alpha}^{*}$})\\

\n (i) Let $\displaystyle{(\mathcal{B}_{\alpha}^{*}\varphi)(u) = \frac{2^{\frac{1}{4}}\alpha^{\frac{5}{4}}}{\pi^{\frac{5}{4}}} \int_{\mathbb{C}}e^{\frac{\alpha}{2}\overline{z}^{2}  - \alpha(u - \overline{z})^{2}}\varphi(z)e^{ - \alpha \vert z \vert^{2}} dxdy}$ then \\

\n $\displaystyle{\mathcal{B}_{\alpha}^{*}\varphi \in L_{2}(\mathbb{R}) \,\, \forall \,\, \varphi \in L_{2} (\mathbb{C} , d\mu_{\alpha})}$. \\

\n (ii) $\displaystyle{<\mathcal{B}_{\alpha}f, \varphi >_{ L_{2} (\mathbb{C} , d\mu_{\alpha})} = < f, \mathcal{B}_{\alpha}^{*}\varphi >_{L_{2}(\mathbb{R})}}$.\\

\n (iii) $\displaystyle{ \mathcal{B}_{\alpha}^{*} :  L_{2} (\mathbb{C} , d\mu_{\alpha}) \longrightarrow  L_{2}(\mathbb{R})}$  is the {\color{red}left inverse }   of  $\displaystyle{ \mathcal{B}_{\alpha} :   L_{2}(\mathbb{R})  \longrightarrow  \mathbb{B}_{\alpha}  \subset   L_{2} (\mathbb{C} , d\mu_{\alpha})}$ \\

\n so $\displaystyle{\mathcal{B}_{\alpha}^{*}\mathcal{B}_{\alpha}f = f \,\, \forall \,\, f \in L_{2}(\mathbb{R}) }$.\\

\end{proposition}

\n {\bf{\color{red} Proof}}\\

\n As $\displaystyle{(\mathcal{B}_{\alpha}^{*}\varphi)(u) = \frac{2^{\frac{1}{4}}\alpha^{\frac{5}{4}}}{\pi^{\frac{5}{4}}} \int_{\mathbb{C}}e^{\frac{\alpha}{2}\overline{z}^{2}  - \alpha(u - \overline{z})^{2}}\varphi(z)e^{ - \alpha \vert z \vert^{2}} dxdy}$ then \\

\n $\displaystyle{ \overline{(\mathcal{B}_{\alpha}^{*}\varphi)(u)} = \frac{2^{\frac{1}{4}}\alpha^{\frac{5}{4}}}{\pi^{\frac{5}{4}}} \int_{\mathbb{C}}e^{\frac{\alpha}{2}z^{2}  - \alpha(u - z)^{2}} \overline{\varphi(z)}e^{ - \alpha \vert z \vert^{2}} dxdy}$ \\

 \n So \\
 
 \n $\displaystyle{\vert\vert (\mathcal{B}_{\alpha}^{*}\varphi)\vert\vert_{L_{2}(\mathbb{R})}^{2} = \int_{\mathbb{R}} \vert (\mathcal{B}\varphi)(u) \vert^{2} du}$\\
  
 \n $\displaystyle{= \frac{2^{\frac{1}{2}}\alpha^{\frac{5}{2}}}{\pi^{\frac{5}{2}}} \int_{\mathbb{R}} \int_{\mathbb{C}^{2}} e^{\frac{\alpha}{2}\overline{z}^{2}  - \alpha(u - \overline{z})^{2}}\varphi(z)e^{ - \alpha \vert z \vert^{2}} e^{\frac{\alpha}{2}z'^{2}  - \alpha(u - z')^{2}} \overline{\varphi(z')}e^{ - \alpha \vert z' \vert^{2}} dxdydx'dy' du}$\\
 
 \n $\displaystyle{= \frac{2^{\frac{1}{2}}\alpha^{\frac{5}{2}}}{\pi^{\frac{5}{2}}} \int_{\mathbb{R}} \int_{\mathbb{C}^{2}} e^{\frac{\alpha}{2}\overline{z}^{2} -  \frac{\alpha}{2}\vert z\vert ^{2}         - \alpha(u - \overline{z})^{2}}\varphi(z)e^{ -  \frac{\alpha}{2} \vert z \vert^{2}} e^{\frac{\alpha}{2}z'^{2} - \frac{\alpha}{2}\vert z'\vert^{2} - \alpha(u - z')^{2}} \overline{\varphi(z')}e^{ - \ \frac{\alpha}{2} \vert z' \vert^{2}} dxdydx'dy' du}$\\
 
  \n $\displaystyle{= \frac{2^{\frac{1}{2}}\alpha^{\frac{5}{2}}}{\pi^{\frac{5}{2}}} \int_{\mathbb{R}} \int_{\mathbb{C}^{2}} e^{\frac{\alpha}{2}\overline{z}^{2} -  \frac{\alpha}{2}\vert z\vert ^{2}         - \alpha(u - \overline{z})^{2}} e^{\frac{\alpha}{2}z'^{2} - \frac{\alpha}{2}\vert z'\vert^{2} - \alpha(u - z')^{2}}\varphi(z)e^{ -  \frac{\alpha}{2} \vert z \vert^{2}} \overline{\varphi(z')}e^{ - \ \frac{\alpha}{2} \vert z' \vert^{2}} dxdydx'dy' du}$\\

  \n $\displaystyle{= \frac{2^{\frac{1}{2}}\alpha^{\frac{5}{2}}}{\pi^{\frac{5}{2}}} \int_{\mathbb{R}} \int_{\mathbb{C}^{2}} e^{\frac{\alpha}{2}\overline{z}^{2} -  \frac{\alpha}{2}\vert z\vert ^{2}         - \alpha(u - \overline{z})^{2}} e^{\frac{\alpha}{2}z'^{2} - \frac{\alpha}{2}\vert z'\vert^{2} - \alpha(u - z')^{2}}\phi(z) \overline{\phi(z')}dxdydx'dy' du}$\\
  
  \n where $\displaystyle{\phi(z) = \varphi(z)e^{ - \frac{\alpha}{2} \vert z \vert^{2}}}$\\

\n Now, as \\

\n {\color{red} $\rhd$} $\displaystyle{e^{\frac{\alpha}{2}\overline{z}^{2} -  \frac{\alpha}{2}\vert z\vert ^{2}- \alpha(u - \overline{z})^{2}} = e^{\frac{\alpha}{2}\overline{z}^{2} -  \frac{\alpha}{2}\vert z\vert ^{2}- \alpha(u^{2} - 2u \overline{z}) - \alpha \overline{z}^{2}}}$\\

\n $\displaystyle{= e^{-\frac{\alpha}{2}\overline{z}^{2} -  \frac{\alpha}{2}\vert z\vert ^{2}- \alpha(u^{2} - 2u \overline{z})}}$\\

 \n $\displaystyle{= e^{-\alpha x^{2} - \alpha u^{2} + 2\alpha ux - i\alpha xy - 2i\alpha uy}}$\\
 
 \n $\displaystyle{= e^{- \alpha (x - u)^{2}  - i\alpha xy - 2i\alpha uy}}$\\
 
\n and\\

\n {\color{red} $\rhd$} $\displaystyle{ e^{\frac{\alpha}{2}z'^{2} - \frac{\alpha}{2}\vert z'\vert^{2} - \alpha(u - z')^{2}} = e^{\frac{\alpha}{2}z'^{2} -  \frac{\alpha}{2}\vert z\vert ^{2} - \alpha(u^{2} - 2u z') - \alpha z'^{2}}}$\\

 \n $\displaystyle{= e^{-\frac{\alpha}{2}z'^{2} -  \frac{\alpha}{2}\vert z\vert ^{2} - \alpha(u^{2} - 2u z')}}$\\

 \n $\displaystyle{= e^{- \alpha (x' - u)^{2}  + i\alpha x'y' - 2i\alpha uy'}}$\\

\n Then we deduce that\\

 \n $\displaystyle{\vert\vert (\mathcal{B}_{\alpha}^{*}\varphi)\vert\vert_{L_{2}(\mathbb{R})}^{2} = \int_{\mathbb{R}} \vert (\mathcal{B}_{\alpha}^{*}\varphi)(u) \vert^{2} du}$\\
  
 \n $\displaystyle{= \frac{2^{\frac{1}{2}}\alpha^{\frac{5}{2}}}{\pi^{\frac{5}{2}}} \int_{\mathbb{R}} \int_{\mathbb{R}^{4}} e^{ - \alpha(x - u)^{2} - \alpha (x' - u)^{2} + i\alpha(x'y' - xy) - 2i\alpha u(y - y')}\phi(z)\overline{\phi(z')} dxdydx'dy' du}$.\\
 \par \hfill { } $\blacksquare$\\
 
\begin{lemma}
 ({\color{red}An explicit computation})\\
 
 \n $\displaystyle{\int_{\mathbb{R}} \int_{\mathbb{R}^{4}} e^{ - \alpha(x - u)^{2} - \alpha (x' - u)^{2} + i\alpha(x'y' - xy) - 2i\alpha u(y - y')}\phi(z)\overline{\phi(z')} dxdydx'dy' du}$\\
 
\n $\displaystyle{=  \int_{\mathbb{R}} \int_{\mathbb{R}^{4}} e^{ - 2\alpha( u - \frac{x + x'}{2}  - i \frac{y - y'}{2})^{2} - \frac{\alpha}{2}(x - x')^{2} - \frac{\alpha}{2}(y - y')^{2} +  i\alpha(x'y - xy') }\phi(z)\overline{\phi(z')} dxdydx'dy' du}$\\

\n $\displaystyle{=   \int_{\mathbb{R}^{4}} e^{ - \frac{\alpha}{2}(x - x')^{2} - \frac{\alpha}{2}(y - y')^{2} +  i\alpha(x'y - xy')} \phi(z)\overline{\phi(z')} ( \int_{\mathbb{R}}e^{ - 2\alpha( u - \frac{x + x'}{2}  - i \frac{y - y'}{2})^{2}}du) dxdydx'dy' }$\\

\n $\displaystyle{=  \frac{\sqrt{\pi}}{\sqrt{2\alpha} } \int_{\mathbb{R}^{4}} e^{ - \frac{\alpha}{2}(x - x')^{2} - \frac{\alpha}{2}(y - y')^{2} +  i\alpha(x'y - xy')} \phi(z)\overline{\phi(z')}  dxdydx'dy' }$\\
 \end{lemma}
\n {\bf{\color{red} Proof}}\\

\n $\displaystyle{- 2\alpha( u - \frac{x + x'}{2}  - i \frac{y - y'}{2})^{2} - \frac{\alpha}{2}(x - x')^{2} - \frac{\alpha}{2}(y - y')^{2} +  i\alpha(x'y - xy') {\color{red}=} }$\\

\n $\displaystyle{- 2\alpha[ (u - \frac{x + x'}{2})^{2}  - 2i (u -\frac{x + x'}{2} )(\frac{y - y'}{2}) -  (\frac{y - y'}{2})^{2}] {\color{red}-}}$\\
 
\n $\displaystyle{  \frac{\alpha}{2}(x^{2} -2xx' + x'^{2}) - \frac{\alpha}{2}(y^{2} - 2yy' +y'^{2}) +  i\alpha(x'y - xy') {\color{red} =} }$\\

\n $\displaystyle{-\alpha {\color{red}\{}2 [u^{2} - (x + x')u +  (\frac{x + x'}{2})^{2}] - 4iu(\frac{y - y'}{2} ) + 4i (\frac{x + x'}{2})(\frac{y - y'}{2}) - 2( \frac{y^{2} - 2yy' +y'^{2}}{4}) {\color{red} +} }$\\

\n $\displaystyle{\frac{x^{2}}{2} - xx' + \frac{x'^{2}}{2} + \frac{y^{2}}{2} - yy' + \frac{y'^{2}}{2} - ix'y + ix y'{\color{red}\}}}$\\

\n $\displaystyle{{\color{red}=} - \alpha[ 2u^{2} - 2xu - 2x'u + \frac{1}{2} (x + x')^{2} -2iu(y - y') + i(x + x')(y - y') {\color{red}+}}$\\

\n $\displaystyle{ \frac{x^{2}}{2} - xx' + \frac{x'^{2}}{2} - ix'y + ix y']}$\\

\n $\displaystyle{{\color{red}=} - \alpha[ 2u^{2} - 2xu - 2x'u + \frac{1}{2} x^{2} + xx'  + \frac{1}{2}x'^{2} + i(y - y')(-2u + (x + x')) + i(xy' - x'y) {\color{red}+}}$\\

\n $\displaystyle{ \frac{x^{2}}{2} - xx' + \frac{x'^{2}}{2}]}$\\

\n $\displaystyle{{\color{red}=} - \alpha[ (u^{2} - 2xu  + x^{2}) + (u^{2} - 2x'u + x'^{2}) + i(-2u(y - y') + (y - y')(x + x') + xy' - x'y] }$\\

\n $\displaystyle{{\color{red}=} - \alpha[ (u - x)^{2} + (u - x')^{2}  + i(-2u(y - y') + xy - x'y'] }$\\

\n $\displaystyle{{\color{red}=} - \alpha(u - x)^{2}  \alpha (u - x')^{2}  + 2i \alpha u(y - y') - i \alpha (xy - x'y') }$\\

\n $\displaystyle{{\color{red}=} - \alpha(u - x)^{2}  \alpha (u - x')^{2}  + 2i \alpha u(y - y') + i \alpha (x'y' - xy) }$\\

\n It follows that \\

\n $\displaystyle{\int_{\mathbb{R}} \int_{\mathbb{R}^{4}} e^{ - \alpha(x - u)^{2} - \alpha (x' - u)^{2} + i\alpha(x'y' - xy) - 2i\alpha u(y - y')}\phi(z)\overline{\phi(z')} dxdydx'dy' du}$\\
 
\n $\displaystyle{=  \int_{\mathbb{R}} \int_{\mathbb{R}^{4}} e^{ - 2\alpha( u - \frac{x + x'}{2}  - i \frac{y - y'}{2})^{2} - \frac{\alpha}{2}(x - x')^{2} - \frac{\alpha}{2}(y - y')^{2} +  i\alpha(x'y - xy') }\phi(z)\overline{\phi(z')} dxdydx'dy' du}$\\

\n And by Fubini's theorem, we deduce that\\

\n $\displaystyle{\int_{\mathbb{R}} \int_{\mathbb{R}^{4}} e^{ - \alpha(x - u)^{2} - \alpha (x' - u)^{2} + i\alpha(x'y' - xy) - 2i\alpha u(y - y')}\phi(z)\overline{\phi(z')} dxdydx'dy' du}$\\
 
\n $\displaystyle{=   \int_{\mathbb{R}^{4}} e^{ - \frac{\alpha}{2}(x - x')^{2} - \frac{\alpha}{2}(y - y')^{2} +  i\alpha(x'y - xy')} \phi(z)\overline{\phi(z')} ( \int_{\mathbb{R}}e^{ - 2\alpha( u - \frac{x + x'}{2}  - i \frac{y - y'}{2})^{2}}du) dxdydx'dy' }$\\

\n Now  as   \\

\n $\displaystyle{\int_{\mathbb{R}}e^{ - 2\alpha( u - \frac{x + x'}{2}  - i \frac{y - y'}{2})^{2}}du  = \frac{\sqrt{\pi}}{\sqrt{2\alpha}}}$\\

\n then\\

\n $\displaystyle{\int_{\mathbb{R}} \int_{\mathbb{R}^{4}} e^{ - \alpha(x - u)^{2} - \alpha (x' - u)^{2} + i\alpha(x'y' - xy) - 2i\alpha u(y - y')}\phi(z)\overline{\phi(z')} dxdydx'dy' du}$\\

\n $\displaystyle{=  \frac{\sqrt{\pi}}{\sqrt{2\alpha} } \int_{\mathbb{R}^{4}} e^{ - \frac{\alpha}{2}(x - x')^{2} - \frac{\alpha}{2}(y - y')^{2} +  i\alpha(x'y - xy')} \phi(z)\overline{\phi(z')}  dxdydx'dy' }$.\hfill { } $\blacksquare$\\

\n This implies that  the following corollary\\

\begin{corollary}
\quad\\

\n $\displaystyle{\vert\vert  \mathcal{B}_{\alpha}^{*}\varphi \vert\vert_{L_{2}(\mathbb{R})^{2}} = \frac{\alpha^{2}}{\pi^{2}} \int_{\mathbb{R}^{4}} e^{ - \frac{\alpha}{2}(x - x')^{2} - \frac{\alpha}{2}(y - y')^{2} +  i\alpha(x'y - xy')} \phi(z)\overline{\phi(z')}  dxdydx'dy' }$\\
\end{corollary}

\begin{remark}
\quad\\

\n Let $\displaystyle{F(z) = \int_{\mathbb{R}^{2}} e^{ - \frac{\alpha}{2}(x - x')^{2} - \frac{\alpha}{2}(y - y')^{2} +  i\alpha(x'y - xy')}\overline{\phi(z')}dx'dy' }$, then we observe that  $F(z) = (G* \phi)(z)$ is the convolution of $\phi$ with the gaussian $\displaystyle{G(z) = e^{-\frac{\alpha}{2}\vert z \vert^{2}}}$\\
\end{remark}

\n From this remark, we deduce the (i) of  above proposition in form of the lemma:\\

\begin{lemma}
\quad\\

\n $\displaystyle{\mathcal{B}_{\alpha}^{*}\varphi \in L_{2}(\mathbb{R}) \,\, \mbox{if}\,\, \varphi \in L_{2}(\mathbb{C}, d\mu_{z})}$
\end{lemma}
\n {\bf{\color{red} Proof}}\\

\n From above remark, we deduce that $\vert F(z)\vert  \leq (G* \vert \phi\vert )(z)$ and by using the Cauchy-Schwarz inequality and Young's convolution inequalities, we obtain \\

\n $\displaystyle{\vert\vert \mathcal{B}_{\alpha}^{*}\varphi\vert\vert_{L_{2}(\mathbb{R})}^{2} \leq  \frac{\alpha^{2}}{\pi^{2}} \vert\vert \phi F \vert\vert_{L_{1}(\mathbb{C})}}$\\

\n $\displaystyle{\leq  \frac{\alpha^{2}}{\pi^{2}} \vert\vert \phi \vert\vert_{L_{2}(\mathbb{C})}  \vert\vert (G* \vert \phi\vert) \vert\vert_{L_{2}(\mathbb{C})}}$\\

\n $\displaystyle{\leq  \frac{\alpha^{2}}{\pi^{2}} \vert\vert G \vert\vert_{L_{1}(\mathbb{C})}  \vert\vert \phi \vert\vert_{L_{2}(\mathbb{C})}}$\\

\n $\displaystyle{\leq  \frac{2\alpha}{\pi} \vert\vert \phi \vert\vert_{L_{2}(\mathbb{C})}^{2}}$ = $\displaystyle{ 2 \vert\vert \varphi \vert\vert_{L_{2}(\mathbb{C} , d\mu_{\alpha})}^{2} < \infty}$. \hfill { } $\blacksquare$\\

\begin{lemma}
\quad\\

$\displaystyle{ < \mathcal{B}_{\alpha}f , \varphi >_{L_{2}(\mathbb{C} , d\mu_{\alpha})} = < f , \mathcal{B}_{\alpha}^{*}\varphi >_{L_{2}(\mathbb{R})} \,\, \mbox{for every} \,\, f \in L_{2}(\mathbb{R}),  \varphi \in L_{2}(\mathbb{C} , d\mu_{\alpha}) }$
\end{lemma}
\n {\bf{\color{red} Proof}}\\

\n {\color{blue}$\rhd$} $\displaystyle{ < \mathcal{B}_{\alpha}f , \varphi >_{L_{2}(\mathbb{C} , d\mu_{\alpha})} = \frac{\alpha}{\pi} \int_{\mathbb{C}}\mathcal{B}_{\alpha}f(z) \overline{\varphi(z)}e^{-\alpha \vert z\vert^{2}}  dxdy}$\\

$\displaystyle{=  2^{\frac{1}{4}} (\frac{\alpha}{\pi})^{\frac{5}{4}}  \int_{\mathbb{C}}\int_{\mathbb{R}}\mathcal{N}(z, u) f(u)du \overline{\varphi(z)}e^{-\alpha \vert z\vert^{2}}  dxdy}$\\

$\displaystyle{=  2^{\frac{1}{4}} (\frac{\alpha}{\pi})^{\frac{5}{4}}  \int_{\mathbb{R}} f(u) \int_{\mathbb{C}}\mathcal{N}(z, u)  \overline{\varphi(z)}e^{-\alpha \vert z\vert^{2}}  dxdydu}$\\

$\displaystyle{=  2^{\frac{1}{4}} (\frac{\alpha}{\pi})^{\frac{5}{4}}  \int_{\mathbb{R}} f(u) \int_{\mathbb{C}}\overline{\mathcal{N}(\overline{z}, u)}  \overline{\varphi(z)}e^{-\alpha \vert z\vert^{2}}  dxdydu}$\\

$\displaystyle{= \int_{\mathbb{R}} f(u)\overline{(\mathcal{B}_{\alpha}^{*} \varphi)(u)}du}$\\

$\displaystyle{= < f , \mathcal{B}_{\alpha}^{*}\varphi >_{L_{2}(\mathbb{R})}}$.\\

\n All these computations are justified because Fubini's theorem can be used to interchange integrations.\hfill { } $\blacksquare$\\

\begin{lemma}

$\displaystyle{ \mathcal{B}_{\alpha}^{*}: L_{2}(\mathbb{C} , d\mu_{\alpha}) \longrightarrow L_{2}(\mathbb{R})}$ is the left inverse of \\ $\displaystyle{\mathcal{B}_{\alpha}:  L_{2}(\mathbb{R}) \longrightarrow  \mathbb{B}_{\alpha} \subset L_{2}(\mathbb{C} , d\mu_{\alpha})}$, so that  $\displaystyle{\mathcal{B}_{\alpha}^{*}\mathcal{B}_{\alpha}f = f}$ for every $f \in L_{2}(\mathbb{R})$.\\
\end{lemma}
\n {\bf{\color{red}Proof}}\\

\n From lemma  (3.11) we have \\

\n $\displaystyle{\vert\vert [\mathcal{B}_{\alpha}f] \vert\vert_{L_{2}(\mathbb{C}, d\mu_{\alpha})} = \vert\vert f \vert\vert_{L_{2}(\mathbb{R})} \,\, \forall \,\, f \in  L_{2}(\mathbb{R})}$ \hfill { } {\color{blue}($\circ_{1}$)}\\

\n and from above lemma we have  \\

\n $\displaystyle{< \mathcal{B}_{\alpha} f , \varphi >_{L_{2}(\mathbb{C} , d\mu_{\alpha})}  =   < f , \mathcal{B}_{\alpha}^{*}\varphi >_{L_{2}(\mathbb{R})}}$.\hfill { } {\color{blue}($\circ_{2}$)}\\

\n This implies that for every $f \in L_{2}(\mathbb{R})$ \\

\n $\displaystyle{\vert\vert \mathcal{B}_{\alpha}f \vert\vert_{L_{2}(\mathbb{C}, d\mu_{\alpha})}^{2} = < \mathcal{B}_{\alpha}f , \mathcal{B}_{\alpha}f >_{L_{2}(\mathbb{C}, d\mu_{\alpha})} = < \mathcal{B}_{\alpha}^{*}\mathcal{B}_{\alpha}f , f >_{L_{2}(\mathbb{R})} = \vert\vert  f \vert\vert_{L_{2}(\mathbb{R})}^{2}}$ \hfill { } {\color{blue}($\circ_{3}$)}\\

\n It follows that $ \displaystyle{ \mathcal{B}_{\alpha}^{*}}$ is the left inverse of $\displaystyle{\mathcal{B}_{\alpha}}$.\hfill { } $\blacksquare$\\

\begin{lemma}
({\color{red}An orthonormal system in $ \mathbb{B}_{\alpha}$  and an explicit formula for $\mathcal{B}_{\alpha}\mathcal{B}_{\alpha}^{*} \varphi$ if $\varphi \in L_{2}(\mathbb{C} , d\mu_{\alpha})$})\\

\n (i) Let $\displaystyle{ \varphi_{n,\alpha}(z) = \sqrt{\alpha^{n}} \frac{z^{n}}{\sqrt{n!}},  \alpha > 0 , n = 1, 2, ....}$ then $\displaystyle{<  \varphi_{n,\alpha},  \varphi_{m,\alpha} >_{L_{2}(\mathbb{C} , d\mu_{\alpha}) }  = \delta_{n,m}}$.\\

\n (ii)  $\displaystyle{\mathcal{B}_{\alpha}\mathcal{B}_{\alpha}^{*} \varphi (z) = \frac{\alpha}{\pi} \int_{\mathbb{R}}\int_{\mathbb{C}}e^ {\alpha z \overline{z'}}\varphi(z') d\mu_{\alpha'} dx'dy'}$ if $\varphi \in L_{2}(\mathbb{C} , d\mu_{\alpha})$, z = x + iy \,\,\mbox{and}\,\, z' = x' + iy'.\\ 

\n (iii) If $\varphi \in \mathbb{B}_{\alpha}$,  then  $\displaystyle{\mathcal{B}\mathcal{B}_{\alpha}^{*} \varphi  = \varphi}$\\
\end{lemma}
\n {\bf{\color{red} Proof}}\\

\n (i) We use the polar coordinates $z = x + iy = re^{i\theta},  r > 0 , \theta \in [0, \pi$ to get :\\

\n $\displaystyle{<  \varphi_{n,\alpha},  \varphi_{m,\alpha} >_{L_{2}(\mathbb{C} , d\mu_{\alpha}) }  = \frac{\alpha}{\pi} \int_{\mathbb{C}} \sqrt{\alpha^{n}} \frac{z^{n}}{\sqrt{n!}}\sqrt{\alpha^{m}} \frac{\overline{z}^{m}}{\sqrt{m!}} e^{-\alpha\vert^{2}}dxdy}$\\

\n $\displaystyle{= \frac{\alpha}{\pi} \frac{\sqrt{\alpha^{n}\alpha^{m}}}{{\sqrt{n! m!}}} \int_{\mathbb{C}} z^{n}\overline{z}^{m} e^{-\alpha \vert z \vert^{2}} dxdy}$\\ 

\n $\displaystyle{=  \frac{\alpha}{\pi} \frac{\sqrt{\alpha^{n}\alpha^{m}}}{ \sqrt{n! m!}} ( \int_{0}^{2\pi} e^{i(n - m) \theta}d\theta)(\int_{0}^{+ \infty} r^{n + m + 1}e^{-\alpha r^{2}}dr)}$. \\

\n This yields zero for $n \neq m$ because,  $\displaystyle{  \int_{0}^{2\pi} e^{i(n - m) \theta}d\theta = 0}$.\\

\n For $n = m$ we have \\

\n $\displaystyle{<  \varphi_{n,\alpha},  \varphi_{m,\alpha} >_{L_{2}(\mathbb{C} , d\mu_{\alpha}) }  =  \frac{\alpha}{\pi} \frac{\alpha^{n}}{n!} \int_{0}^{+ \infty}  2\pi r^{2n + 1}e^{-\alpha r^{2}}dr)}$.\\

\n Setting $\displaystyle{r = \frac{1}{\sqrt{\alpha}}u}$ so  $\displaystyle{dr = \frac{1}{\sqrt{\alpha}}du}$ then we get\\

 \n $\displaystyle{ \int_{0}^{+ \infty}  2\pi r^{2n + 1}e^{-\alpha r^{2}}dr =  \frac{1}{\alpha^{n}} \int_{0}^{+ \infty}  2\pi u^{2n + 1}e^{-u^{2}}  \frac{du}{\alpha} = \frac{\pi n!}{\alpha^{n+1}}}$.\\
 
 \n and \\
 
 \n $\displaystyle{<  \varphi_{n,\alpha},  \varphi_{n,\alpha} >_{L_{2}(\mathbb{C} , d\mu_{\alpha}) }  = 1}$.\\

\n (ii)   \n $\displaystyle{\mathcal{B}_{\alpha}\mathcal{B}_{\alpha}^{*} \varphi (z) = \frac{2^{\frac{1}{2}} \alpha^{\frac{3}{2}}}{\pi^{\frac{3}{2}}} e^{\frac{1}{2} z^{2}}\int_{\mathbb{R}}e^{-\alpha(z - u)^{2}}}$ $\displaystyle{ \int_{\mathbb{C}}e^{\frac{1}{2} \overline{z}'^{2} - \alpha(\overline{z' }- u)^{2} - \alpha \vert z' \vert^{2}} \varphi(z')dx'dy' du}$\\

\n $\displaystyle{ =  \frac{2^{\frac{1}{2}} \alpha^{\frac{3}{2}}}{\pi^{\frac{3}{2}}} e^{-\frac{1}{2} z^{2}}\int_{\mathbb{C}}e^{-\frac{3}{2} \alpha x'^{2} - \frac{1}{2}\alpha y'^{2} - i\alpha x'y'}\varphi(z') {\color{red}[} \int_{\mathbb{R}} e^{-2\alpha u^{2} + 2\alpha u(x - iy) + 2\alpha u (x' + i y')}du{\color{red}]} dx'dy'}$\\ 

\n $\displaystyle{ = \frac{\alpha}{\pi} \int_{\mathbb{C}} e^{\alpha(z - z')\overline{z'}}\varphi(z') dx'dy'}$\\

\n $\displaystyle{ = \frac{\alpha}{\pi} \int_{\mathbb{C}} e^{\alpha z\overline{z'} - \alpha \vert z' \vert^{2}}\varphi(z') dx'dy'}$\\

\n $\displaystyle{ = \frac{\alpha}{\pi} \int_{\mathbb{C}} e^{\alpha z\overline{z'}}\varphi(z') d\mu_{\alpha'}dx'dy'}$\\

\n i.e.\\

\n $\displaystyle{\mathcal{B}_{\alpha}\mathcal{B}_{\alpha}^{*} \varphi (z) = \frac{\alpha}{\pi} \int_{\mathbb{C}} e^{\alpha z\overline{z'}}\varphi(z') d\mu_{\alpha'}dx'dy'}$\hfill { } {\color{blue} ($\circ_{4}$)}\\

\n (iii) Every entire function $\varphi$ can be represented by the Taylor series which converges uniformly and absolutely for every $z \in \mathbb{C}$ :\\

\n $\displaystyle{ \varphi(z) = \sum_{n=0}^{\infty} \frac{\varphi^{(n)}(0)}{n!} z^{n}}$. \hfill { } {\color{blue} ($\circ_{5}$)}\\

\n If $\displaystyle{\varphi \in \mathbb{B}_{\alpha} \subset L_{2}(\mathbb{C} , d\mu_{\alpha})}$, this Taylor series becomes the decomposition over the orthonormal
basis of monomials $\displaystyle{\varphi_{n,\alpha}(z) = \frac{\sqrt{\alpha^{n}}z^{n}}{\sqrt{n!}}}$ given by :\\

\n $\displaystyle{\varphi(z) = \sum_{n=0}^{\infty}< \varphi , \varphi_{n,\alpha} >_{L_{2}(\mathbb{C} , d\mu_{\alpha})} \varphi_{n,\alpha}(z)}$ \hfill { } {\color{blue} ($\circ_{6}$)}\\

\n or\\

\n $\displaystyle{\varphi(z) = \sum_{n=0}^{\infty} \frac{\alpha}{\pi} \int_{\mathbb{C}} \varphi(z') \overline{\varphi_{n,\alpha}(z')} d\mu_{\alpha'} dx'dy' \varphi_{n,\alpha}(z)}$ \hfill { } {\color{blue} ($\circ_{7}$)}\\

\n Or in the form of representation :\\

\n $\displaystyle{\varphi(z) = < \varphi , \mathcal{V}_{z} >_{L_{2}(\mathbb{C} , d \mu_{\alpha})}}$   \hfill { } {\color{blue} ($\circ_{8}$)}\\

\n where $\displaystyle{ \mathcal{V}_{z} \in \mathbb{B}_{\alpha}}$ given by  $\displaystyle{ \mathcal{V}_{z}(z') = e^{\alpha zz'}}$.\\

\n Now, from {\color{blue} ($\circ_{4}$)} and if $\varphi \in \mathbb{B}_{\alpha}$ then we have :\\

\n $\displaystyle{\mathcal{B}_{\alpha}\mathcal{B}_{\alpha}^{*} \varphi (z) = \frac{\alpha}{\pi} \int_{\mathbb{C}} e^{\alpha z\overline{z'}}\varphi(z') d\mu_{\alpha'}dx'dy'}$\\

\n $\displaystyle{= \frac{\alpha}{\pi} \int_{\mathbb{C}} \sum_{n=0}^{\infty} \frac{\alpha^{n} z^{n}\overline{z'}^{n}}{n!}\varphi(z') d\mu_{\alpha'}dx'dy'}$\\

\n $\displaystyle{= \frac{\alpha}{\pi} \int_{\mathbb{C}} \sum_{n=0}^{\infty} \frac{\sqrt{\alpha^{n}} z^{n}\sqrt{\alpha^{n}}\overline{z'}^{n}}{\sqrt{n!}\sqrt{n!}}\varphi(z') d\mu_{\alpha'}dx'dy'}$\\

\n $\displaystyle{= \sum_{n=0}^{\infty} (\frac{\alpha}{\pi} \int_{\mathbb{C}}\varphi(z') \overline{\varphi_{n,\alpha}(z')} d\mu_{\alpha'}dx'dy')}\varphi_{n, \alpha}(z)$\\ 

\n $\displaystyle{= \sum_{n=0}^{\infty} < \varphi , \varphi_{n,\alpha} >_{L_{2}(\mathbb{C}, d\mu_{\alpha})} \varphi_{n, \alpha}(z)}$\\

\n $\displaystyle{= \varphi(z)}$\\

\n Hence, If $\varphi \in \mathbb{B}_{\alpha}$,  then  $\displaystyle{\mathcal{B}\mathcal{B}_{\alpha}^{*} \varphi  = \varphi}$. \hfill { } $\blacksquare$\\

\begin{remark}

\n From {\color{blue} ($\circ_{5}$)} and {\color{blue} ($\circ_{6}$)}, we deduce that\\

\n $\displaystyle{\varphi^{(n)}(0) = \frac{\alpha^{n+1}}{\pi} \int_{\mathbb{C}} \varphi(z) \overline{z}^{n}e^{- \alpha \vert z \vert^{2}}dxdy}$  \hfill { } {\color{blue} ($\circ_{9}$)}\\

\n In particular\\

\n $\displaystyle{\varphi(0) = \frac{\alpha}{\pi} \int_{\mathbb{C}} \varphi(z)e^{- \alpha \vert z \vert^{2}}dxdy}$ \hfill { } {\color{blue} ($\circ_{10}$)}\\
\end{remark}

\n We conclude  this subsection by summarizing the properties  presented in it :\\

\n {\color{red}$\star_{1}$} The global Bargmann transform $\displaystyle{\mathcal{B} : L_{2}(\mathbb{R}) \longrightarrow \mathbb{B}_{\alpha} \subset L_{2}(\mathbb{C}, d\mu_{\alpha})}$ is a Hilbert space isomorphism:\\

\n  $\displaystyle{< \mathcal{B}f ,  \mathcal{B}g >_{L_{2}(\mathbb{C} , d\mu_{\alpha})} = < f, g >_{L_{2}(\mathbb{R})} \,\, f, g \in L_{2}(\mathbb{R})}$.\\

\n  {\color{red}$\star_{2}$} The adjoint $\displaystyle{\mathcal{B}_{\alpha}^{*} :  L_{2}(\mathbb{C}, d\mu_{\alpha})  \longrightarrow L_{2}(\mathbb{R})}$ of  $\mathbb{B}_{\alpha}$ is defined by:\\

 \n $\displaystyle{(\mathcal{B}_{\alpha}^{*}\varphi)(u) = \frac{2^{\frac{1}{4}}\alpha^{\frac{5}{4}}}{\pi^{\frac{5}{4}}} \int_{\mathbb{C}}e^{\frac{\alpha}{2}\overline{z}^{2}  - \alpha(u - \overline{z})^{2}}\varphi(z)e^{ - \alpha \vert z \vert^{2}} dxdy}$\\
 
 \n  {\color{red}$\star_{3}$} \n $\displaystyle{\mathcal{B}_{\alpha}\mathcal{B}_{\alpha}^{*} \varphi (z) = \frac{\alpha}{\pi} \int_{\mathbb{C}} e^{\alpha z\overline{z'}}\varphi(z') d\mu_{\alpha'}dx'dy'}$\\
 
  \n  {\color{red}$\star_{4}$} $\displaystyle{\varphi_{n,\alpha}(z) = \frac{\sqrt{\alpha^{n}}z^{n}}{\sqrt{n!}}}$ is a complete orthonormal system of $\mathbb{B}_{\alpha}$.\\
  
   \n  {\color{red}$\star_{5}$}  $\displaystyle{\mathcal{B}_{\alpha}^{*}\varphi_{n,\alpha}(z) = \frac{\sqrt{\alpha^{n}}z^{n}}{\sqrt{n!}}}$ is a {\color{red}complete orthonormal system} of $L_{2}(\mathbb{R})$.\\
   
   $\displaystyle{ \varphi(z) = < \varphi , \mathcal{V}_{z} >}$ where $\displaystyle{\mathcal{V}_{z} \in \mathbb{B}_{\alpha}}$ given by $\displaystyle{\mathcal{V}_{z}(z') = e^{\alpha z\overline{z'}}}$.\\
   
   \n  {\color{red}$\star_{6}$}   $\displaystyle{\mathcal{B}_{\alpha}^{*} : L_{2} (\mathbb{C} , d\mu_{\alpha}) \longrightarrow   L_{2} (\mathbb{R})}$ is the {\color{red} left inverse} of $\displaystyle{\mathcal{B}_{\alpha} : L_{2} (\mathbb{R}) \longrightarrow  \mathbb{B}_{\alpha} \subset L_{2} (\mathbb{C} , d\mu_{\alpha})}$.\\
    
 \n  {\color{red}$\star_{7}$} The adjoint Bargmann transform  $\mathcal{B}_{\alpha}^{*} $ is {\color{red}not a unitary transformation} from $ L_{2} (\mathbb{C} , d\mu_{\alpha}) $ to $ L_{2}(\mathbb{R})$. {\bf In fact:} \\
 
 \n We consider the nonzero function $\varphi(z) = \overline{z} $ which belongs to $L_{2} (\mathbb{C} , d\mu_{\alpha}) $ then $\displaystyle{\vert\vert \varphi \vert\vert_{ L_{2} (\mathbb{C} , d\mu_{\alpha})} \neq 0}$. \\
 
 \n  As $\mathcal{B}_{\alpha}^{*} \overline{z} \in L_{2}(\mathbb{R})$ computing $\displaystyle{\vert\vert \mathcal{B}_{\alpha}^{*} \overline{z}\vert\vert_{L_{2}(\mathbb{R})}}$ :\\
 
 \n $\displaystyle{\mathcal{B}_{\alpha}^{*} \overline{z} = \frac{2^{\frac{1}{4}} \alpha^{\frac{5}{4}}}{ \pi^{\frac{5}{4}}} \int_{\mathbb{C}} (x - iy)e^{-\frac{3}{2}\alpha x^{2} - \frac{1}{2}\alpha y^{2} - \alpha u^{2} + 2\alpha xu - i\alpha y(2u - x)}dxdy}$\\
 
\n $\displaystyle{ = \frac{2^{\frac{1}{4}} \alpha^{\frac{5}{4}}}{ \pi^{\frac{5}{4}}} \int_{\mathbb{C}} (x - iy)e^{-2\alpha (x - u)^{2}  - \alpha u^{2} - \frac{1}{2}\alpha (y - i(x - 2u)^{2}} dxdy}$\\
 
 \n $\displaystyle{ = \frac{2^{\frac{5}{4}} \alpha^{\frac{3}{4}}}{ \pi^{\frac{3}{4}}} \int_{\mathbb{R}} (x - u)e^{-2\alpha (x - u)^{2}  - \alpha u^{2}} dx}$\\
 
  \n $\displaystyle{ = \frac{2^{\frac{5}{4}} \alpha^{\frac{3}{4}}}{ \pi^{\frac{3}{4}}} \frac{- e^{\alpha u^{2}}}{4\alpha}\int_{\mathbb{R}} -4\alpha(x - u)e^{-2\alpha (x - u)^{2} } dx}$\\
  
  \n $\displaystyle{ = \frac{2^{\frac{5}{4}} \alpha^{\frac{3}{4}}}{ \pi^{\frac{3}{4}}} [e^{-2\alpha t^{2}}]_{_{-\infty}}^{^{+\infty}}}$ where $t = x -u$.\\
  
  \n $\displaystyle{ = 0}$\\
  
 \n Hence \\
 
\n  $\displaystyle{\vert\vert \varphi \vert\vert_{ L_{2} (\mathbb{C} , d\mu_{\alpha})^{2}} \neq \vert\vert \mathcal{B}_{\alpha}^{*} \varphi \vert\vert^{2}_{L_{2} (\mathbb{R})}}$\\
 
 \n and\\
  
 \n  {\color{red}$\star_{8}$}  $\displaystyle{\mathcal{B}_{\alpha}^{*}  :  L_{2} (\mathbb{C} , d\mu_{\alpha}) \longrightarrow  L_{2} (\mathbb{R})}$ {\color{red}is not the right inverse} of
   $\displaystyle{\mathcal{B}_{\alpha} :  L_{2} (\mathbb{R}) \longrightarrow  \mathbb{B}_{\alpha} \subset L_{2} (\mathbb{C} , d\mu_{\alpha})}$\\
   
 \n  {\color{red}$\star_{9}$}   $\displaystyle{\mathcal{B}_{\alpha}^{*} :   \mathbb{B}_{\alpha} \subset L_{2} (\mathbb{C} , d\mu_{\alpha}) \longrightarrow  L_{2} (\mathbb{R})}$ is a {\color{red}unitary transformation}.\\
 
\n  {\color{red}$\star_{10}$} the orthogonal projection operator of $\displaystyle{L_{2} (\mathbb{C} , d\mu_{\alpha}) \longrightarrow  \mathbb{B}_{\alpha} \subset L_{2} (\mathbb{C} , d\mu_{\alpha})}$ is given by $\mathcal{B}_{\alpha}\mathcal{B}_{\alpha}^{*} $.\\
 
\n  {\color{red}$\star_{11}$} It is well known that every function $\varphi \in \mathbb{B}_{\alpha}$ must satisfy the growth condition\\

\n $\displaystyle{\lim \limits_{z \longrightarrow \infty}e^{-\alpha \vert z \vert^{2}} \varphi(z) = 0}$ ( see page 38 of {\color{blue}[49]} for example). \\

\subsection{ {\color{blue} The relationship between Bargmann transform,  Gabor transform and FBI transform}}

\n We recall the definition and properties of Gabor transform({\color{blue}[7], [8]}).\\

\begin{definition}

 Gabor transform $W_{ \phi}(f)(p, q)$ is defined as follows:\\

$$ \displaystyle{W_{\phi}(f) (p, q) = \int_{\mathbb{R}}\overline{\phi_{p,q}(u)} f(u) du \quad (f \in L_{2}(\mathbb{R}), u, p, q \in \mathbb{R})}$$

$\displaystyle{\phi(u) = \frac{1}{\pi^{\frac{1}{4}}} e^{-\frac{u^{2}}{2}}}$ is Gaussian and $\displaystyle{\phi_{p,q}(u) = \frac{1}{\pi^{\frac{1}{4}}}  e^{ip u} e^{-\frac{(u - q)^{2}}{2}}}$ is Gabor function.
\end{definition}

\begin{proposition} (Inversion formula of Gabor transform)\\

$$ \displaystyle{ f(u) = \frac{1}{2\pi} \int_{\mathbb{R}^{2}} \phi_{p,q}(u) W_{\phi}(f) (p, q) dpdq}$$

\end{proposition}

\n {\bf{\color{red} Proof}}\\

$ \displaystyle{ \int_{\mathbb{R}^{2}} \phi_{p,q}(u) W_{\phi}(f) (p, q) dpdq}$\\

$ \displaystyle{ =  \int_{\mathbb{R}^{2}} \phi_{p,q}(u)  \int_{\mathbb{R}} e^{ipv}\phi(v - q) f(v) dvdpdq   }$\\

$ \displaystyle{ =  \int_{\mathbb{R}^{3}} e^{-ip u}\phi(u - q)  e^{ipv}\phi(v - q) f(v) dvdpdq   }$\\

$ \displaystyle{ =  \int_{\mathbb{R}^{2}} \{\int_{\mathbb{R}}e^{-ip (u - v)}dp\}\phi(u - q) \phi(v - q) f(v) dvdq }$\\

$ \displaystyle{ =  2\pi \int_{\mathbb{R}^{2}} \delta(u - v)\phi(u - q) \phi(v - q) f(v) dvdq }$\\

$ \displaystyle{ =  2\pi \int_{\mathbb{R}}\phi(u - q) \phi(u - q) f(u) dq }$\\

$ \displaystyle{ =  2\pi < \phi, \phi > f(u) }$\\

$ \displaystyle{ =  2\pi f(u) }$. \hfill { } $\blacksquare$\\

\begin{proposition} (Unitarity of Gabor Transform))\\

$$\displaystyle{< W_{\phi} (f), W_{\phi} (f) > = \frac{1}{2\pi} < f, g >}$$
\end{proposition}

\n Gabor transform is closely related to FBI (Fourier-Bros-Iagolnitzer) transform and Bargmann transform ({\color{blue}[13]}).\\

\begin{definition} (FBI transform)\\

FBI transform $\displaystyle{P^{t}(f)(p, q)}$ is defined by \\

$$\displaystyle{P^{t}(f)(p, q) = \int_{\mathbb{R}} e^{-ipu}e^{-t(u - q)^{2}} f(u) du}$$
\end{definition}

\n From above definitions we deduce the following relationship between FBI transform, Bargmann transform and Gabor transform :\\

\n {\bf {\color{red}$\bullet_{1}$}} FBI transform is related to Gabor transform as follows \\

$$\displaystyle{P^{\frac{1}{2}}(f)(p, q) = \int_{\mathbb{R}} e^{-ipu}e^{-\frac{1}{2}(u - q)^{2}} f(u) du}$$

\n {\bf {\color{red}$\bullet_{2}$}} Bargmann transform is related to Gabor transform as follows \\

$$\displaystyle{ \mathcal{B}(f)(z) = \frac{1}{\pi^{\frac{1}{4}}} e^{\frac{1}{4}(p^{2} + q^{2} + 2ipq} \int_{\mathbb{R}}e^{- i pu}e^{-\frac{1}{2} (u - q)^{2}} f(u) du \quad (z = \frac{q + ip}{\sqrt{2}})}$$ 

\n Gabor transform is used for iris identification and signal analysis of human voice.\\

\subsection{ {\color{blue}The standard creation and annihilation operators  on Segal-Bargmann space}}

\n The mapping  $\mathcal{B}$, establishes a unitary isomorphism between the linear operators on $ \mathbb{B}$, and those on $L_{2}(\mathbb{R})$, namely, \\

 \begin{equation}
  \hat{T} =  \mathcal{B}^{-1} T  \mathcal{B}
   \end{equation} 
   
  \n where \\
  
\n $T$ is an operator on $\mathbb{B}$, and $\hat{T}$ the corresponding operator on $L_{2}(\mathbb{R})$ \\

\n The domains $ \mathcal{D}(T)$ and  $ \mathcal{D}(\hat{T})$  are related by \\

 \begin{equation}
  \mathcal{D}(\hat{T}) =  \mathcal{B}^{-1} \mathcal{D}(T) 
   \end{equation} 
 
\n and \\

\n The inverse Segal-Bargmann transform $\displaystyle{ \mathcal{B}^{-1} :  \mathbb{B} \longrightarrow L_{2}(\mathbb{R})}$ of a function, $\varphi$, is given by \\

\begin{equation}
\displaystyle{(\mathcal{B}^{-1}\varphi) (z) = c \int_{\mathbb{C}} \varphi(z) e^{-\frac{1}{2}(u^{2} + \overline{z}^{2}) + \sqrt{2} \overline{z}u}e^{- \vert z \vert^{2}}dxdy}
\end{equation}
  
\n We have two unbounded operators  densely defined which are important in the study of $L_{2}(\mathbb{R})$:\\

\n $\displaystyle{Rf(u) = u f(u)}$ (operator of multiplication by $u$), $\displaystyle{Sf(u) = f'(u)}$ (operator of differentiation).\\

\n We will identify the operators on $\mathbb{B}$ that correspond to these two operators\\ under the Bargmann transform.\\

\begin{theorem}
{\color{blue} K. Zhu [50]}\\
\n (i)  For $\varphi \in \mathbb{B}$, we have\\

\begin{equation}
\displaystyle{(\mathcal{B}R\mathcal{B}^{-1}\varphi)(z) = \frac{1}{2}(\varphi^{'}(z)  + z\varphi(z)) = \frac{1}{2}(\frac{\partial}{\partial z} + z)\varphi(z)}
\end{equation}
\n(ii) For $\varphi \in \mathbb{B}$, we have\\
\begin{equation}
\displaystyle{(\mathcal{B}S\mathcal{B}^{-1} \varphi)(z) = \varphi^{'}(z) - z\varphi(z) = (\frac{\partial}{\partial z} - z)\varphi(z)}
\end{equation}
\end{theorem}

\n {\bf{\color{red} Proof}}\\

\n  (i) Let $\varphi \in \mathbb{B}$ and the Bargmann transform $\mathcal{B}$ is the operator from $L_{2}(\mathbb{R}) \longrightarrow \mathbb{B}$ defined by (3.9) as follows\\

\begin{equation}
\displaystyle{(\mathcal{B}f)(z) = c' \int_{\mathbb{R}} e^{2uz - u^{2} - \frac{z^{2}}{2}} f(u) du \quad \mbox{by taking}  \quad c' = (\frac{2}{\pi})^{\frac{1}{4}} }
\end{equation}
Furthermore, the inverse of $\mathcal{B}^{-1}$ is also an integral operator:\\

\begin{equation}
\displaystyle{(\mathcal{B}^{-1}\varphi)(u) = c' \int_{\mathbb{R}} e^{2u\overline{z} - u^{2} - \frac{\overline{z}^{2}}{2}} \varphi(z) d\mu(z) } 
\end{equation}

\n Then we have \\

\n $\displaystyle{R(\mathcal{B}^{-1}\varphi)(u) = c' u \int_{\mathbb{C}} e^{2u\overline{w} - u^{2} - \frac{\overline{w}^{2}}{2}} \varphi(w) d\mu(w) }$\\
 
 \n and so\\
 
 \n $\displaystyle{\mathcal{B}R(\mathcal{B}^{-1}\varphi)(u) = c'^{2}  \int_{\mathbb{R}}u e^{2uz  - u^{2} - \frac{\overline{z}^{2}}{2}} du \int_{\mathbb{C}} e^{2u\overline{w} - u^{2} - \frac{\overline{w}^{2}}{2}} \varphi(w) d\mu(w) }$ \\
 
 \n $\displaystyle{ = c'^{2} e^{- \frac{z^{2}}{2}} \int_{\mathbb{C}}e^{  \frac{\overline{w}^{2}}{2}}  \varphi(w) d\mu(w) \int_{\mathbb{R}} u e^{- 2u^{2}  + 2u(z + \overline{w})} du}$ (by Fubini's theorem) \\

  \n $\displaystyle{ = \frac{ c'^{2} }{2}e^{- \frac{z^{2}}{2}} \int_{\mathbb{C}}e^{  \frac{\overline{w}^{2}}{2}}  \varphi(w) d\mu(w) \int_{\mathbb{R}} v e^{- v^{2}  + \sqrt{2}v(z + \overline{w})} dv}$ (by a change variable $\displaystyle{ u = \frac{v}{\sqrt{2}})}$\\
 
 \n $\displaystyle{ = \frac{ c'^{2} }{2} \int_{\mathbb{C}}e^{ z\overline{w}}  \varphi(w) d\mu(w) \int_{\mathbb{R}} v e^{- [ v -  \frac{z + \overline{w}}{\sqrt{2}}]^{2}} dv}$
 
 \n $\displaystyle{ = \frac{ c'^{2} }{2} \int_{\mathbb{C}}e^{ z\overline{w}}  \varphi(w) d\mu(w) \int_{\mathbb{R}}( v +  \frac{z + \overline{w}}{\sqrt{2}})e^{- v^{2}} dv}$ 
  
\n $\displaystyle{ = \frac{ c'^{2} }{2}\int_{\mathbb{C}} \frac{z + \overline{w}}{\sqrt{2}}e^{ z\overline{w}}  \varphi(w) d\mu(w) \int_{\mathbb{R}}e^{- v^{2}} dv}$ (by using $\displaystyle{\int_{\mathbb{R}}v e^{- v^{2}} dv = 0}$)
 
\n $\displaystyle{ = \frac{ 1}{2} \int_{\mathbb{C}} (z + \overline{w}) e^{ z\overline{w}}  \varphi(w) d\mu(w)}$ (by using the value $\sqrt{\pi}$ of $\displaystyle{\int_{\mathbb{R}} e^{- v^{2}} dv }$ and the value of the constant $c'$).\\
\n $\displaystyle{ = \frac{ 1}{2}[ z\int_{\mathbb{C}}e^{ z\overline{w}}  \varphi(w) d\mu(w) +  \frac{\partial}{\partial z} \int_{\mathbb{C}} e^{ z\overline{w}}  \varphi(w) d\mu(w)]}$\\

\n $\displaystyle{ = \frac{ 1}{2}[ z\varphi(z) + \varphi^{'}(z)]}$ \quad (by using  $\displaystyle{\varphi(z) = \int_{\mathbb{C}}e^{ z\overline{w}}\varphi(w) d\mu(w)}$).\\

 \n  (i) Let $\varphi \in \mathbb{B}$ and $\displaystyle{c' = (\frac{2}{\pi})^{\frac{1}{4}}}$ again. We have\\
 
 \n $\displaystyle{S(\mathcal{B}^{-1}\varphi)(u) = c'  \frac{\partial}{\partial u}\int_{\mathbb{C}} e^{2u\overline{w} - u^{2} - \frac{\overline{w}^{2}}{2}} \varphi(w) d\mu(w) }$\\
 
 \n $\displaystyle{= c' \int_{\mathbb{C}} (2 \overline{w} - 2u)e^{2u\overline{w} - u^{2} - \frac{\overline{w}^{2}}{2}} \varphi(w) d\mu(w) }$\\
 
 \n $\displaystyle{= 2c' \int_{\mathbb{C}}  \overline{w} e^{2u\overline{w} - u^{2} - \frac{\overline{w}^{2}}{2}} \varphi(w) d\mu(w) - 2u (\mathcal{B}^{-1}\varphi)(u)}$\\ 
 
 \n and so\\
 
 \n $\displaystyle{\mathcal{B}S(\mathcal{B}^{-1}\varphi)(u) = 2c'^{2}  \int_{\mathbb{R}}e^{2uz - u^{2} - \frac{z^{2}}{2}} du \int_{\mathbb{C}}  \overline{w} e^{2u\overline{w} - u^{2} - \frac{\overline{w}^{2}}{2}} \varphi(w) d\mu(w) - 2\mathcal{B}R(\mathcal{B}^{-1}\varphi)(u)}$\\ 
 
 \n $\displaystyle{=  2c'^{2}e^{-\frac{z^{2}}{2}} \int_{\mathbb{C}}  \overline{w} e^{-\frac{\overline{w}^{2}}{2}}\varphi(w)d\mu(w) \int_{\mathbb{R}}e^{- 2u^{2} + 2u(z + \overline{w})}du - 2\mathcal{B}R(\mathcal{B}^{-1}\varphi)(u)}$\\ 
 
  \n $\displaystyle{=  \sqrt{2}c'^{2}e^{-\frac{z^{2}}{2}} \int_{\mathbb{C}}  \overline{w} e^{-\frac{\overline{w}^{2}}{2}}\varphi(w)d\mu(w) \int_{\mathbb{R}}e^{- v^{2} + \sqrt{2}v(z + \overline{w})}dv - 2\mathcal{B}R(\mathcal{B}^{-1}\varphi)(u) ; (v = \sqrt{2}u)}$\\ 
  
   \n $\displaystyle{=  \sqrt{2}c'^{2}e^{-\frac{z^{2}}{2}} \int_{\mathbb{C}}  \overline{w} e^{-\frac{\overline{w}^{2}}{2}}\varphi(w)d\mu(w) \int_{\mathbb{R}}e^{- [v - \frac{z + \overline{w}}{\sqrt{2}}]^{2}}dv - 2\mathcal{B}R(\mathcal{B}^{-1}\varphi)(u)}$\\ 
 
 \n $\displaystyle{=  \sqrt{2\pi}c'^{2} \int_{\mathbb{C}}  \overline{w} e^{z\overline{w}}\varphi(w)d\mu(w) - 2\mathcal{B}R(\mathcal{B}^{-1}\varphi)(u) }$\\ 
 
 \n $\displaystyle{= 2 \varphi^{'}(z) - 2\mathcal{B}R(\mathcal{B}^{-1}\varphi)(u) }$\\ 
 
  \n $\displaystyle{= 2 \varphi^{'}(z) - 2[ \frac{1}{2}(z \varphi(z) + \varphi^{'}(z))}$ (by using (i))\\
  
 \n  It follows that \\
 
  \n $\displaystyle{\mathcal{B}S\mathcal{B}^{-1}\varphi(z) = \varphi^{'}(z) - z \varphi(z)}$. \hfill { } $\blacksquare$\\

 \begin{corollary} 
 
 (i) Let $\displaystyle{R_{1} = \mathcal{B}S\mathcal{B}^{-1} }$ and  $\displaystyle{S_{1} = \mathcal{B}S\mathcal{B}^{-1} }$, then $\displaystyle{S_{1}R_{1} - R_{1}S_{1} = I}$.\\
 
 \n (ii) Let $\displaystyle{a = \frac{1}{\sqrt{2}}(\frac{d}{du} + u )}$ then $\displaystyle{\frac{1}{\sqrt{2}}(\frac{d}{du} + u ) =   \mathcal{B}^{-1}  \mathfrak{M} \mathcal{B}}$ where $\mathfrak{M}$ is the operator of multiplication by $z = x + iy ; (x, y) \in \mathbb{R}^{2}$ and $i^{2} = -1$ (i.e $\mathfrak{M}\varphi = z\varphi, \varphi \in O(\mathbb{C})$) \\
 
 \n (iii) Let $\displaystyle{a^{*} = \frac{1}{\sqrt{2}}(-\frac{d}{du} + u )}$ then $\displaystyle{\frac{1}{\sqrt{2}}(-\frac{d}{du} + u ) =   \mathcal{B}^{-1}  \mathfrak{D} \mathcal{B}}$ where $\mathfrak{D}$ is the operator of derivative in order $z$ (i.e   $\displaystyle{\mathfrak{D}\varphi = \frac{\partial}{\partial z}\varphi, \varphi \in O(\mathbb{C})}$)\\
 
 \n We observe that on $O(\mathbb{C})$ the operators $\mathfrak{D}$  and $\mathfrak{M}$ satisfy : \\

\begin{equation}
\displaystyle{[\frac{\partial}{\partial z}, z] = I} 
\end{equation}
 \end{corollary}
 
 \n Define the operators $A$ and $B$ on Segal-Bargmann space $\mathbb{B}$ as follows: \\
 
 \begin{equation}
 A\varphi (z) = \frac{\partial}{\partial z}\varphi(z)  \quad \mbox{if} \quad \varphi \in \mathbb{B}
 \end{equation}
 \begin{equation}
 B\varphi (z) = z\varphi(z) \quad  \mbox{if} \quad \varphi \in \mathbb{B}
 \end{equation}
 
 \n Then we have\\
 
 \begin{proposition} 
 (i) $A$ and $B$ are closed  \\
 
 \n (ii) $D(A) = D(B)$ i.e. $\displaystyle{\{ \varphi \in   \mathbb{B} ; \frac{\partial}{\partial z}\varphi \in  \mathbb{B}\} = \{ \varphi \in   \mathbb{B} ; z \varphi \in  \mathbb{B}\}}$  \\
 
\n (iii) $ A^{*}  = B \quad \mbox{and} \quad B^{*} = A$  \\
 
 \n (iv) $D(A)$ is dense in $\mathbb{B}$\\
 
  \end{proposition} 
  
 \n {\bf{\color{red}Proof}}\\
   
 \n (i) {\color{red}$\rhd$} Let $\displaystyle{\varphi_{n} \longrightarrow \phi}$ in $\mathbb{B}$ and  $\displaystyle{ \frac{\partial}{\partial z}\varphi_{n} \longrightarrow \psi}$ in $\mathbb{B}$. Then, for every $z$, \\
 $\displaystyle{\psi(z) = \lim\limits_{n \longrightarrow + \infty}  \frac{\partial}{\partial z}\varphi_{n}(z)}$ and   $\displaystyle{\frac{\partial}{\partial z}\phi(z) =  \lim\limits_{n \longrightarrow + \infty}  \frac{\partial}{\partial z}\varphi_{n}(z)}$ (by (ii) proposition 2.11), hence $\displaystyle{\psi = \frac{\partial}{\partial z}\phi(z)}$.\\
 
\n  {\color{red}$\rhd$} Let next $\displaystyle{\varphi_{n} \longrightarrow \phi}$ in $\mathbb{B}$ and  $\displaystyle{z\varphi_{n} \longrightarrow \psi}$ in $\mathbb{B}$. Then, for every $z$, \\
$\displaystyle{\psi(z) = \lim\limits_{n \longrightarrow + \infty} z\varphi_{n}(z) =  z\phi(z)}$, hence $\displaystyle{\psi = z\phi(z)}$.\\

\n (ii) Let  \\
$$\displaystyle{\varphi(z) = \sum_{n=0}^{\infty} a_{n}z^{n}} \quad \mbox{and} \quad \displaystyle{\vert\vert \varphi\vert\vert^{2} = \sum_{n=0}^{\infty} n! \vert a_{n} \vert^{2}}$$
\n Then\\
$$\displaystyle{\vert\vert z\varphi\vert\vert^{2} = \sum_{n=0}^{\infty} (1 + n)\,n! \vert a_{n} \vert^{2}}$$
\n and\\
$$\displaystyle{\vert\vert \frac{\partial}{\partial z}\varphi\vert\vert^{2} = \sum_{n=0}^{\infty}  n. n! \vert a_{n} \vert^{2}}$$
\n Hence\\
\begin{equation}
\displaystyle{\vert\vert z\varphi\vert\vert^{2} = \vert\vert \varphi\vert\vert^{2} + \vert\vert \frac{\partial}{\partial z}\varphi\vert\vert^{2}}
\end{equation}
\n In (3.26) either both sides are infinite, or they have the same finite value, which proves (ii) of this proposition.\\

\n (iii) Let $ \displaystyle{\varphi(z) = \sum_{n=0}^{\infty} a_{n}z^{n}} \quad \mbox{and}$ \quad  $ \displaystyle{\phi(z) = \sum_{n=0}^{\infty} b_{n}z^{n}} $, then  \\

\n {\color{red}$\bullet$} $\displaystyle{\frac{\partial}{\partial z}\varphi(z) = \sum_{n=0}^{\infty} n a_{n}z^{n-1} =   \sum_{n=0}^{\infty} (n +1) a_{n+1}z^{n} }$. This implies

\n $\displaystyle{< \frac{\partial}{\partial z}\varphi, \phi > = \sum_{n=0}^{\infty} n! (n +1) a_{n+1} \overline{ b_{n}} = \sum_{n=0}^{\infty} (n +1)!a_{n+1} \overline{ b_{n}} = \sum_{n=0}^{\infty} n! a_{n} \overline{ b_{n-1}} ; \,\,\, b_{-1} = 0}$\\

\n $\displaystyle{ = \, <  \varphi , [\frac{\partial}{\partial z}]^{*}(\phi) >}$.\\

\n {\color{red}$\bullet$} $\displaystyle{z\phi(z) = \sum_{n=0}^{\infty}b_{n}z^{n+1} =   \sum_{n=0}^{\infty} b_{n-1}z^{n} ; \,\,\, b_{-1} = 0}$. This implies $ A^{*} = B$.

\n Similar, we have $ B^{*} = A$. \\

\n (iv) As the set of poly,nomials $\mathcal{P}$ is dense in $\mathbb{B}$ and $\displaystyle{\mathcal{P} \subset D(A)}$, we deduce that $D(A)$ is dense in $\mathbb{B}$. \hfill { } $\blacksquare$\\

\begin{remark}
({\color{red}Fock-Sobolev space})\\

\n  (i) The identity $\displaystyle{\mid\mid z\varphi\mid\mid^{2} = \mid\mid \varphi\mid\mid^{2} + \mid\mid \frac{d}{dz}\varphi\mid\mid^{2}}$ hold for all entire $\varphi$. This identity is the standard bosonic commutation relation for the annihilation operator and the creation operator in quantum field theory; see, for example, ({\color{blue}[20]}, Section 14.4.1) and the space $\mathfrak{B}_{1} = \{\varphi: \mathbb{C}\longrightarrow  \mathbb{C}\, entire ;\mid\mid \varphi\mid\mid + \mid\mid \frac{d}{dz}\varphi\mid\mid < +\infty\}$ is is naturally referred to as the Fock-Sobolev space of order $1$, because of the similarity to the way the classical Sobolev spaces are defined. The Fock-Sobolev spaces of order $m$ have been studied by Cho and Zhu in [6].\\
\n (ii) Let $\displaystyle{\phi(z) = \sum_{k=1}^{\infty}\frac{z^{k}}{\sqrt{k!}k^{2}}}$, we note that $\phi \in \mathbb{B}_{0}$ but $\phi\quad /\!\!\!\!\!\in \mathfrak{B}_{1}$\\
\end{remark}

\n {\color{red}$\rhd$} The realization in $\mathbb{B}_{s}$ of operators :\\

\n $\displaystyle{A\varphi(z) = \frac{d}{dz}\varphi(z)}$ with domain $\displaystyle{D(A) = \{\varphi \in \mathbb{B}; \frac{d}{dz}\varphi \in \mathbb{B}\}}$

\n and \\

\n $\displaystyle{A^{*}\varphi(z) = z\varphi(z)}$ with domain $\displaystyle{D(A^{*}) = \{\varphi \in \mathbb{B}; z\varphi \in \mathbb{B}\}}$\\

\n  where $\displaystyle{\varphi(z) = \sum_{n=0}^{\infty}a_{n}z^{n}}$  is given by :\\

\begin{equation}
\n \left\{\begin{array}[c]{l}A(a_{n})_{n=0}^{\infty}=((n+1)a_{n+1})_{n=0}^{\infty}\\ \quad \\ with \quad domain:\\

\n D(A) = \{(a_{n})_{n=0}^{\infty} \in \mathbb{B}_{s}; \displaystyle{\sum_{n=0}^{\infty}n!n\mid a_{n}
\mid^{2}} < \infty\} \\ \end{array}\right.
\end{equation}
\n and\\
\begin{equation}
\n \left\{\begin{array}[c]{l}A^{*}(a_{n})_{n=0}^{\infty}=(a_{n-1})_{n=0}^{\infty}\quad;\quad a_{-1} = 0 \\ \quad \\ with \quad domain: \\
D(A^{*}) = \{(a_{n})_{n=0}^{\infty} \in \mathbb{B}_{s};\displaystyle{\sum_{n=0}^{\infty}n!\mid a_{n-1}\mid^{2}} < \infty\} \\ \end{array}\right. 
\end{equation}

\n In $\mathbb{B}_{s}$ realization, the first result on the domains of $A$ and $A^{*}$ is\\

\begin{theorem}
\n $D(A) \hookrightarrow \mathbb{B}_{s}$ is compact.\\
\end{theorem}
\n {\color{red}{\bf Proof }}\\

\n Let $C_{s}$ be the unit ball of $D(A)$ equipped with graph norm:\\

\n $C_{s} =\{(a_{n})_{n=0}^{\infty} \in D(a)$ such that $\mid\mid(a_{n})_{n=0}^{\infty}\mid\mid_{s}^{2} + \mid\mid a((a_{n})_{n=0}^{\infty})\mid\mid_{s}^{2} \leq 1\}$ i.e. \\

\n $C_{s} =\{(a_{n})_{n=0}^{\infty} \in D(a)$ such that $\displaystyle {\mid a_{0}\mid^{2} +\sum_{n=1}^{\infty}(n+1)!\mid a_{n}\mid^{2} \leq 1} \}$\\

\n To prove that $C_{s}$ is relatively compact for the norm associated to $B_{s}$, we observe that $B_{s}$ is complete space then
it suuffices to show that $C_{s}$ can be covering by finite number of unit balls of radius $\frac{1}{p}$ with arbitrary $p\in \mathbb{N}-\{0\}$ \\

\n Writing $\displaystyle {\sum_{n=1}^{\infty}(n+1)!\mid a_{n}\mid^{2}}$ as follows\\

 \n  $\displaystyle {\sum_{n=1}^{\infty}(n+1)!\mid a_{n}\mid^{2} = 2!\mid a_{1}\mid^{2} + 3!\mid a_{2}\mid^{2}+4!\mid a_{3}\mid^{2}+ ............} $\\

 \n  $ = $ $1!\mid a_{1}\mid^{2} + 2!\mid a_{2}\mid^{2}+3!\mid a_{3}\mid^{2}+ ............................+$\\

\n   $\quad$ $1!\mid a_{1}\mid^{2} + 2!\mid a_{2}\mid^{2}+3!\mid a_{3}\mid^{2}+ ............................+$\\

\n  $\quad$ $0 \quad \quad \quad + 2!\mid a_{2}\mid^{2}+3!\mid a_{3}\mid^{2}+ .............................+$\\

\n   $\quad$ $0 \quad \quad \quad +  0 \quad \quad \quad +  3!\mid a_{3}\mid^{2}+ ............................+$\\

\n   $\quad$ $............................................................................... +$\\

\n $ = 2 \displaystyle {\sum_{n=1}^{\infty}n!\mid a_{n}\mid^{2}}$ + $\displaystyle {\sum_{n=2}^{\infty}n!\mid a_{n}\mid^{2}}$ + ........ + $\displaystyle{\sum_{n=p}^{\infty}n!\mid a_{n}\mid^{2}}$ + .......\\

\n then we deduce that $(p+1)\displaystyle{\sum_{n=p}^{\infty}n!\mid a_{n}\mid^{2}} \leq 1$
i.e. $(0,0, ......, 0,a_{p}, a_{p+1}, ......)$ is in $C_{s}^{0}(0,\frac{1}{p})$ the ball
with radius $\frac{1}{p}$ around the origin of $B_{s}$.\\

\n As the set $\{(a_{1}, a_{2}, .....,a_{p-1}) \in \mathbb{C}^{p-1}$ ; $\displaystyle {\sum_{n=1}^{p-1}n!\mid a_{n}\mid^{2}}\leq 1\}$ is compact then they exist $m$ balls $K(x_{i},\frac{1}{p})$ with center $x_{i}\in \mathbb{C}^{p-1}$ of radius $\frac{1}{p}$ such that
$\{(a_{1}, a_{2}, .....,a_{p-1} \in \mathbb{C}^{p-1}$;$\displaystyle {\sum_{n=1}^{p-1}n!\mid a_{n}\mid^{2}}\leq 1\} = \bigcup_{i=1}^{m} K(x_{i},\frac{1}{p})$\\

\n then we get\\

\n $C_{s}\subset \displaystyle {\bigcup_{i=1}^{m} }K(x_{i},\frac{1}{p})\bigcup C_{s}^{0}(0,\frac{1}{p})$
and we deduce the property iii).\hfill { } $\blacksquare$ \\

 \n We can also prove iii) by applying a following classic proposition\\

\begin{proposition} 
({\bf {\color{blue} [Fr\'echet]}})\\

\n  Let $K \subset \mathbb{B}_{s}$ such that\\

\n  i) $K$ is bounded and closed.\\

\n  ii) $\forall \varepsilon > 0$, $\exists N_{\varepsilon} > 0$ such that $\displaystyle{\sum_{n= N_{\varepsilon}}^{\infty}n!\mid a_{n}\mid^{2} \leq \varepsilon \quad \forall \{a_{n}\}_{n=1}^{\infty}} \in K$\\

\n then $ K $ is compact.$\hfill { } {\large\diamond} $\\
\end{proposition} 
\n \begin{lemma}

 ({\bf {\color{red}Fundamental lemma}})\\

\n (i) For all integer  $k \neq 0$ and for all polynomial  $P(A, A^{*})$  of degree  $r$ such that $ r < 2k$ then  we have:\\
\begin{equation}
\displaystyle{\forall \,\, \epsilon > 0, \exists C_{\epsilon} > 0 ; \mid < P(A, A^{*})\varphi, \varphi >\mid \leq \epsilon < A^{*k}A^{k}\varphi, \varphi > + C_{\epsilon}\mid\mid \varphi \mid\mid^{2}}
\end{equation}
 \n $\displaystyle{\forall \varphi \in D(A^{2k})}$.\\
 
\n (ii) $\forall\, k \in \mathbb{N}$, we have:\\
\begin{equation}
\displaystyle{\forall \,\, \epsilon > 0, \exists C_{\epsilon} > 0 ; \mid \mid A^{k}\varphi \mid \mid^{2} \leq \epsilon \mid \mid A^{k+1}\varphi \mid \mid^{2} + C_{\epsilon}\mid\mid \varphi \mid\mid^{2} \,\,  \forall \varphi \in D(A^{k+1}).}
\end{equation}
\end{lemma}
\n {\bf {\color{red}Proof}}\\

\n Let $\varphi(z) = \displaystyle{\sum_{n=0}^{+\infty}a_{n}e_{n}(z)}$ then :\\

\n $A^{*m}A^{l}\varphi = \displaystyle{\sum_{n=0}^{+\infty}n(n-1)......(n-l)a_{n}\frac{z^{n+m-l}}{\sqrt{n!}}}$\\

\n $= \displaystyle{\sum_{n=0}^{+\infty}(n+l-m)(n+l -m-1)......(n-m+1)a_{n+l-m}\frac{z^{n+m-l}}
{\sqrt{n!}}}$\\

{\bf ***} if $l \geq m$ we have :\\

\n $\sqrt{(n+l-m)!} = \sqrt{n!}\sqrt{n+1}.........\sqrt{n +l-m}$.\\

\n and\\

\n $ \displaystyle{A^{*m}A^{l}\varphi = \sum_{n=0}^{+\infty}\sqrt{n+l-m}\sqrt{n+l-m-1}........(n+1).....(n - m+1)a_{n+l-m}e_{n}(z)}$\\

\n Let $u(n) = \sqrt{n+l-m}\sqrt{n+l-m-1}........(n+1).....(n-m+1).$ then we deduce that\\

\n $u(n)\simeq n^{\frac{l-m}{2}}n^{m} = n^{\frac{l+m}{2}}.$\\

\n in particular\\

\n $u(n) = O(n^{\frac{l+m}{2}}).$\\

\n {\bf ***} if $l \leq m$, similarly we get :\\

\n $u(n) = O(n^{\frac{l+m}{2}}).$\\

\n Now calculate the expression $< A^{*m}A^{l}\phi, \phi >$ :\\

\n $\displaystyle{< A^{*m}A^{l}\phi, \phi > = \sum_{n=0}^{+\infty}u(n)a_{n+l-m} \overline{a_{n}}}$\\

\n then\\

\n $\displaystyle{\mid < A^{*m}A^{l}\phi, \phi > \mid \leq \sum_{n=0}^{+\infty}u(n)\mid a_{n+l-m}\overline{a_{n}} \mid}$\\

\n $\displaystyle{\leq \frac{1}{2}\sum_{n=0}^{+\infty}u(n)[\mid a_{n+l-m}\mid^{2} + \mid\overline{a_{n}} \mid^{2}]}$\\

\n $\displaystyle{\leq \frac{1}{2}\sum_{n=0}^{+\infty}u(n)\mid a_{n+l-m}\mid^{2}} + \frac{1}{2}\displaystyle{\sum_{n=0}^{+\infty}u(n) \mid\overline{a}_{n} \mid^{2}}$\\

\n As:\\

\n $\frac{1}{2}\displaystyle{\sum_{k=0}^{+\infty}u(n)\mid a_{n+l-m} \mid^{2}} = \frac{1}{2}\displaystyle{\sum_{n=0}^{+\infty}u(n + l - m)\mid a_{n}\mid^{2}}$\\

\n we deduce that :\\ 

\n $\mid < A^{*m}A^{l}\phi, \phi > \mid \leq \frac{1}{2}\displaystyle{\sum_{n=0}^{+\infty}[u(n) + u(n+l-m)]\mid a_{n}\mid^{2}}$\\

\n and as $u(n) = O(n^{\frac{l+m}{2}})$ then $\exists C_{0} > 0$ and $\exists C_{1} > 0$ such that :\\

\n $ u(n) = C_{0} + C_{1}n^{\frac{l+m}{2}}.$\\

\n For $ m+l < 2k$, by applying Youn inequality we get :\\

\n $\forall \delta > 0, \exists C_{\delta} > 0 ; n^{\frac{l+m}{2}} \leq \delta n^{k} + C_{\delta}$ \\

\n It follows that :\\

\n $\mid < A^{*m}A^{l}\phi, \phi > \mid \leq \frac{\delta C_{1}}{2}\displaystyle{\sum_{n=0}^{+\infty}\mid a_{n}\mid^{2}} + \frac{C_{0} + C_{\delta}}{2}\displaystyle{\sum_{n=0}^{+\infty}\mid a_{n}\mid^{2}}$\\

\n and\\

\n $\forall \epsilon > 0, \exists C_{\epsilon} > 0 ; \mid < P(A, A^{*})\varphi, \varphi >\mid \leq \epsilon < A^{*k}A^{k}\varphi, \varphi > + C_{\epsilon}\mid\mid \varphi \mid\mid^{2} \\ \forall \varphi \in D(A^{2k}).$\\

\n (ii) From (i) we deduce (ii). \hfill { } $\blacksquare$\\

\begin{corollary}

Let $k \in \mathbb{N}$ and $\displaystyle{P(A, A^{*}) = \sum_{i + j \leq 2k}a_{i,j}A^{*^{i}}A^{j},\,\, a_{i,j} \in \mathbb{C}}$  with minimal domain:\\

$\displaystyle{D_{min} := D_{min}(P(A^{*} , A)) = \{\varphi \in \mathbb{B} \,\, \mbox{such that there exist} \,\, p_{n} \,\, \mbox {in} \,\, \mathcal{P} \,\, \mbox{(polynomials set) }}$\\
\begin{equation}
\n \mbox{and} \,\, \displaystyle{\psi \in \mathbb{B} \,\, \mbox{with}\,\,  p_{n} \longrightarrow \varphi \,\, \mbox{and} \,\, P(A^{*} , A)p_{n} \underset {n \longrightarrow + \infty}{ \longrightarrow} \psi \}}
\end{equation}

\n then we have:\\

\begin{equation}
\displaystyle{\forall \,\, \epsilon > 0, \exists C_{\epsilon} > 0 ; \mid <  \sum_{i + j \leq 2k}a_{i,j}A^{*^{i}}A^{j},\varphi, \varphi >\mid \leq \epsilon < A^{*k}A^{k}\varphi, \varphi > + C_{\epsilon}\mid\mid \varphi \mid\mid^{2}},
\end{equation}
\n for all $\displaystyle{\varphi \in D(A^{2k})}$.\\

\end{corollary}

\begin{proposition} 
\quad \\

\n Let   $T$ be a closed linear operator in a Hilbert space $\mathcal{H}$ with domain $D(T)$ dense in $\mathcal{H}$ {\bf with}  $\rho(T) \neq \emptyset$ where $\displaystyle{\rho(T) = \{ \lambda \in \mathbb{C} ; (T - \lambda I)^{-1} \,\,  \mbox{exisits}  \,\, \}}$. \\

\n Then the resolvent $\displaystyle{\mathcal{R}_{\lambda} =  (T - \lambda I)^{-1} }$ of $T$ is compact if and only if the {\bf injection} of $D(T)$ equipped with graph norm in $\mathcal{H}$ is compact.\\
\end{proposition}

\n {\bf {\color{red}Proof}}\\

\n Let $\{\varphi_{n}\} \subset D(T)$ be a bounded sequence with respect graph norm, i.e. $\exists \, C > 0$ such that $\mid\mid T\varphi_{n} \mid\mid + \mid\mid \varphi_{n} \mid\mid \leq C$. Let $\lambda \in \rho(T)$ then we have :\\

\n $\mid\mid (\lambda I - T)\varphi_{n} \mid\mid  \leq \mid\mid T\varphi_{n} \mid\mid  + \mid \lambda \mid \mid\mid \varphi_{n} \mid\mid  \leq C(1 + \mid \lambda)$.\\

\n  It follows that the sequence $(\lambda I - T)\varphi_{n}$ is bounded in $\mathcal{H}$. since $(\lambda I - T)^{-1}$ is compact then the sequence $\varphi_{n} = (\lambda I - T)^{-1}(\lambda I - T)\varphi_{n}$ admit a subsequence convergent in  $\mathcal{H}$. Hence the {\bf injection} of $D(T)$ equipped with graph norm in $\mathcal{H}$ is compact.\\
\n Conversely, let $\lambda \in \rho(T)$ and $i$ is the canonical injection of $D(T)$ in $\mathcal{H}$ which is compact. Since $(\lambda I -T)^{-1} = io(\lambda I -T)^{-1}$ then $R_{\lambda}$ is compact as product of a compact operator $i$ with a bounded operator $R_{\lambda}$. \hfill { } $\blacksquare$\\

\begin{definition}
( {\color{red}Schatten-von Neumann spaces})\\

\n For $0 < p < +\infty$, the quasi-normed operator ideal $\mathfrak{C}_{p}$ (known as the Schatten-von Neumann ideal) is the space of all compact operators $\mathbb{T}$ such that $\displaystyle{\sum_{n=1}^{\infty}s_{n}^{p}(\mathbb{T}) < + \infty}$  associated to quasi-norm $\mid\mid \mathbb{T}\mid\mid_{p} = \displaystyle{(\sum_{n=1}^{\infty} s_{n}^{p})^{\frac{1}{p}}}$ for $ p > 0$\\
where $\displaystyle{s_{n}(\mathbb{T}):= s_{n}}$ are the eigenvalues of the compact selfadjoint operator $\displaystyle{\sqrt{\mathbb{T}^{*}\mathbb{T}}}$ (i.e.  the singular values of the operator $\mathbb{T}$). \\
\end{definition}

\n {\color{red}$\rhd_{1}$} The functional $\vert\vert \mathbb{T} \vert\vert_{p}$ is a norm on the vector space  $\mathfrak{C}_{p}$ when $p \geq 1$.\\

\n {\color{red}$\rhd_{2}$} The set  $\mathfrak{C}_{p}$ is a two-sided ideal of the algebra of bounded operators $\mathcal{L}(\mathcal{H})$.\\

\n {\color{red}$\rhd_{3}$} If $\displaystyle{\mathbb{T} \in \mathfrak{C}_{p}}$, then $\displaystyle{\mathbb{T}^{*} \in \mathfrak{C}_{p}}$\\

\n {\color{red}$\rhd_{4}$} If $\displaystyle{\mathbb{T} \in \mathfrak{C}_{p}}$ and  $\displaystyle{\mathbb{S} \in \mathfrak{C}_{q}}$, then $\displaystyle{\mathbb{T}\mathbb{S} \in \mathfrak{C}_{r}}$, where $\displaystyle{\frac{1}{r} = \frac{1}{p} + \frac{1}{q}}$.\\

\n \n {\color{red}$\rhd_{3}$} Operators in the class  $\mathfrak{C}_{1}$ are called {\color{red}trace-class} operators, and  the norm $\vert\vert . \vert\vert_{1}$ is called the {\color{red}trace-class} (or {\color{red}nuclear}) norm.\\

\n The proof of above properties can be found in the books  {\color{blue}[11]}, {\color{blue}[16]} and {\color{blue}[29]}.\\

\n For others informations about singular numbers and the Schatten-von Neumann ideals we can see {\color{blue} [38]} and {\color{blue}[28]} and references therein.\\

\begin{lemma}
\n Let $\displaystyle{\mathcal{N} = A^{*}A = z\frac{\partial}{\partial z}}$ with domain $\displaystyle{D(\mathcal{N}) = \{\varphi \in \mathbb{B} ; A\varphi \in D(A^{*})\}}$ then we have \\

\n (i) $\displaystyle{\mathcal{N}(\frac{z^{n}}{\sqrt{n!}}) = n (\frac{z^{n}}{\sqrt{n!}})}$ \\

\n (ii) $\displaystyle{A^{*}A}$ is self-adjoint operator with resolvent in $\displaystyle{\mathfrak{C}_{1+\epsilon}}$  $\forall \epsilon > 0$\\

 \n (iii) For a natural $k = 1, 2, ....$, $\displaystyle{A^{*^{k}}A^{k}}$ is self-adjoint operator with resolvent in  $\displaystyle{\mathfrak{C}_{p}}$ $\forall  p > \frac{1}{k}$, in particular this resolvent is nuclear i.e. it is in $\displaystyle{\mathfrak{C}_{1}}$ if $k \geq 2$.\\
 
 \n  The eigenvalues  of $\displaystyle{A^{*^{k}}A^{k}}$ are  $ \displaystyle{\sigma_{n,k} = n(n-1) ......(n -(k-1))}$ if $k \leq n$  \\and $0$ if $k > n$  corresponding to eigenfunctions $\displaystyle{(e_{n}(z))_{n\geq 0} = (\frac{z^{n}}{\sqrt{n!}})_{n \geq 0}}$ \\
\end{lemma}
\n {\color{red}{\bf Proof}}\\

\n (i) is trivial\\

\n (ii) By using the Neumann theorem (see theorem 2.6), we deduce that $\displaystyle{A^{*}A}$ is self-adjoint operator, positive and  $\displaystyle{(A^{*}A + \sigma I)^{-1}}$ exists for all $\sigma > 0$. In particular, $\displaystyle{ \sigma \in \rho(\mathcal{N})}$.  \\

\n This resolvent of $\mathcal{N}$ is compact because the {\bf injection} of $D(\mathcal{N})$ equipped with graph norm in $D(A)$ is continuous and as  the {\bf injection} of $D(A)$ equipped with graph norm in $\mathbb{B}$ is compact (see theorem 3.25) then by using the proposition 3.26, we deduce that  the {\bf injection} of $D(\mathcal{N})$ equipped with graph norm in $\mathbb{B}$ is compact.\\

\n The eigenvalues  of $\displaystyle{A^{*}A}$ are  $ \displaystyle{\sigma_{n} = n}$ corresponding to eigenfunctions \\

\n $\displaystyle{(e_{n}(z))_{n\geq 0} = (\frac{z^{n}}{\sqrt{n!}})_{n \geq 0}}$.\\

\n  As  $\displaystyle{ \sum_{n=1}^{+\infty}\frac{1}{n^{1 + \epsilon}} < \infty }$  $\forall   \epsilon > 0$  then the resolvent  of $A^{*}A$ is in  $\displaystyle{\mathfrak{C}_{1 + \epsilon}}$ $\forall   \epsilon > 0$ \\

\n (iii) {\color{red}$\rhd$} By using the Neumann theorem, we deduce that $\displaystyle{A^{*^{k}}A^{k}}$ is positive  self-adjoint operator and  $\displaystyle{(A^{*^{k}}A^{k} + \sigma I)^{-1}}$ exists for all $\sigma > 0$. In particular, $\displaystyle{\rho(A^{*^{k}}A^{k}) \neq \emptyset}$. \\

\n{\color{red}$\rhd$}  If $k \leq  n$ we have \\

\n $\displaystyle{A\frac{z^{n}}{\sqrt{n!}} = n\frac{z^{n-1}}{\sqrt{n!}} \Longrightarrow A^{*}A\frac{z^{n}}{\sqrt{n!}} = n\frac{z^{n}}{\sqrt{n!}}}$,\\

\n  $\displaystyle{A^{2}\frac{z^{n}}{\sqrt{n!}} = n(n-1)\frac{z^{n-2}}{\sqrt{n!}} \Longrightarrow A^{*^{2}}A^{2}\frac{z^{n}}{\sqrt{n!}} = n(n-1)\frac{z^{n}}{\sqrt{n!}}}$\\

\n .........................................................\\

\n .........................................................\\

\n  $\displaystyle{A^{k}\frac{z^{n}}{\sqrt{n!}} = n(n-1)....(n -(k-1))\frac{z^{n-k}}{\sqrt{n!}} \Longrightarrow A^{*^{k}}A^{k}\frac{z^{n}}{\sqrt{n!}} = n(n-1)....(n-(k-1))\frac{z^{n}}{\sqrt{n!}}}$\\

\n{\color{red}$\rhd$}  If $k >  n$ we have \\

\n  $\displaystyle{A^{k}\frac{z^{n}}{\sqrt{n!}} = 0 \Longrightarrow A^{*^{k}}A^{k}\frac{z^{n}}{\sqrt{n!}} = 0 \times \frac{z^{n}}{\sqrt{n!}}}$\\

\n Hence\\

\n {\color{red}$\rhd$} The eigenvalues of $\displaystyle{A^{*^{k}}A^{k}}$ are  $ \displaystyle{\sigma_{n,k} = n(n-1) ......(n -(k-1))}$ if $k \leq n$  and $0$ if $k > n$  associated to eigenfunctions $\displaystyle{(e_{n}(z))_{n\geq 0} = (\frac{z^{n}}{\sqrt{n!}})_{n \geq 0}}$ \\

\n and\\

\n{\color{red}$\rhd$} $\displaystyle{\sigma_{n,k} \sim n^{k}}$, $\displaystyle{\sigma_{n,k}^{p} \sim n^{pk}}$\\

\n{\color{red}$\rhd$} The {\bf injection} of $\displaystyle{D(A^{*^{k}}A^{k})}$ (equipped with graph norm) in $D(A^{*}A)$ is continuous (by $k$-compositions of continuous injections) and as the {\bf injection} of $D(\mathcal{A^{*}A})$ (equipped with graph norm) in $D(A)$ is continuous and  the {\bf injection} of $D(A)$ (equipped with graph norm ) in $\mathbb{B}$ is compact (see theorem 3.25) then by using the proposition 3.26, we deduce that  the {\bf injection} of $D(A^{*^{k}}A^{k})$ equipped with graph norm in $\mathbb{B}$ is compact. As $\rho(A^{*^{k}}A^{k}) \neq \emptyset$, it follows that $\displaystyle{ \mathcal{R}_{\sigma} = (A^{*^{k}}A^{k} - \sigma I)^{-1}}$ is compact for every $\displaystyle{ \sigma \in \rho( A^{*^{k}}A^{k})}$.\\

\n  {\color{red}$\rhd$} For any $k$, the operator :\\
\begin{center}
$\displaystyle{A^{*^{k}}A^{k} = \sum_{n=1}^{\infty} \sigma_{n,k} < . ,  e_{n} > e_{n} ,\,\, D(A^{*^{k}}A^{k}) = \{\varphi \in \mathbb{B} ; \sum_{n=1}^{\infty} \sigma_{n,k}^{2} \vert < \varphi  , e_{n} >\vert^{2} < + \infty \}}$
\end{center}
\n where  $\displaystyle{ (e_{n}(z) )_{n=1}^{+\infty}}$ =  $\displaystyle{ (\frac{z^{n}}{\sqrt{n!}})_{n=1}^{+\infty}}$  is  orthonormal complete system in $\mathbb{B}$,
\n have a resolvent in $\displaystyle{\mathfrak{C}_{p} \,\, \forall \,\, p > \frac{1}{k}}$. {\bf{In fact}}:\\

\n \n  $\displaystyle{\sum_{n=1}^{\infty} \frac{1}{\sigma_{n,k}^{p}} \sim  \sum_{n=1}^{\infty}\frac{1}{n^{pk}} < +\infty  \,\, \forall \,\, p > \frac{1}{k}}$, in particular this resolvent is nuclear i.e. it is in $\displaystyle{\mathfrak{C}_{1}}$. \hfill { } $\blacksquare$\\
\begin{lemma}
\n  {\color{red}$\rhd$} Let $\displaystyle{\Delta\sigma_{n,k} = \sigma_{n+1,k} - \sigma_{n,k}}$ then from the definition of $\sigma_{n,k}$ we have:\\
$$\displaystyle{\Delta\sigma_{n,k} = k \sigma_{n,k-1} } \quad (1 \leq k \leq n)$$
\end{lemma}

\n {\bf {\color{red}Proof}}\\

\n Let  $ \displaystyle{\sigma_{n,k} = n(n-1) .....(n - (k-2)(n -(k-1))}$ then \\

\n $ \displaystyle{\sigma_{n+1,k} = (n+1)n(n-1) .....(n - (k-2))}$\\

\n  and \\

\n $\displaystyle{\Delta\sigma_{n,k} = (n + 1- (n -k+1) \sigma_{n,k-1} = k \sigma_{n,k-1}}$. \hfill { } $\blacksquare$\\

\n   {\color{red}$\rhd$} Let $\displaystyle{ S_{k} = A^{*^{k}}A^{k}, k = 1, 2, ...}$ and $\displaystyle{P_{m}(A^{*}, A) = \sum_{i+j \leq m} a_{ij}A^{*^{i}}A^{j} ; a_{ij} \in \mathbb{C}}$ such that deg $P_{m}(A^{*}, A) = m $ with $m \leq 2k-1$.\\

\begin{remark}

\n {\color{red} $\bullet_{1}$} For $k = 1$, we have $S_{1} = A^{*}A$ and $\displaystyle{P_{1}(A^{*}, A) = a_{00}I + a_{01}A + a_{10}A^{*}}$\\

\n {\color{red} $\bullet_{2}$} For $k = 2$, we have $S_{2} = A^{*^{2}}A^{2}$ and \\$\displaystyle{P_{2}(A^{*}, A) = a_{00}I + a_{01}A +  a_{02}A^{2} + a_{10}A^{*} + a_{11} A^{*}A + a_{12}A^{*}A^{2} + a_{20} A^{*^{2}} + a_{21}A^{*^{2}}A}$\\

\n {\color{red} $\bullet_{3}$} The operator $\displaystyle{H_{\mu,\lambda} = P_{2}(A^{*}, A) }$ where  $\displaystyle{a_{00} = a_{01} = a_{02} = a_{10} = a_{20} = 0}$, \\$a_{11} = \mu $ and $\displaystyle{a_{12}= a_{21} = i \lambda}$ where $\mu$ is the intercept of Pomeron and $\lambda$ is the triple coupling of Pomeron (these  parameters play an essential role in Reggeon field theory) and $i^{2} = -1$.\\
\end{remark}

\begin{proposition}

\n Let   $\displaystyle{H_{\mu,\lambda} =  \mu A^{*}A + i\lambda A^{*}(A + A^{*})A}$ and  $\displaystyle{H_{\lambda^{'}} = \lambda^{'}S_{2} + H_{\mu,\lambda}}$ with respective domains $D(H_{\mu,\lambda})$, $D(S_{2})$ and $D(H_{\lambda'})$ in Bargmann space then we get: \\

\n (i) $\forall \epsilon > 0, \exists C_{\epsilon} > 0 ; \mid < H_{\mu,\lambda}\phi, \phi >\mid \leq \epsilon < S_{2}\phi, \phi > + C_{\epsilon}\mid \mid \phi \mid\mid^{2} \quad \forall \phi \in D(S_{2})$.\\

\n (ii) $\forall \lambda' > 0, \exists C > 0 ; \Re e < H_{\lambda'}\phi, \phi > \geq -C\mid \mid \phi \mid\mid^{2} \quad \forall \phi \in D_{min}(H_{\lambda'})$.\\

\n (iii) $\forall \lambda' < 0, \exists C > 0 ; Re < H_{\lambda'}\phi, \phi > \leq C\mid \mid \phi \mid\mid^{2} \quad \forall \phi \in D_{min}(H_{\lambda'})$.\\

\n (iv) For $\lambda'  \neq 0, \exists \beta_{0} \in \mathbb{R}; H_{\lambda'}^{min} + \beta_{0}I$ is invertible from $D_{min}(H_{\lambda'})$ to $\mathbb{B}$.\\
\end{proposition}

 \n {\bf{\color{red} Proof}}\\
 
\n (i)  Let  $\displaystyle{H_{\mu,\lambda} =  \mu A^{*}A + i\lambda A^{*}(A + A^{*})A}$ then \\
 
 \n $\displaystyle{\mid < H_{\mu, \lambda}\phi, \phi >\mid \, \leq \, \mid \mu \mid. \mid\mid A\phi\mid\mid^{2} + 2\mid \lambda \mid .\mid\mid A^{2}\phi\mid\mid. \mid\mid A\phi\mid\mid}$.\\
 
 \n Now, as $\forall \delta > 0$ we have \\

\n $\displaystyle{\mid\mid A^{2}\phi\mid\mid \mid\mid \phi\mid\mid \leq \delta \mid\mid A^{2}\phi\mid\mid^{2} + \frac{1}{\delta} \mid\mid \phi\mid\mid^{2}}$.\\

\n then we deduce that :\\

\n $\displaystyle{\mid < H_{\mu,\lambda}\phi, \phi >\mid \leq (\mid \mu \mid  + \frac{2\mid \lambda \mid}{\delta})\mid\mid A\phi\mid\mid^{2} + 2\delta\mid \lambda \mid \mid\mid A^{2}\phi\mid\mid^{2}}$.\\

\n Let's apply the fundamental lemma to get:\\

\n $\forall \delta_{1} > 0, \exists C_{\delta_{1}} > 0 ; \mid\mid A\phi\mid\mid^{2} \leq \delta \mid\mid A^{2}\phi\mid\mid^{2} + C_{\delta_{1}}\mid\mid\phi\mid\mid^{2}$\\

\n i.e. \\

\n $\displaystyle{\mid < H_{\mu,\lambda}\phi, \phi >\mid \leq \delta_{1}C \mid\mid A^{2}\phi\mid\mid^{2} + C_{\delta_{1}}\mid\mid\phi\mid\mid^{2}}$ where  $C$ is a constant.\\

\n Or\\

\n $\forall \epsilon > 0, \exists C_{\epsilon} > 0 ; \mid < H_{\mu,\lambda}\phi, \phi >\mid \leq \epsilon < S_{2}\phi, \phi > + C_{\epsilon}\mid \mid \phi \mid\mid^{2} \,\,\forall \phi \in D(S_{2})$.\\

\n (ii) and (iii) are an immediate consequence of (i).\\

\n (iv) Let $\beta > 0$, we consider $H_{\lambda'} = \lambda' S_{2} + H_{\mu,\lambda}$ in following form
$H_{\lambda'} = \lambda' (S_{2} + \frac{1}{\lambda'}H_{\mu,\lambda})$ then  it follows that $S_{2} + \frac{1}{\lambda'}H_{\mu,\lambda} + \beta I = [I + \frac{1}{\lambda'}H_{\mu,\lambda}(S_{2} + \beta I)^{-1}](S_{2} + \beta I)$.\\

\n Now as $(S_{2} + \beta I)$ is invertible then so just show that $\mid\mid \frac{1}{\lambda'}H_{\mu, \lambda}(S_{2} + \beta I)^{-1}\mid\mid < 1$: \\
\n Let $\psi \in \mathbb{B}$, then we have :\\

\n $\forall \epsilon > 0, \exists C_{\epsilon} > 0$;  $\mid\mid \frac{1}{\lambda'}H_{\mu,\lambda}(S_{2} + \beta I)^{-1}\psi\mid\mid \leq \epsilon \mid\mid S_{2}(S_{2} + \beta I)^{-1}\psi\mid\mid  + \\C_{\epsilon}\mid\mid(S_{2} + \beta I)^{-1}\psi\mid\mid$.\\

\n $\leq \epsilon \mid\mid (S_{2} + \beta I - \beta I)(S_{2} + \beta I)^{-1}\psi\mid\mid  + C_{\epsilon}\mid\mid(S_{2} + \beta I)^{-1}\psi\mid\mid$.\\

\n $\leq \epsilon \mid\mid \psi\mid\mid  + (\epsilon\beta + C_{\epsilon})\mid\mid(S_{2} + \beta I)^{-1}\psi\mid\mid$.\\

\n As $\mid\mid(S_{2} + \beta I)^{-1}\mid\mid \leq \frac{1}{\beta}$ then we deduce that \\

\n  $\mid\mid \frac{1}{\lambda'}H(S_{2} + \beta I)^{-1}\psi \mid\mid \leq (2\epsilon + \frac{C_{\epsilon}}{\beta})\mid\mid \psi \mid\mid$.\\

\n  in particular  for $ \epsilon < \frac{1}{2}$ and  $ \beta > \frac{C_{\epsilon}}{1-2\epsilon}$ we get\\

 \n $\mid\mid \frac{1}{\lambda'}H_{\mu, \lambda}(S_{2} + \beta I)^{-1}\mid\mid < 1$.\hfill { } $\blacksquare$\\
 \subsection{ {\color{blue} Some applications to differential operators arising in diffusion problem or in Reggeon field theory}} 

\n {\color{red} (a)} We consider the evolution problem :\\
\begin{equation}
\left \{ \begin{array} {c} \displaystyle{\frac{\partial}{\partial t}u(t, x) = - \mathbb{H}_{\mu, \lambda} u(t, x), (t, x) \in \mathbb{R}_{+}^{*} \times \mathbb{R}}. \quad\quad\quad  \quad\quad\quad \quad\quad\quad \quad\quad\quad \quad\quad\quad \quad\quad\quad\quad\\
\quad\\
\displaystyle{u(0, x) = u_{0}(x), \quad u_{0} \in L_{2}(\mathbb{R})} \quad\quad\quad \quad\quad\quad \quad\quad\quad \quad\quad\quad \quad\quad\quad \quad\quad\quad \quad\quad\quad \quad \quad\\
\quad\\
\mbox{where} \quad\quad\quad \quad\quad\quad \quad\quad\quad \quad\quad\quad \quad\quad\quad \quad\quad\quad \quad\quad\quad \quad\quad\quad \quad\quad\quad \quad\quad\quad \quad\quad\quad \quad\\
\quad\\
\displaystyle{\mathbb{H}_{\mu,\lambda} = -(\frac{\mu}{2}+ \frac{i\lambda x}{\sqrt{2}}) \frac{\partial^{2}}{\partial x^{2}} -\frac{i\lambda}{\sqrt{2}}\frac{\partial}{\partial x} + \frac{\mu}{2}(x^{2} - 1) + \frac{i\lambda}{\sqrt{2}} x(x^{2} - 2)\,\, \mu , \lambda \in \mathbb{R} \,\,\mbox{and}\,\, i^{2} = -1}\\
\end{array} \right.\\
\end{equation}

\n {\color{red} $\rhd$} For $\lambda = 0$, we observe that \\

\n $\displaystyle{-\mathbb{H}_{\mu,0} = \frac{\mu}{2}[ \frac{\partial^{2}}{\partial x^{2}} + (1 - x^{2})]}$ is the harmonic oscillator in $L_{2}(\mathbb{R})$-representation. In this case  (3.35) is the diffusion equation with harmonic potential: \\

\begin{equation}
\left \{ \begin{array} {c} \displaystyle{\frac{\partial}{\partial t}u(t, x) = \frac{\mu}{2}[\frac{\partial^{2}}{\partial x^{2}} + (1 - x^{2})] u(t, x);  (t, x) \in \mathbb{R}_{+}^{*} \times \mathbb{R} \,\, \mbox{and} \,\, \mu \in \mathbb{R}} \\
\quad\\
\quad\\
\displaystyle{u(0, x) = u_{0}(x), \quad u_{0} \in L_{2}(\mathbb{R})} \quad\quad\quad \quad\quad\quad \quad\quad\quad \quad\quad\quad \quad \quad\quad\\
\end{array} \right.
\end{equation}

\n We use the Bargmann transform as an intertwining operator that transports the harmonic oscillator to a complex $\mathcal{N}$ operator to solve  the above equation by considering its equivalent form in Bargmann space $\mathbb{B}$:\\

\begin{equation}
\left \{ \begin{array} {c} \displaystyle{\frac{\partial}{\partial t}\phi(t, z) = -\mu \mathcal{N}\phi(t, z) ; \quad (t, z) \in \mathbb{R}_{+}^{*} \times \mathbb{C} \,\, \mbox{and} \,\, \mu \in \mathbb{R}}   \quad\quad\quad\\
\quad\\
\quad\\
\displaystyle{\phi(0, z) = \varphi(z), \quad  \varphi \in \mathbb{B}} \quad\quad\quad \quad\quad\quad \quad\quad\quad \quad\quad\quad \quad \quad\quad\\
\end{array} \right.
\end{equation}

\n {\color{red}$\rhd$} The diffusion equation (3.36) after Bargmann transform  can be solved in the general form $\displaystyle{\phi(t; z) = \varphi(ze^{-\mu t})}$, where $\varphi \in \mathbb{B}$ also represents the initial
data $\phi(0; z) = \varphi(z)$. In fact :\\

\n $\displaystyle{\frac{\partial}{\partial t}\phi(t; z) = \frac{\partial}{\partial t}\varphi(ze^{-\mu t}) = \frac{\partial Z}{\partial t} \frac{\partial}{\partial Z}\varphi(Z) = -\mu Z \frac{\partial}{\partial Z}\varphi(Z)= - \mu \mathcal{N} \varphi(Z) }$\\

\n  where $Z = ze^{-\mu t}$.\\

\n {\color{red}(b)} We consider the evolution problem :\\

\begin{equation}
\left \{ \begin{array} {c} \displaystyle{\frac{\partial}{\partial t}u(t, x) =  \mathbb{H} u(t, x), (t, x) \in \mathbb{R}_{+}^{*} \times \mathbb{R}}.  \quad\quad\quad \quad\quad\quad \quad\quad\\
\quad\\
\displaystyle{u(0, x) = u_{0}(x), \quad u_{0} \in L_{2}(\mathbb{R})} \quad\quad \quad\quad\quad \quad\quad\quad \quad\quad\quad \\
\quad\\
\mbox{where} \quad \quad\quad\quad \quad\quad\quad \quad\quad\quad \quad\quad\quad \quad\quad\quad \quad\quad\quad \quad\quad\\
\quad\\
\displaystyle{\mathbb{H} = -\frac{\partial^{3}}{\partial x^{3}} + (x^{2} - 1) \frac{\partial}{\partial x} + 2x = \frac{\partial}{\partial x}[-\frac{\partial^{2}}{\partial x^{2}} + x^{2} - 1] }\\
\end{array} \right.\\
\end{equation}

\n To reformulate the Cauchy problem (3.38) in the Bargmann space $\mathbb{B}$, we need to recall  the properties (3.21) and (3.22) of theorem 3.26 of  action of Bargmann transform $\mathcal{B}$ on to the derivative $\displaystyle{\frac{\partial}{\partial x}}$ and on the multiplication by $x$:\\

\n For $\varphi \in \mathbb{B}$, we have\\

$$ {\color{red}(i)} \quad \displaystyle{(\mathcal{B}x\mathcal{B}^{-1}\varphi)(z) = \frac{1}{2}(\varphi^{'}(z)  + z\varphi(z)) = \frac{1}{2}(\frac{\partial}{\partial z} + z)\varphi(z)}$$

\n $$ {\color{red}(ii)} \quad \displaystyle{(\mathcal{B}\frac{\partial}{\partial x}\mathcal{B}^{-1} \varphi)(z) = \varphi^{'}(z) - z\varphi(z) = (\frac{\partial}{\partial z} - z)\varphi(z)} \quad $$

\n It follows that:\\
\begin{equation}
\displaystyle{\hat{\mathbb{H}} = \mathcal{B}\mathbb{H}\mathcal{B}^{-1} = \mathcal{B}\frac{\partial}{\partial x}\mathcal{B}^{-1}\mathcal{B}[-\frac{\partial^{2}}{\partial x^{2}} + x^{2} - 1] \mathcal{B}^{-1}}
\end{equation}

\n Now as \\

\n  $\displaystyle{(\mathcal{B}\frac{\partial}{\partial x}\mathcal{B}^{-1} \varphi)(z) = (\frac{\partial}{\partial z} - z)\varphi(z)}$ and $\displaystyle{(\mathcal{B}[-\frac{\partial^{2}}{\partial x^{2}} + x^{2} - 1] \mathcal{B}^{-1} \varphi)(z) = z\frac{\partial}{\partial z}\varphi(z)}$\\

\n Then we deduce that $\mathbb{H}$ is transformed  to $\hat{\mathbb{H}}$ in Bargmann space $\mathbb{B}$ :\\

\begin{equation}
\displaystyle{\hat{\mathbb{H}} \varphi(z) = z \frac{\partial^{2}}{\partial z} \varphi(z) - (z^{2} -1)\frac{\partial}{\partial z}\varphi(z)}
\end{equation}

\n Or\\

\begin{equation}
\displaystyle{\hat{\mathbb{H}} = A + A^{*}(A - A^{*})A}
\end{equation}

where\\

\n\n{\color{red}$\bullet$} $\displaystyle{Ae_{n} = \sqrt{n}e_{n-1}}$\\

\n \n{\color{red}$\bullet$} $\displaystyle{A^{*}e_{n} = \sqrt{n+1}e_{n+1}}$\\

\n \n{\color{blue}$\bullet$} $\displaystyle{A^{*}A e_{n} = ne_{n}}$\\

\n{\color{red}$\bullet$} $\displaystyle{A^{*^{2}}Ae_{n} = n\sqrt{n+1}e_{n+1}}$\\

\n\n{\color{blue}$\bullet$} $\displaystyle{A^{2}e_{n} = \sqrt{n}\sqrt{n-1}e_{n-2}}$\\

\n \n{\color{red}$\bullet$} $\displaystyle{A^{*}A^{2}e_{n} = \sqrt{n}\sqrt{n-1}\sqrt{n-1}e_{n-1} = (n-1)\sqrt{n} e_{n-1}}$\\

\n and\\

\n $\displaystyle{\hat{\mathbb{H}}e_{n} = A + A^{*}(A -A^{*}Ae_{n} = [(n-1)\sqrt{n} + \sqrt{n}]e_{n-1} -   n\sqrt{n+1}e_{n+1}}$\\

\n $\displaystyle{ = n\sqrt{n}e_{n-1} -  n\sqrt{n+1}e_{n+1}}$\\

\n The equivalent form of  Cauchy problem  (3.38) is given by\\

\begin{equation}
\left \{ \begin{array} {c} \displaystyle{\frac{\partial}{\partial t}U(t, z) = \hat{ \mathbb{H}} U(t, z), (t, z) \in \mathbb{R}_{+}^{*} \times \mathbb{C}}.  \quad\quad\quad \quad\quad\quad \quad\quad\\
\quad\\
\displaystyle{U(0, z) = \varphi(z), \quad \varphi \in \mathbb{B}} \quad\quad \quad\quad\quad \quad\quad\quad \quad\quad\quad \quad\quad\quad\\
\quad\\
\mbox{where} \quad \quad \quad\quad\quad \quad\quad\quad \quad\quad\quad \quad\quad\quad \quad\quad\quad \quad\quad\quad \quad\quad\\
\quad\\
\displaystyle{\hat{\mathbb{H}} = z \frac{\partial^{2}}{\partial z^{2}} - (z^{2} -1)\frac{\partial}{\partial z}} \quad \quad\quad\quad \quad\quad\quad\quad\quad\quad \quad\quad\quad\\
\end{array} \right.\\
\end{equation}
\quad\\
\n where \\

\n $\displaystyle{\varphi = \mathcal{B}f}$ and $\displaystyle{U(t, z) = \mathcal{B}u(t, x)}$.\\

\n Using the following decomposition $U(t, z)$ on the Bargmann basis $\displaystyle{e_{n}(z) = \frac{z^{n}}{\sqrt{n!}}}$ :\\

\n $\displaystyle{U(t,z) = \sum_{n=0}^{\infty}a_{n}(t)e_{n}(z) = \sum_{n=0}^{\infty} a_{n}(t)\frac{z^{n}}{\sqrt{n!}}}$\\

\n we deduce that\\

\n $\displaystyle{\frac{\partial}{\partial t}U(t,z) = \sum_{n=0}^{\infty}\frac{\partial}{\partial t}a_{n}(t)e_{n}(z) = \sum_{n=0}^{\infty} \frac{\partial}{\partial t}a_{n}(t)\frac{z^{n}}{\sqrt{n!}}}$\\

\n Let $\displaystyle{U(t,z) = a_{0}(t) + \sum_{n=1}^{\infty} a_{n}(t)\frac{z^{n}}{\sqrt{n!}}}$ Then we have :\\

\n $\displaystyle{\frac{\partial}{\partial z}U(t,z) =  \sum_{n=1}^{\infty} n a_{n}(t)\frac{z^{n-1}}{\sqrt{n!}} = \sum_{n=0}^{\infty} \sqrt{n+1} a_{n+1}(t)\frac{z^{n}}{\sqrt{n!}} }$\\

\n $\displaystyle{z^{2}\frac{\partial}{\partial z}U(t,z) =  \sum_{n=1}^{\infty} n a_{n}(t)\frac{z^{n+1}}{\sqrt{n!}} = \sum_{n=2}^{\infty} (n-1) \sqrt{n} a_{n-1}(t)\frac{z^{n}}{\sqrt{n!}} }$\\

\n $\displaystyle{\frac{\partial^{2}}{\partial z^{2}}U(t,z) =  \sum_{n=1}^{\infty} n(n-1) a_{n}(t)\frac{z^{n-2}}{\sqrt{n!}} = \sum_{n=0}^{\infty} n\sqrt{n+1}a_{n+1}(t)\frac{z^{n-1}}{\sqrt{n !}} }$\\

\n $\displaystyle{z\frac{\partial^{2}}{\partial z^{2}}U(t,z) =  \sum_{n=1}^{\infty} n(n-1) a_{n}(t)\frac{z^{n-1}}{\sqrt{n!}} = \sum_{n=0}^{\infty} n\sqrt{n+1}a_{n+1}(t)\frac{z^{n}}{\sqrt{n !}} }$\\

\n It follows that\\

\n {\color{red}$\bullet$} $\displaystyle{\hat{\mathbb{H}} U(t, z) = z \frac{\partial^{2}}{\partial z^{2}}U(t,z) - (z^{2} -1)\frac{\partial}{\partial z}U(t,z)} $

\n $\displaystyle{ = \sum_{n=1}^{\infty} n\sqrt{n+1}a_{n+1}(t)\frac{z^{n}}{\sqrt{n !}}  -  \sum_{n=2}^{\infty} (n-1)\sqrt{n} a_{n-1}(t)\frac{z^{n}}{\sqrt{n!}}  + \sum_{n=0}^{\infty} \sqrt{n+1} a_{n+1}(t)\frac{z^{n}}{\sqrt{n!}} }$\\

\n $\displaystyle{ = a_{1}(t) + 2 \sqrt{2}a_{2}(t)z +  \sum_{n=2}^{\infty}[(n +1) \sqrt{n+1}a_{n+1}(t) - (n-1) \sqrt{n} a_{n-1}(t)]\frac{z^{n}}{\sqrt{n!}} }$\\

\n {\color{red}$\bullet$} $\displaystyle{\frac{\partial}{\partial t}a_{0}(t) = a_{1}(t)}$\\

\n {\color{red}$\bullet$} $\displaystyle{ \frac{\partial}{\partial t}a_{1}(t) = 2 \sqrt{2}a_{2}(t) }$\\

\n{ \color{red}$\bullet$} $\displaystyle{\frac{\partial}{\partial t}a_{n}(t) = (n +1) \sqrt{n+1}a_{n+1}(t) - (n -1) \sqrt{n} a_{n-1}(t)}$, $n \geq 2$ \\

\n Setting $\displaystyle{a_{n} = \frac{i^{n}}{\sqrt{n}}\tilde{a}_{n}}$ to get \\
\begin{equation}
\displaystyle{\frac{i^{n}}{\sqrt{n}}\frac{\partial}{\partial t}\tilde{a}_{n}(t) = \frac{i^{n+1}}{\sqrt{n+1}}(n +1) \sqrt{n+1}\tilde{a}_{n+1}(t) + \frac{i^{n+1}}{\sqrt{n-1}}(n -1) \sqrt{n} \tilde{a}_{n-1}(t)} 
\end{equation}
\begin{equation}
\displaystyle{\frac{1}{\sqrt{n}}\frac{\partial}{\partial t}\tilde{a}_{n}(t) = i[(n +1) \tilde{a}_{n+1}(t) + \sqrt{n -1} \sqrt{n} \tilde{a}_{n-1}(t)]} 
\end{equation}

\begin{equation}
\displaystyle{\frac{\partial}{\partial t}\tilde{a}_{n}(t) = i[\sqrt{n}(n +1) \tilde{a}_{n+1}(t) + n \sqrt{n -1} \tilde{a}_{n-1}(t)]} 
\end{equation}

\n one can rewrite system (3.45) in the form:\\

\begin{equation}
\left \{ \begin{array} {c} \displaystyle{\frac{\partial}{\partial t}\tilde{a}_{n}(t) = \color{red}{i}(\mathbb{J}\tilde{a})_{n}(t)} \quad \quad \quad \quad \quad \quad \quad \quad \quad \quad \quad \quad\quad\\
\quad\\
\text{where} \quad \quad \quad \quad \quad \quad \quad \quad \quad \quad \quad \quad\quad \quad \quad \quad \quad \quad\quad\\
\displaystyle{(\mathbb{J}\tilde{a})_{n}(t) = \sqrt{n}(n +1) \tilde{a}_{n+1}(t) + n \sqrt{n -1} \tilde{a}_{n-1}(t)}\\
\quad\\
\text{or} \quad \quad \quad \quad \quad \quad \quad \quad \quad \quad \quad \quad \quad\quad \quad \quad \quad \quad \quad\quad\\
\quad\\
\displaystyle{(\mathbb{J}\tilde{a})_{n}(t) =\omega_{n-1} \tilde{a}_{n-1}(t) + \omega_{n} \tilde{a}_{n+1}(t)}  \quad \quad \quad \quad\\
\quad\\
\text{with}  \quad \quad \quad \quad \quad \quad \quad \quad \quad \quad \quad \quad\quad \quad \quad \quad \quad \quad\quad\\
\displaystyle{\omega_{n} =  \sqrt{n}(n +1)} \quad \quad \quad \quad \quad \quad \quad \quad \quad \quad \quad \quad \quad \\
\end{array} \right.\\
\end{equation}

\n $\mathbb{J}$ is called the {\color{red}Jacobi operator}. We refer to Teschl in {\color{blue}[39]} and  to Yafaef in  {\color{blue}[45]}
 for a systematical spectral study of Jacobi Operators. \\

\n Below, we formulate some well-known facts concerning  eigenvalue problems for finite Jacobian matrices.\\
\begin{definition}
\quad\\

\n {\color{red}$\bullet$}  An $n\times n$ matrix $A = [a_{ij} ]$ is called a diagonal matrix if $a_{ij} = 0$ whenever $i \neq j$.\\

\n {\color{red}$\bullet$} An $n\times n$ matrix $A = [a_{ij} ]$ is referred to as a {\color{red}tridiagonal} matrix if $a_{ij} = 0$
whenever$\vert i - j \vert > 1$.\\

\n The entries of $A = [a_{ij} ]$ that lie in positions $(i, i+1) i = 1, . . .n - 1$ are referred to as the {\color{red}superdiagona}l, and the entries
of $A$ in positions $(i, i - 1) i = 2, . . . , n $ are called {\color{red}subdiagonal}.\\

\n Hence a matrix is tridiagonal if the only nonzero entries of A are contained on its sub-, main, and superdiagonal.\\
\end{definition}

\n Consider the (n + 1)dimensional Jacobian matrix:\\
$L =$ $\left ( \begin{array} {ccccccc}0&c_{0}&0&...&0& 0\\
\quad\\
c_{0}&0&c_{1}&...&0&0\\
\quad\\
...&...&...&...&...&...\\
\quad\\
0&0&0&...&0\,&c_{n-1}\\
\quad\\
...&...&...&...&c_{n-1}&0\\
\end{array} \right )$\\

\n which has zeros on the main diagonal; positive numbers $\displaystyle{c_{0}, c_{1}, ...., c_{n-1}}$ on two adjacent diagonals; and zeros everywhere else. Consider the difference equation\\

\n Consider the difference equation\\
\begin{equation}
\displaystyle{c_{n-1} y_{n-1} + c_{n}y_{n+1} = \lambda y_{n}, n = 0, 1, ..., n, \,\, c_{-1} = c_{n} = 1}
\end{equation}

\n where $\lambda$ is the spectral parameter. Denote by $P_{n}(\lambda)$ the solution to Eq. (3.47) with the initial conditions $ P_{-1} (\lambda) = 0$ and $P_{0} (\lambda) = 1$. It is well known that the eigenvalues of $L$ are real, simple, and coincide with the zeros of the polynomial $P_{n+1} (\lambda)$. Moreover, these eigenvalues are symmetric about the point $\lambda = 0$.\\

\n Let $\lambda_{0}, ..., \lambda_{n}$ be the eigenvalues of $L$. The normalizing coefficients $\alpha_{k}$ are defined as

\n $\displaystyle{\alpha_{k} = \sum_{j=0}^{n}P_{j}^{2}(\lambda_{k}) , k = 0, 1, ..., n}$.\\

\n Moreover, symmetric eigenvalues correspond to identical normalizing coefficients and
\begin{equation}
\displaystyle{ \sum_{k=0}^{n}\frac{1}{\alpha_{k}} = 1}.
\end{equation}

\n {\color{red}$\bullet$} The collection $\displaystyle{\{\lambda_{k} ; \alpha_{k} > 0\}_{k=0}^{n}}$ is called the {\color{red}spectral data} of the matrix $L$.\\

\n {\color{red}$\bullet$} Given an n-by-n nonsingular tridiagonal matrix $T$:\\

\begin{equation}
T = \left ( \begin{array} {cccccc}a_{1}&b_{1}& \,&.\,&\,\\
\quad\\
c_{1}&a_{2}& b_{2}&...,&..\\
\quad\\
\,&c_{2}& ...,&...&...\\
\quad\\
...&...& ...&..&b_{n-1}\\
\quad\\
...&...& ...&c_{n-1}&a_{n}\\
\end{array} \right )\\
\end{equation}

\n Usmani {\color{blue}[12, 40, 41]} gave an elegant and concise formula for the inverse:\\

\n $\displaystyle{(T_{ij})^{-1}}$ = $\left \{ \begin{array} {c} (-1)^{i+j}b_{i}....b_{j-1}\theta_{i-1}\phi_{j+1}/\theta_{n} \quad \text{if} \,\, i \leq j.\\
\quad\\
 (-1)^{i+j}c_{j}....c_{i-1}\theta_{i-1}\phi_{j+1}/\theta_{n} \quad \text{if} \,\, i > j.\\
 \end{array} \right.$\\
 
\n  where $\theta_{i}$'s verify the recurrence relation :\\

 \begin{equation}
\displaystyle{\theta_{i} = a_{i}\theta_{i-1} - b_{i-1}c_{i-1}\theta_{i-2}, \,\, \text{for}\,\, i=2, ......, n}
\end{equation}
\n with initial conditions $\theta_{0} = 1$ and $\theta_{1 }= a_{1}$,\\

\n and\\
 
\n $\phi_{i}$'s verify the recurrence relation :\\

 \begin{equation}
\displaystyle{\phi_{i} = a_{i}\phi_{i+1} - b_{i}c_{j}\theta_{i+2}, \,\, \text{for}\,\, i = n-1, ......, 1}
\end{equation}
\n with initial conditions $\phi_{n+1} = 1$ and $\phi_{n} = a_{n}$,\\

\n {\color{red}$\bullet$} Observe that a tridiagonal matrix $A$ in the form (3.49) is called {\color{red}symmetric} if $b_{i} = c_{i}, i = 1, . . . , n - 1$ and is called {\color{red}irreducible} if both $b_{i} \neq 0$ and $c_{i} \neq 0$, for$ i = 1, . . . , n - 1$.\\

\n {\color{red}$\bullet$} Given  $\lambda \in \mathbb{C}$ the tridiagonal matrix $T - \lambda I$:\\

$T - \lambda I$ = $\left ( \begin{array} {cccccc}a_{1} - \lambda&b_{1}& \,&.\,&\,\\
\quad\\
c_{1}&a_{2} - \lambda& b_{2}&...,&..\\
\quad\\
\,&c_{2}& ...,&...&...\\
\quad\\
...&...& ...&..&b_{n-1}\\
\quad\\
...&...& ...&c_{n-1}&a_{n}-\lambda\\
\end{array} \right )$\\

\n Then $\displaystyle{P_{n}(\lambda) = det(T - \lambda I} )$ verify the following recurrence relation:\\

\begin{equation}
\left \{ \begin{array}{c}P_{0}(\lambda) = 1\quad\quad\quad \quad\quad\quad \quad\quad\quad \quad\quad\quad \quad\quad\quad \quad\quad\quad \quad\\
\quad\\
P_{1}(\lambda) = a_{1} - \lambda \quad\quad\quad \quad\quad\quad \quad\quad\quad \quad\quad\quad \quad\quad\quad\quad\quad\\
\quad\\
P_{i}(\lambda) =  a_{i}P_{i-1}(\lambda) - b_{i-1}c_{i-1}P_{i-2}(\lambda), \,\, \text{for}\,\, i = 2, ......, n\\
\end{array} \right.\\
\end{equation}

\n Observe that $\theta_{n} = det T$ . See also {\color{blue} [25]} and the references {\color{blue}[30]} and {\color{blue}[43]} are useful to consult.\\

\n {\color{red}$\bullet$} The following result on the eigenvalues of a tridiagonal matrix is well known, although we present a proof here for completeness.\\

\begin{lemma}

\n Let $T$ be an $n\times n$  {\color{red}tridiagonal} matrix in the form \\

\begin{equation}
T = \left ( \begin{array} {cccccc}a_{1}&b_{1}& \,&.\,&\,\\
\quad\\
c_{1}&a_{2}& b_{2}&...,&..\\
\quad\\
\,&c_{2}& ...,&...&...\\
\quad\\
...&...& ...&..&b_{n-1}\\
\quad\\
...&...& ...&c_{n-1}&a_{n}\\
\end{array} \right )\\
\end{equation}

\n If {\color{red}$b_{i}c_{i} > 0$} for $i = 1, 2, . . . , n - 1$, then the eigenvalues of $T$ are {\color{red}real} and have algebraic multiplicity one (i.e., are simple). Moreover, $T$ is similar (via a positive diagonal matrix) to a symmetric nonnegative tridiagonal matrix.\\
\end{lemma}

\n {\color{red}Proof}\\

\n Let $D = [d_{ij}]$ be the $ n \times n$ diagonal matrix where $d_{11} = 1$, and for $\displaystyle{k > 1, d_{kk} = \frac{\sqrt{b_{1}b_{2}......b_{k-1}}}{\sqrt{c_{1}c_{2}......c_{k-1}}}}$. Then it is readily verified that\\

\begin{equation}
DTD^{-1} = \left ( \begin{array} {cccccc}a_{1}&\sqrt{b_{1}c_{1}}& \,&.\,&\,\\
\quad\\
\sqrt{b_{1}c_{1}}&a_{2}&\sqrt{b_{2} c_{2}}&...,&..\\
\quad\\
\,&\sqrt{b_{2}c_{2}}& ...,&...&...\\
\quad\\
...&...& ...&..&\sqrt{b_{n-1}c_{n-1}}\\
\quad\\
...&...& ...&\sqrt{b_{n-1}c_{n-1}}&a_{n}\\
\end{array} \right )\\
\end{equation}

\n Since $DTD^{-1}$ is symmetric and $T$ and $DTD^{-1}$ have the same eigenvalues, this implies that the eigenvalues of $T$ are real.\\

\n  Suppose $\lambda$ is an eigenvalue of $T$, then $T - \lambda I$ is also of the form (3.53). If the first row and last column of $T - \lambda I$ are deleted, then the resulting matrix is an $(n -1)\times (n- 1)$ upper triangular matrix with no zero entries on its main diagonal, since $b_{i}c_{i} > 0$, for $i = 1, 2, . . . , n - 1$. Hence this submatrix has rank $n - 1$. It follows that $T - \lambda I$ has rank at least $n -1$. However, $T - \lambda I$ has rank at most $n - 1$ since $ \lambda$ is an eigenvalue of $T$ . So by definition $\lambda$ has geometric multiplicity one. This completes the proof, \hfill { } $\blacksquare$\\

\n {\color{red}$\bullet$} Much work has been done to study the bounds of the eigenvalues of {\color{red}real symmetric} matrices. Methods discovered in the mid-nineteenth century reduce the original matrix to a {\color{red}tridiagonal matrix} whose eigenvalues are the {\color{red}same} as those of the original matrix. \\

\n Exploiting this idea, Golub in $1962$ determined {\color{red}lower bounds} on tridiagonal matrices of the form:\\

$S$ = $\left ( \begin{array} {cccccc}a_{1}&b_{1}& \,&.\,&\,\\
\quad\\
b_{1}&a_{2} & b_{2}&...,&..\\
\quad\\
\,&b_{2}& ...,&...&...\\
\quad\\
...&...& ...&..&b_{n-1}\\
\quad\\
...&...& ...&b_{n-1}&a_{n}\\
\end{array} \right )$\\

\begin{proposition} (Golub {\color{blue}[17]}, Corollary 1.1)\\

\n Let $A$ be an $n \times n$ matrix with real entries $a_{ij} = a_{i}$ for $i = j, a_{ij} = b_{m}$ for $\vert i - j \vert = 1$ where $m = min(i; j)$, and $a_{ij} = 0 $otherwise. Then the interval $\displaystyle{[a_{k} - \sigma_{k} , a_{k} + \sigma_{k}] }$ where $\displaystyle{\sigma_{k}^{2} = b_{k}^{2} + b_{k-1}^{2}}$
contains at least one eigenvalue.\\
\end{proposition}

\begin{theorem}
\quad\\

\n Every Jacobi matrix $S$ of the following form:\\
\begin{equation}
S = \left ( \begin{array} {cccccc}a_{1}&b_{1}& \,&.\,&\,\\
\quad\\
b_{1}&a_{2} & b_{2}&...,&..\\
\quad\\
\,&b_{2}& ...,&...&...\\
\quad\\
...&...& ...&..&b_{n-1}\\
\quad\\
...&...& ...&b_{n-1}&a_{n}\\
\end{array} \right )
\end{equation}

\n has $n$ distinct real eigenvalues, $\displaystyle{\lambda_{1},......, \lambda_{k}, ....., \lambda_{n}}$ corresponding to eigenvectors  $\displaystyle{\mathcal{U}^{1},......, \mathcal{U}^{k}, .....,\mathcal{U}_{n}}$ with $\displaystyle{\vert\vert \mathcal{U}^{k} \vert\vert = 1}$ and the first component $u_{1}^{k} > 0 $ of eigenvector $\displaystyle{ \mathcal{U}^{k} = (u_{1}^{k}, ....., u_{j}^{k}, ..... u_{n}^{k})}$  for all $k =1, ...., n$\\
\n In addition, $\gamma_{k} = u_{1}^{k} $, called {\color{red}norming constants}, satisfy $\displaystyle{\sum_{k=1}^{n}\gamma_{k}^{2} = 1}$.\\
\end{theorem}

\n {\bf {\color{red}Proof}}\\

\n Every eigenvector $\mathcal{U}^{k}$ of $S$  corresponding to an eigenvalue $\lambda_{k}$ satisfies,\\

\n $\displaystyle{a_{1}u_{1}^{k}  + b_{1}u_{2}^{k} = \lambda_{k}u_{1}^{k}}$\\

\n $\displaystyle{b_{1}u_{1}^{k}  + a_{2}u_{2}^{k}  + b_{2}u_{3}^{k} = \lambda_{k}u_{2}^{k}}$\\
...................................................\\
...................................................\\
...................................................\\
\n $\displaystyle{b_{n-2}u_{n-2}^{k}  + a_{n-1}u_{n-1}^{k}  + b_{n-1}u_{n}^{k} = \lambda_{k}u_{n-1}^{k}}$\\
\n $\displaystyle{b_{n-1}u_{n-1}^{k}  + a_{n}u_{n}^{k} = \lambda_{k}u_{n}^{k}}$\\

\n {\color{red}equivalently}\\
\begin{equation}
\n \left \{ \begin{array} {c} \displaystyle{ b_{1}u_{2}^{k} = \lambda_{k}u_{1}^{k} - a_{1}u_{1}^{k}} \quad\quad\quad \quad\quad\quad \quad\quad\quad\\
\quad\\
\n \displaystyle{ b_{2}u_{3}^{k} = \lambda_{k}u_{2}^{k} - b_{1}u_{1}^{k}  - a_{2}u_{2}^{k} }\quad\quad\quad\quad\quad\quad\\
\quad\\
...................................................\quad\quad\quad\\
...................................................\quad\quad\quad\\
...................................................;\quad\quad\quad\\
\quad\\
\n \displaystyle{ b_{n-1}u_{n}^{k} = \lambda_{k}u_{n-1}^{k} - b_{n-2}u_{n-2}^{k}  - a_{n-1}u_{n-1}^{k} }\\
\n \displaystyle{0 = \lambda_{k}u_{n}^{k} - b_{n-1}u_{n-1}^{k}  - a_{n}u_{n}^{k} }\quad\quad\quad\quad\quad\quad\\
\end{array} \right.\\
\end{equation}

\n Suppose by contradiction that $u_{1}^{k} = 0$ then it follows from the first equation that $u_{2}^{k} = 0$, then from the second that $u_{3}^{k} = 0$, etc., hence $\mathcal{U}^{k} = 0$ which is a contradiction. Thus, $u_{1}^{k} \neq 0$ for all $k = 1, ..., n$ , so replacing, if necessary, the eigenvector  $\mathcal{U}^{k}$ by the eigenvector $\displaystyle{\frac{\vert u_{1}^{k}\vert \mathcal{U}^{k}}{u_{1}^{k} \vert\vert \mathcal{U}^{k}\vert\vert}}$ we obtain $u_{1}^{k} > 0$ and  $\vert\vert \mathcal{U}^{k} \vert\vert = 1$.\\

\n Since the Jacobi matrix $S$ is self-adjoint, its eigenvalues are all real. Suppose by contradiction that there are less than $n$ distinct eigenvalues. Then some eigenvalue $\lambda_{k}$ must have multiplicity more than $1$, so there are at least two linearly independent eigenvectors $\mathcal{U}^{k}$ and $\mathcal{U}^{k}$ corresponding to the eigenvalue $\lambda_{k}$. Then $\displaystyle{\mathcal{W}^{k} = v_{1}^{k}\mathcal{U}^{k} - u_{1}^{k}\mathcal{V}^{k}}$ is also an eigenvector corresponding to the eigenvalue $\lambda_{k}$, but $w_{1}^{k} = 0$ which is impossible. \\

\n Since eigenvectors corresponding to different eigenvalues of any self-adjoint matrix are necessarily orthogonal, we have that the square matrix $\displaystyle{\mathbb{U } = [\mathcal{U}^{1}, ....., \mathcal{U}^{k}, .....,\mathcal{U}^{n}]}$ has orthonormal columns and, hence, satisfies $\mathbb{U}\mathbb{U}^{*} = I $. iv)), that is, the rows of $\mathbb{U}$ are also orthonormal so, in particular, $\displaystyle{\sum_{k=1}^{n}\gamma_{k}^{2} = \sum_{k=1}^{n} \vert u_{1}^{k} \vert^{2} = 1}$. \hfill { } $\blacksquare$\\

\n {\color{red}$\bullet$} We come back to Cauchy problem (3.46) associated to the matrix:

\begin{equation}
\left ( \begin{array} {ccccccc}0&\omega_{1}&0&...&0& 0\\
\quad\\
\omega_{1}&0&\omega_{2}&...&0&0\\
\quad\\
...&...&...&...&...&...\\
\quad\\
0&0&0&...&0\,&\omega_{n-1}\\
\quad\\
...&...&...&...&\omega_{n-1}&0\\
\end{array} \right )
\end{equation}

\n where $\displaystyle{\omega_{n} = \sqrt{n}(n + 1)}$ with $n \longrightarrow +\infty$.\\

\n i.e.\\

\begin{equation}
\left ( \begin{array} {ccccccc}0&\omega_{1}&0&...&0& 0&...\\
\quad\\
\omega_{1}&0&\omega_{2}&...&0&0&...\\
\quad\\
...&...&...&...&...&...&...\\
\quad\\
0&0&0&...&0\,&\omega_{n-1}&...\\
\quad\\
...&...&...&...&\omega_{n-1}&0&...\\
\quad\\
...&...&...&...&...&...&...\\
\end{array} \right )
\end{equation}

 \n {\color{red}$\bullet$} We consider the Jacobi operator acting on $\displaystyle{\tilde{a} = \{\tilde{a}_{n} \}_{n \in \mathbb{N}}}$\\
\begin{equation}
\displaystyle{(\mathbb{J}\tilde{a})_{n} = \omega_{n-1}\tilde{a}_{n-1} + \omega_{n}\tilde{a}_{n+1} ; \, \omega_{n} = (n+1)\sqrt{n}, n \geq 1 }
\end{equation}

 \n {\color{red}$\rhd\bullet$} Let us consider a second-order difference equation\\

\begin{equation}
\displaystyle{\omega_{n-1}\tilde{a}_{n-1} + \omega_{n}\tilde{a}_{n+1} = z\tilde{a}_{n}, n \geq 1}
\end{equation}

\n associated with the operator $\mathbb{J}$.\\

\n Clearly, values $\tilde{a}_{1}$ and $\tilde{a}_{2}$ determine uniquely a solution $\tilde{a}_{n}$ of equation (3.60). In particular, the solutions $P_{n}(z)$ and $Q_{n}(z)$ are distinguished by boundary conditions.\\

\begin{equation}
\displaystyle{P_{1}(z) = 1}, \quad \displaystyle{Q_{1}(z) = 0}
\end{equation}
 
\n and\\

\begin{equation}
\displaystyle{P_{1}(z) = \frac{z}{\omega_{1}}}, \quad \displaystyle{Q_{1}(z) = \frac{1}{\omega_{1}}}
\end{equation}

\n  {\color{red}$\bullet$}  We consider Jacobi operators $\mathbb{J}$ acting in the space $\ell_{2}(\mathbb{N})$ with the scalar product $\displaystyle{ < \tilde{a}, \tilde{b} > = \sum_{n=1}^{\infty} \tilde{a}_{n}\overline{\tilde{b}_{n}}}$

\begin{definition} 

\n For the analysis of linear second order difference equations, the Wronskian, also called the Casoratian see {\color{blue}[1]} play an important role. It defined as follows : \\

\n $\displaystyle{W(\tilde{a}, \tilde{b})(n) = \tilde{a}_{n}\overline{\tilde{b}_{n+1}} - \tilde{a}_{n+1}\overline{\tilde{b}_{n}}}$ for every $n \in \mathbb{N}$.\\

\n and we define\\

\n $\displaystyle{[\tilde{a}, \tilde{b}]_{n} = \omega_{n}W(\tilde{a}, \tilde{b})(n)}$
\end{definition} 
\n It is clear that the wronskian is a bilinear skew-symmetric map. In addition we have \\

\n $\displaystyle{W(\tilde{c}\tilde{a}, \tilde{c}\tilde{b})(n) = \tilde{c}_{n}\tilde{c}_{n+1} {W(\tilde{a}, \tilde{b})(n)}}$ for every $n \in \mathbb{N}$.\\

\n Moreover, $\displaystyle{W(\tilde{a}, \tilde{b}) = 0}$ when $\tilde{a}$ and $\tilde{b}$ are linearly dependent.\\

\n Then we verify easily that\\

\n $\displaystyle{\sum_{k=1}^{n} (\mathbb{J}\tilde{a})_{n}\overline{\tilde{b}_{n}} -  \tilde{a}_{n}(\mathbb{J}\overline{\tilde{b}})_{n} = -[\tilde{a}, \tilde{b}]_{n}}$\\

\n Let $\displaystyle{D_{max} = \{ \tilde{a} \in \ell_{2}(\mathbb{N} ; \mathbb{J}\tilde{a} \in \ell_{2}(\mathbb{N} \}}$ be the maximal domain of $\mathbb{J}$  and we denote $\mathbb{J}$ acting on $D_{max}$ by $\mathbb{J}_{max}$. \\

\n By using the arsenal of Difference Operator Theory {\color{blue}[5]} Chapter VII] and the results of {\color{blue} [26]} or {\color{blue} [44]} on self-adjoint extensions of circle-boundary Jacobi matrices we deduce that \\

\begin{lemma}

\n If $\tilde{a} \in D_{max}$ and $\tilde{b} \in D_{max}$ then the limit of $[\tilde{a} , \tilde{b}]_{k}$ exists as $k \longrightarrow +\infty$. This limit is denoted $[\tilde{a}, \tilde{b}]_{\infty}$\\
\end{lemma}

\n {\bf {\color{red}Proof}}\\

\n Let $\tilde{a} \in D_{max}$ and $\tilde{b} \in D_{max}$ then $\mathbb{J}\tilde{a} \in \ell_{2}$ and $\mathbb{J}\tilde{b} \in \ell_{2}$ and it follows that $\displaystyle{(\mathbb{J}\tilde{a})_{j}\overline{\tilde{b}_{j}}}$ and $\displaystyle{\tilde{a}_{j}\overline{(\mathbb{J}\tilde{b}_{j})}}$ are in $\ell_{1}$ so it follows that $\displaystyle{lim \, \sum_{j=1}^{k}[(\mathbb{J}\tilde{a})_{j}\overline{\tilde{b}_{j}} - \tilde{a}_{j}(\mathbb{J}\overline{\tilde{b}})_{j}]}$ exists as $k \longrightarrow + \infty$.\\

\n {\bf Consequently} \\

\n If we put $\displaystyle{ lim \, [\tilde{a} , \tilde{b}]_{k} = [\tilde{a} , \tilde{b}]_{\infty} }$ as $k \longrightarrow + \infty$ then we get \\
\begin{equation}
\displaystyle{< \mathbb{J}_{max}\tilde{a} , \tilde{b} > - < \tilde{a} , \mathbb{J}_{max}\tilde{b} > =  -[\tilde{a} , \tilde{b}]_{\infty} }
\end{equation}

\n {\color{red}$\rhd$} Let $\mathcal{D}_{0}$ be the set of  $\tilde{a}$ in $\ell_{2}$ having a non-zero finite number of components, then $\displaystyle{\mathbb{J}_{\mid \mathcal{D}_{0}}}$ is a symmetric operator and it admits a closure that will be denoted by $\mathbb{J}_{min } = \mathbb{J}$ of domain \\

\begin{equation}
\displaystyle{D_{min} = \{\tilde{a} \in \ell_{2} ; \exists \tilde{a}_{k} \in \mathcal{D}_{0}, lim\, \tilde{a}_{k} = \tilde{a} \, and \, \exists \tilde{b}  \in \ell_{2} ; lim \, (\mathbb{J}\tilde{a}_{k}) = \tilde{b} , k \longrightarrow +\infty \}}
\end{equation}

\n {\color{red}$\rhd$} Let $\lambda$ be a complex number with nonzero imaginary part, $\Im m \lambda  \neq 0$. Assume that $\Im m \lambda > 0$. By $\mathcal{R}_{\lambda}$ and $\mathcal{R}_{\overline{\lambda}}$ we denote the ranges of the operators $\mathbb{J} - \lambda I$ and $\mathbb{J} - \overline{\lambda} I$, respectively, and by $\mathcal{N}_{\lambda}$ and $\mathcal{N}_{\overline{\lambda}}$ we denote their orthogonal complements in the space $\ell_{ 2}$.\\

 \n The spaces $\mathcal{N}_{\lambda}$ and $\mathcal{N}_{\overline{\lambda}}$ are called the {\color{red}defect subspaces} corresponding to the numbers  $\lambda$ and $\overline{\lambda}$. It is known that the dimensions $ n_{+ } = dim \mathcal{N}_{\lambda}$ and $n_{-} = dim  \mathcal{N}_{\overline{\lambda}} $ are the same in the upper and lower half-planes. The numbers  $n_{+ }$ and  $n_{-}$ are called {\color{red}defect numbers}, and the pair $( n_{+},  n_{-})$ is the {\color{red}defect index} of the operator $\mathbb{J}$. The operator $\mathbb{J}$ is self-adjoint if and only if its defect numbers satisfy the equality $ n_{+} =  n_{-} = 0$.\\

\n Using Theorem 1.5, Ch. VII, {\color{blue}[5]} we deduce the following properties\\

\n ({\color{red}$\bullet_{1}$}) $\displaystyle{D_{min} = \{\tilde{a}  \in D_{max} ; [\tilde{a}  , \tilde{b} ]_{\infty} = 0 \,\quad  \forall \, \psi \in D_{max}\}}$.\\ 

\n ({\color{red}$\bullet_{2}$}) $\mathbb{J}_{min} $ is a closed operator symmetric with indices of default (1, 1).\\

\n ({\color{red}$\bullet_{3}$}) $\mathbb{J}_{max}$ is the adjoint of $\mathbb{J}_{min}$.\\

\n ({\color{red}$\bullet_{4}$})  If $u$ and $v$ are  two independent solutions then at infinity, the minimal domain of  $\mathbb{J}_{min}$ is caracterized by \\

\n $\displaystyle{D_{min} \{\tilde{a} \in D_{max}; [\tilde{a}, u]_{\infty} =  [\tilde{a}, u]_{\infty} = 0\}}$\\

\begin{definition} 
({\color{red}limit circle at infinity}) [42]\\

\n The Jacobi operator is said to have a {\color{red}limit circle at  infinity} if a solution $\tilde{a}$ of $\mathbb{J}\tilde{a} = z\tilde{a}$ with $\tilde{a}_{1} =1$ belongs to $\ell_{2}(\mathbb{N})$ for some $z \in \mathbb{C}$.
\end{definition}

\begin{lemma}

\n Consider $\mathbb{J}\tilde{a} = 0$ with $\tilde{a}_{1} =1$, then $ \tilde{a} \in \ell_{2}(\mathbb{N})$\\

\end{lemma}

\n {\bf {\color{red}Proof}}

\n $\displaystyle{n\sqrt{n-1}\tilde{a}_{n-1}  + (n+1)\sqrt{n} \tilde{a}_{n+1} = 0 , n \ge 2 \iff  \tilde{a}_{n+1} = - \frac{n\sqrt{n-1}}{(n+1)\sqrt{n} }  \tilde{a}_{n-1} , n \geq 2}$\\

\n $\displaystyle{\iff \tilde{a}_{n+1} = - \frac{\sqrt{n(n-1)}}{n+1}  \tilde{a}_{n-1}}$

\n We observe that $\tilde{a}_{2}  = 0$ , $\tilde{a}_{4}  = 0$ then for every even $n = 2p; p \in \mathbb{N}$,  we have $\tilde{a}_{n}  = 0$.\\

 \n For every odd $n = 2p + 1; p \in \mathbb{N}$, we have :\\
 
 \n $\displaystyle{\tilde{a}_{2p+1}  = - \frac{\sqrt{2p(2p-1)}}{2p + 1}\tilde{a}_{2p-1} ; p \in \mathbb{N}}$. It follows  that\\
 
  \n $\displaystyle{\tilde{a}_{2p+1}  = (-1)^{p} \prod_{j=1}^{p} \frac{\sqrt{2j(2j-1)}}{2j + 1} ; p \in \mathbb{N}}$.\\
  
   \n $\displaystyle{\vert \tilde{a}_{2p+1}\vert  = \prod_{j=1}^{p} \frac{\sqrt{2j(2j-1)}}{2j + 1} ; p \in \mathbb{N}}$.\\
   
 \n $\displaystyle{= exp \{ln[\prod_{j=1}^{p} \frac{\sqrt{2j(2j-1)}}{2j + 1}]\}}$\\
 
 \n $\displaystyle{= exp \{\sum_{j=1}^{p}[ln \sqrt{2j(2j-1)} - ln(2j + 1)]\}}$\\
 
  \n $\displaystyle{= exp \{\sum_{j=1}^{p}[\frac{1}{2}ln 2j(2j-1) - ln(2j + 1)]\}}$\\
  
  \n $\displaystyle{= exp \{\sum_{j=1}^{p}[\frac{1}{2}ln 2j + \frac{1}{2}ln (2j-1) - ln(2j + 1)]\}}$\\
  
  \n $\displaystyle{= exp \{\sum_{j=1}^{p}[\frac{1}{2}ln (1 - \frac{1}{2j}) - ln 2j - ln(1 + \frac{1}{2j})]\}}$\\
  
   \n $\displaystyle{= exp \{\sum_{j=1}^{p}[\frac{1}{2}ln 2j + \frac{1}{2}ln 2j + \frac{1}{2}ln (1 - \frac{1}{2j}) - ln 2j - ln(1 + \frac{1}{2j})]\}}$\\
   
    \n $\displaystyle{= exp \{ \frac{1}{2}\sum_{j=1}^{p}ln (1 - \frac{1}{2j}) - \sum_{j=1}^{p} ln(1 + \frac{1}{2j})\}}$\\
    
    \n $\displaystyle{= exp \{ \frac{1}{2}\sum_{j=1}^{p}- \frac{1}{2j} + \epsilon(\frac{1}{j})(\frac{1}{j}) - \sum_{j=1}^{p}  \frac{1}{2j} + \epsilon(\frac{1}{j}) (\frac{1}{j} )\}}$\\
    
      \n $\displaystyle{= exp \{ \sum_{j=1}^{p}- \frac{1}{4j}  - \frac{1}{2j} + \epsilon(\frac{1}{j})(\frac{1}{j}) \}}$\\
      
           \n $\displaystyle{= exp \{ \sum_{j=1}^{p}- \frac{3}{4j} + \epsilon(\frac{1}{j})(\frac{1}{j}) \}}$\\
 
  \n $\displaystyle{= exp \{ - \frac{3}{4}\sum_{j=1}^{p} \frac{1}{j} + O(\frac{1}{j}) \}}$\\
  
  \n Let $\displaystyle{S_{p} = \sum_{j=1}^{p} \frac{1}{j} }$ then $\displaystyle{\vert \tilde{a}_{2p+1}\vert  \sim \frac{1}{e^{\frac{3}{4}S_{p}}}}$ ; $p \longrightarrow +\infty$.\\
    
    \begin{lemma}
     \n $\displaystyle{\frac{1}{e^{\frac{3}{4}S_{p}}} \leq \frac{1}{p^{\frac{3}{4}}}}$
     \end{lemma}
     
     \n {\bf{\color{red} Proof}}\\
     
     \n We begin by observing that:\\
     
     \n  $\displaystyle{\frac{1}{e^{\frac{3}{4}S_{p}}} \leq \frac{1}{p^{\frac{3}{4}}}\iff}$  $\displaystyle{e^{\frac{3}{4}S_{p}} \geq p^{\frac{3}{4}}\iff}$ $\displaystyle{e^{\frac{3}{4}S_{p}} \geq e^\frac{3}{4} ln(p)}\iff$ $\displaystyle{S_{p} \geq  ln(p)\iff}$  $\displaystyle{S_{p} -  ln(p) {\color{red}\geq} 0}$.\\
  
     \n Calculus students can approximate the integral $\displaystyle{\int_{1}^{p} \frac{dx}{x} = ln(p)}$ by inscribed and circumscribed rectangles, and hence obtain the inequalities (for $p > 1$)\\
    
     \n $\displaystyle{ \frac{1}{p} < \sum_{j=1}^{p}\frac{1}{j}  - ln(p) < 1}$\\
     
    \n   Better sub-estimations $\displaystyle{ \frac{1}{2(p+1} + \gamma {\color{red} < }S_{p} -  ln(p) {\color{red}< } \frac{1}{2p} + \gamma}$ (see {\color{blue}[47]} ) where $\gamma$ is Euler's constant given by:\\
    
    $\displaystyle{\gamma = \lim\limits_{p \longrightarrow + \infty}(\sum_{j=1}^{p} - ln(p))}$. This limit is close to $0.5$ \\
    
 \n   Hence, there is $C > 0$ such that\\
    
    \begin{equation}
    \displaystyle{\vert \tilde{a}_{2p + 1} \leq \frac{C}{p^{\frac{3}{4}}} \,\,  \text{ in particular}  \vert \tilde{a}_{2p + 1}\vert^{2} \leq \frac{C^{2}}{p^{\frac{3}{2}}}}
    \end{equation}
    
    \n so that $\displaystyle{\vert \tilde{a} \in \ell(\mathbb{N})}$ \hfill{ } $\blacksquare$\\
    
  \n As a consequence of this Lemma, there exists a self-adjoint extension of the Jacobi operator $\mathbb{J}$ subject to a boundary condition at $\infty$, so that the spectrum of $\mathbb{J}$ in $\ell_{2}(\mathbb{N})$ consists of a countable set of simple real isolated eigenvalues see {\color{blue}[39]}. \\
  
\n   The existence of the self-adjoint extension of $\mathbb{J}$ guarantees local well-posedness of the Cauchy problem (3.42)\\
       
\begin{remark}
({\color{blue}CMP 1998})\\
 \quad  \\
In an article in Communications in Mathematical Physics, 199,  (1998), \\we have presented a complete scattering analysis of the cubic Heun's operator $\displaystyle{\mathbb{H} = A^{*} (A + A^{*})A}$ acting on Bargmann space $\mathbb{B}$ defined by:\\
\begin{equation}
\left \{ \begin{array} {c} \displaystyle{(\mathbb{H} \varphi)_{1} = \sqrt{2} \varphi_{2}} \quad \quad \quad \quad \quad \quad \quad \quad \quad\quad \quad \quad \quad \quad \quad \quad \quad \quad  \\
\quad\\
\displaystyle{(\mathbb{H} \varphi)_{n} = (n -1) \sqrt{n} \varphi_{n-1} + n\sqrt{n +1} \varphi_{n+1}; \,\, n \geq 2} \quad  \quad \quad\\
\end{array} \right.\\
\end{equation}
where\\
$\displaystyle{\varphi(z) = \sum_{n=1}^{\infty} \varphi_{n}e_{n}(z)} \text{and} \,\, \displaystyle{e_{n}(z) = \frac{z^{n}}{\sqrt{n!}}}$ \\

\n In particular, the boundary conditions at infinity are used in a description of all maximal dissipative extensions in Bargmann space of the minimal cubic Heun's operator.The characteristic functions of the dissipative extensions are computed.Completeness theorems are obtained for the system of generalized eigenvectors associated to it.\\
\n This operator play an important role in Reggeon field theory.\\

\end{remark}

 \section{ {\color{red}Domination of $\displaystyle{\sum_{i+ j < 2k}a_{i,j}A^{*^{i}}A^{j}}$ by $\displaystyle{A^{*^{k}}A^{k}; \, a_{i,j} \in \mathbb{C}}$ }}
\subsection{{\color{blue} On the domination of $\displaystyle{\sum_{i+ j < 2k}a_{i,j}A^{*^{i}}A^{j}}$ by $\displaystyle{A^{*^{k}}A^{k}}$}}
 \n Recalling some useful definitions and classical lemmas:\\
 
 \begin{lemma} (classical lemma)\\

\n Let  $p, q \geq 1$ $\displaystyle{\frac{1}{p} + \frac{1}{q} = 1.}$ Then we have\\

\n (i) $\displaystyle{ab \leq \frac{1}{p}a^{p} + \frac{1}{q}b^{q} \quad \forall a \geq 0, \quad \forall b \geq 0.}$\\

\n (ii) $\displaystyle{\forall \epsilon > 0, \exists C_{\epsilon} > 0 ;  ab \leq \epsilon a^{p} + C_{\epsilon}b^{q} \quad \forall a \geq 0, \quad \forall b \geq 0.}$\\
\end{lemma}

\begin{definition}

 ({\color{red}relative bounded between two operators})\\

\n Let $T_{1}$ and $T_{2}$ are two operators acting on Hilbert space  $\mathcal{H}$ with respective domain $D(T_{1})$ and $D(T_{2})$.\\

\n  We say that $T_{2}$ {\color{red}dominated} by $T_{1}$ ( $T_{2} << T_{1}$)  or $T_{2}$ is {\color{red}relatively bounded} with respect $T_{1}$ or ( $T_{1}$-bounded) if, \\

\n (i) $D(T_{1}) \subset D(T_{2})$.\\

\n (ii) $\exists \, a > 0$ and $ b>0$ such that $ \mid\mid T_{2}\phi \mid\mid \leq a\mid\mid T_{1}\phi \mid\mid + b\mid\mid \phi \mid\mid \quad \forall \phi \in D(T_{1}).$ \\
\end{definition}

\begin{lemma} ({\color{red}equivalence conditions}\\

\n An equivalence condition of \\
\begin{equation}
\exists \, a > 0 \,\, \text{and} \,\,b > 0\,\, \text{such that}\,\, \displaystyle{ \mid\mid T_{2}\phi \mid\mid \leq a\mid\mid T_{1}\phi \mid\mid + b\mid\mid \phi \mid\mid \,\, \forall \phi \in D(T_{1}).}
\end{equation}
\n is\\
\begin{equation}
\exists \, a'> 0 \, \text{and}\, \,b'> 0;\,\, \displaystyle{\mid\mid T_{2} \phi\mid\mid^{2} \leq a'^{2}\mid\mid T_{1}\phi \mid\mid^{2} + b'^{2} \mid\mid \phi \mid\mid^{2}\,\,\forall \phi \in D(T_{1}).}
\end{equation}
where the constants $a'$ and $b'$ are, of corse, in general different from $a$, $b$.\\
\end{lemma}
\n {\bf{\color{red} Proof}}\\
 
\n {\bf {\color{red}$\bullet$}} It is easily seen that (4.2) implies (4.1)  with $a = a'$ and $b = b'$. {\bf In fact:}\\

\n As  $\displaystyle{ \mid\mid T_{2}\phi \mid\mid^{2} \leq a'^{2}\mid\mid T_{1}\phi \mid\mid^{2} + b'^{2}\mid\mid \phi \mid\mid^{2} }$ then \\

\n $\displaystyle{ \mid\mid T_{2}\phi \mid\mid \leq \sqrt{a'^{2}\mid\mid T_{1}\phi \mid\mid^{2} + b'^{2}\mid\mid \phi \mid\mid^{2}}}$ \\

\n $\displaystyle{ \leq \sqrt{(a' \mid\mid T_{1}\phi \mid\mid + b' \mid\mid \phi \mid\mid)^{2}}}$ \\

\n $\displaystyle{ \leq a' \mid\mid T_{1}\phi \mid\mid + b'\mid\mid \phi \mid\mid}$.\\

\n {\bf {\color{red}$\bullet$}} Whereas  (4.1) implies (4.2)  with $a'^{2} = (1 + \frac{1}{\epsilon})a^{2}$ and $b'^{2} = (1 + \epsilon)b^{2}$ with an arbitrary $\epsilon > 0$. {\bf In fact:}\\

\n As $\displaystyle{ \mid\mid T_{2}\phi \mid\mid \leq a\mid\mid T_{1}\phi \mid\mid + b\mid\mid \phi \mid\mid }$ then\\

\n $\displaystyle{ \mid\mid T_{2}\phi \mid\mid^{2} \leq (a\mid\mid T_{1}\phi \mid\mid + b\mid\mid \phi \mid\mid)^{2} =  a^{2}\mid\mid T_{1}\phi \mid\mid^{2} + b^{2}\mid\mid \phi \mid\mid^{2} + 2\frac{a}{\sqrt{\epsilon}}\vert\vert T_{1}\phi \vert\vert \sqrt{\epsilon}b\vert\vert \phi \vert\vert}$ \\

\n and as $\displaystyle{2\frac{a}{\sqrt{\epsilon}}\vert\vert T_{1}\phi \vert\vert \sqrt{\epsilon}b\vert\vert \phi \vert\vert \leq \frac{a^{2}}{\epsilon}\vert\vert T_{1}\phi \vert\vert^{2} + \epsilon b^{2} \vert\vert \phi \vert\vert^{2}}$, we deduce that\\

\n $\displaystyle{ \mid\mid T_{2}\phi \mid\mid^{2} \leq (1 + \frac{1}{\epsilon})a^{2}\vert\vert T_{1}\phi \vert\vert^{2} + (1 + \epsilon)b^{2}\vert\vert\phi \vert\vert^{2}}$.  \hfill { } $\blacksquare$\\

\begin{definition} ({\color{red}$T_{2}$ is strictly dominated by $T_{1}$})\\

\n  We say that $T_{2}$ is {\color{red}strictly dominated} by $T_{1}$ ( $T_{2} <<<T_{1}$) if, \\

\n (i) $D(T_{1}) \subset D(T_{2})$.\\

\n (ii)$_{bis}$ $ \forall \epsilon > 0 \quad \exists C_{\epsilon} > 0 $ such that $ \mid\mid T_{2}\phi \mid\mid \,\, \leq \epsilon\mid\mid T_{1}\phi \mid\mid + \,C_{\epsilon}\mid\mid \phi \mid\mid \,\, \forall \phi \in D(T_{1}).$ \\
\end{definition}

\begin{definition}({\color{red}$T_{2}$ is $T_{1}$-compact})\\

\n We say hat $T_{2}$ is {\color{red}$T_{1}$-compact} if, for all sequence $\{\phi_{k}\}$ in $D(T_{1})$ such that the sequences 
$\{\phi_{k}\}$ are $\{T_{1}\phi_{k}\}$ are bounded and we can extract from $\{T_{2}\phi_{k}\}$ a convergent subsequence.\\
\end{definition}

\begin{proposition}
.\\

\n Let $T_{1}$ be a closed operator on $\mathcal{H}$ with {\color{red}non-empty} resolvent set. Let $T_{2}$ be an operator on $\mathcal{H}$. We say that $T_{2}$ is $T_{1}$-compact (or relatively compact with respect to $T_{1}$) if $D(T_{1}) \subset D(T_{2})$ and one of the following equivalent assertions is satisfied.\\

\n (i) For all $\sigma \in \rho(T_{1})$, the operator $\displaystyle{T_{2}(T_{1} - \sigma I)^{-1}}$ is compact.\\

\n (ii) For any sequence $(\varphi_{n})$ bounded in $D(T_{1})$  (i.e. $(\varphi_{n})$ and $(T_{1}\varphi_{n})$ are bounded in $\mathcal{H}$) then
$(T_{2}\varphi_{n})$ has a {\color{red}convergent subsequence}.\\
\end{proposition}
\n {\bf{\color{red}Proof}}\\

\n {\color{blue}$(ii) \Longrightarrow (i)$} Let  $\sigma \in \rho(T_{1})$. Let $(\psi_{n})$ be a bounded sequence in $\mathcal{H}$. Then $\displaystyle{((T_{1} - \sigma I)^{-1} \psi_{n})}$ is bounded in $D(T_{1})$ and hence $\displaystyle{(T_{2}(T_{1} - \sigma I)^{-1} \psi_{n})}$ has a convergent subsequence in $\mathcal{H}$. This proves that $\displaystyle{T_{2}(T_{1} - \sigma I)^{-1}}$ is compact.\\

\n {\color{blue}$(i) \Longrightarrow (ii)$} Conversely, assume that $\displaystyle{T_{2}(T_{1} - \sigma I)^{-1}}$ is compact for some $\displaystyle{ \sigma \in \rho(T_{1})}$ and consider $(\psi_{n})$ bounded in $D(T_{1})$. Then $\displaystyle{(T_{1} - \sigma I)^{-1}\psi_{n}}$ is bounded in $\mathcal{H}$. Then $\displaystyle{(T_{2}(T_{1} - \sigma I)^{-1}(T_{1} - \sigma I) \psi_{n})}$ has a convergent subsequence in $\mathcal{H}$. This proves (ii).  \hfill { } $\blacksquare$\\

\begin{proposition}
\n Let $T_{1}$ and $T_{2}$ are two operators acting on Hilbert space  $\mathcal{H}$ with respective domain $D(T_{1})$ and $D(T_{2})$.\\

\n If $T_{1}$ and $T_{2}$ satisfy the following properties :\\

\n (i) $T_{2}$ is strictly dominated by $T_{1}$.\\

\n (ii) The resolvent of $T_{1}$ is compact.\\

\n Then\\

\n $T_{2}$ is $T_{1}$-compact.\\
\end{proposition}

\n {\bf{\color{red}Proof}}\\

\n Let $T_{2}$ strictly dominated by $T_{1}$.i.e  $D(T_{1}) \subset D(T_{2})$ and \\

$ \forall \epsilon > 0 \quad \exists C_{\epsilon} > 0 $ such that $ \mid\mid T_{2}\phi \mid\mid \,\, \leq \epsilon\mid\mid T_{1}\phi \mid\mid + \,C_{\epsilon}\mid\mid \phi \mid\mid \,\, \forall \phi \in D(T_{1}).$ \\

\n Then  for any $\psi \in \mathcal{H}$ , $\sigma \in \rho(T_{1})$ we have\\

\n $\displaystyle{ \mid\mid T_{2}(T_{1} - \sigma I)^{-1}\psi \mid\mid \,\, \leq \epsilon\mid\mid T_{1}(T_{1} - \sigma I)^{-1}\psi \mid\mid + \,C_{\epsilon}\mid\mid (T_{1} - \sigma I)^{-1}\psi \mid\mid}$\\

\n $\displaystyle{ \leq   \epsilon \mid\mid \psi \mid\mid + \,\tilde{C}_{\epsilon, \mid \sigma\mid}\mid\mid (T_{1} - \sigma I)^{-1}\psi \mid\mid}$\\

\n Let $\psi_{n} \in \mathcal{H}$ and $\vert\vert \psi_{n}\vert\vert \leq M$. Since $\displaystyle{T_{1} - \sigma I)^{-1}}$ is compact, then there exists a subsequence $\{\psi_{n_{k}}\}_{k=1}^{\infty}$ such that  $\displaystyle{T_{1} - \sigma I)^{-1}\psi_{n_{k}}}$  converges in $\mathcal{H}$. Then from\\

\n $\displaystyle{ \mid\mid T_{2}(T_{1} - \sigma I)^{-1}\psi_{n_{k}} - T_{2}(T_{1} - \sigma I)^{-1}\psi_{n_{m}}  \mid\mid \leq   \epsilon \mid\mid \psi_{n_{k}}  - \psi_{n_{m}} \mid\mid + }$\\

\n $\displaystyle{ \,\tilde{C}_{\epsilon, \mid \sigma\mid}\mid\mid (T_{1} - \sigma I)^{-1}\psi_{n_{k}} -  (T_{1} - \sigma I)^{-1}\psi_{n_{m}} \mid\mid}$\\

\n It follows that  $\,\forall \epsilon > 0$,\,, we can choose $N_{\epsilon}$ such that $\forall \,\, k > N_{\epsilon}$ and  $\forall \,\, m > N_{\epsilon}$ we have\\

\n $\displaystyle{ \mid\mid T_{2}(T_{1} - \sigma I)^{-1}\psi_{n_{k}} - T_{2}(T_{1} - \sigma I)^{-1}\psi_{n_{m}}  \mid\mid \leq  (2M C_{\epsilon,\mid \sigma\mid} + 1) \epsilon }$\\

\n So the subsequence $\displaystyle{(T_{2}(T_{1} - \sigma I)^{-1}\psi_{n_{k}})}$ converges in $\mathcal{H}$ and $\displaystyle{(T_{2}(T_{1} - \sigma I)^{-1}}$ is compact i.e. $T_{2}$ is $T_{1}$-compact. \hfill { } $\blacksquare$\\

\begin{definition}

\n  Let $T$ be an operator acting on a Hilbert space $\mathcal{H}$ with dense domain $D(T)$.\\

\n We say that :\\

\n $T$ is accretive if   $\Re e <T\phi, \phi > \,\, \geq 0 \,\, \forall \phi \in D(T)$.\\

\n $T$ is $\beta_{0}$-accretive if  $\exists \, \beta_{0} \in \mathbb{R}$ such that $ T + \beta_{0}I$ is accretive.\\

\n We will say that :\\

\n  $T$ is maximal $\beta_{0}$-accretive  if it does not exist an operator  $\beta_{0}$-accretive which strictly extends $T$.\\
\end{definition}

\begin{theorem}
 \n  Let $T$ be an operator acting on a Hilbert space $\mathcal{H}$ with dense domain $D(T)$ and $\beta_{0} \in \mathbb{R}$ then  the following properties are equivalent.\\

\n (i) $T$ is maximal $\beta_{0}$-accretive.\\

\n (ii) $T^{*}$ is maximal $\beta_{0}$-accretive.\\

\n (iii) $T$ is $\beta_{0}$-accretive and for all $\displaystyle{\beta \in \mathbb{C} ; \Re e\beta > \beta_{0}}$, the operator $ T + \beta I$ applies $D(T)$ on $\mathcal{H}$ and we have : \\
 
\n $\displaystyle{\mid\mid \phi \mid\mid \leq  \frac{1}{\Re e\beta - \beta_{0}}\mid\mid ( T + \beta I)\phi \mid\mid}$\\

\n (iv)There exist a strongly continuous bounded semigroup $(e^{-tT})_{t\geq 0}$ on $\mathcal{H}$ with $-T$ is its infinitesimal generator satisfy :\\

\n $\displaystyle{\mid\mid e^{-tT}\phi \mid\mid \leq e^{\beta_{0}t}\mid\mid \phi \mid\mid \quad \forall t \geq 0 , \phi \in\mathcal{H}}$. \\
\end{theorem}

\n {\color{red}{\bf Proof}}\\

\n This form of  Hille-Yosida theorem and the above definition are given by  Bardos appendix (C) of  {\color{blue}[6]}. \hfill { } $\blacksquare$\\

\begin{theorem} (see: theorem 111 p.194 of  book's Kato {\bf {\color{blue}[22]}})\\

\n  Let $T_{1}$ and $T_{2}$ two operators acting  Hilbert  space $\mathcal{H}$ with respective  $D(T_{1})$ et $D(T_{2})$ such that $T_{2}$ is $T_{1}$-compact.\\

\n Then\\

\n (i) If $T_{1}$ is closable,  $T = T_{1} + T_{2}$ is also closable.\\

\n (ii) The respective closures of $T_{1}$ and of $T$ have same domain  $D(T) = D(T_{1})$ \\

\n (iii) $T_{2}$ is $T$-compact.\\

\n (iv) $T$ is closable if $T_{1}$ is closable.\\
\end{theorem}

\n In following, we begin to recall some fundamental inequalities.\\

\begin{definition}

\n  Let $T_{1}$ and $T_{2}$ be linear operators in a Hilbert space $\mathcal{H}$. We say that $T_{2}$ is {\color{red}$p$-subordinate} $(p \in [0, 1])$  to $T_{1}$  if $D(T_{1}) \subset D(T_{2})$. and there exists a strictly positive constant $C$ such that \\
\begin{equation}
\displaystyle{\mid\mid T_{2} \phi \mid\mid \leq C\mid\mid T_{1} \phi \mid\mid^{p}\mid\mid  \phi \mid\mid^{1-p}} \,\, \text{ for every} \,\,\phi \in D(T_{1}).
\end{equation}
\n  For $p =1$, we say that $T_{2}$ is subordinate to $T_{1}$.\\
\end{definition}

\begin{definition}
\n  Let $T_{1}$ be a linear operator in a Hilbert space $\mathcal{H}$. The operator $T_{2}$ is said to be {\color{red}$T_{1}$-compact} with {\color{red}order $ p \in [0, 1]$} if $D(T_{1}) \subset D(T_{2})$ and for any $\epsilon > 0$, there exists a constant $C_{\epsilon} > 0$ such that \\
\begin{equation}
\displaystyle{\mid\mid T_{2} \phi \mid\mid \leq \epsilon \mid\mid T_{1} \phi \mid\mid^{p}\mid\mid  \phi \mid\mid^{1-p} + C_{\epsilon}\mid\mid \phi \mid\mid} \,\, \text{ for every} \,\, \phi \in D(T_{1}). 
\end{equation}
\n The operator $T_{2}$ is called {\color{red}$T_{1}$-compact}, if $T_{2}$ is $T_{1}$-compact with {\color{red}unit order}.\\
\end{definition}
 
\n According to results on comparison operators established in {\color{blue}[27]},  then from the above  definitions, we deduce the following classical results:\\

\begin{corollary}

\n Let $T_{1}$ be an operator in $\mathcal{H}$ with a {\color{red}dense domain} $D(T_{1})$. and at least one regular point $\sigma$.\\

\n We suppose that:\\

\n  {\color{red}($\alpha$)} $ D(T_{1}) \subset D(T_{2})$.\\

\n  {\color{red} ($\beta$)}  $\displaystyle{T_{2}(T_{1}- \sigma I)^{-1}}$ is a compact. operator\\

\n Then for any $\epsilon > 0$ , there exists a constant $C_{\epsilon} > 0$ such that \\ 

\n $\displaystyle{\mid\mid T_{2} \phi \mid\mid \leq \epsilon \mid\mid T_{1} \phi \mid\mid + C_{\epsilon} \mid\mid  \phi \mid\mid}$  for every $\phi \in D(T_{1})$\\ 
\end{corollary}

\begin{proposition}

\n Let $T_{1}$ be an operator in $\mathcal{H}$. We suppose there exists {\color{red}$p \in (0, 1]$} such that \\
\begin{equation}
\displaystyle{\vert\vert(T_{1} - \sigma I)^{-1} \vert\vert \leq \frac{C}{\mid \sigma \mid^{p}}} , \,\, \sigma \in \Omega \,\, \quad \vert \sigma \vert \longrightarrow \infty
\end{equation}
\n Where $\Omega$ is an unbounded set of the plane complex.\\

\n Let $T_{2}$ be an operator in $\mathcal{H}$ such that $D(T_{1}) \subset D(T_{2}) $ and\\
\begin{equation}
\forall \,\, \epsilon > 0,\,\, \,\, \displaystyle{ \vert\vert T_{2} \phi \vert\vert \leq \epsilon \vert\vert T_{1}\phi \vert\vert^{p} \vert\vert \phi \vert\vert^{1- p} + C_{\epsilon} \vert\vert \phi \vert\vert, \,\,\, \phi \in D(T_{1})}
\end{equation}

\n Then \\

\begin{equation}
\displaystyle{\vert\vert(T_{1} + T_{2}- \sigma I)^{-1} \vert\vert \leq \frac{C}{\mid \sigma \mid^{p}}} , \,\, \sigma \in \Omega \,\, \quad \vert \sigma \vert \longrightarrow \infty
\end{equation}
\end{proposition}

\n {\bf{\color{red} Proof}}\\

\n For $\sigma \in \rho(T_{1})$, we write \\

$$\displaystyle{T_{1}(T_{1} - \sigma I)^{-1} = (T_{1} - \sigma I + \sigma I)(T_{1} - \sigma I)^{-1} = I + \sigma (T_{1} - \sigma I)^{-1}}$$ 

\n then for $\sigma \in \Omega ,  \quad \vert \sigma \vert \longrightarrow \infty$ we have\\

\begin{equation}
\displaystyle{\vert\vert T_{1}(T_{1} - \sigma I)^{-1}\vert\vert \leq C \vert \sigma \vert^{1-p}}
\end{equation} 

\n For $\phi \in \mathcal{H}$ we apply (4.6) to the element $(T_{1} - \sigma I)^{-1}\phi$ to get:\\

\n $\displaystyle{ \vert\vert T_{2} (T_{1} - \sigma I)^{-1}\phi \vert\vert \leq \epsilon \vert\vert T_{1}(T_{1} - \sigma I)^{-1}\phi \vert\vert^{p} \vert\vert (T_{1} - \sigma I)^{-1}\phi \vert\vert^{1- p} + C_{\epsilon} \vert\vert (T_{1} - \sigma I)^{-1}\phi \vert\vert}$\\

\n $\displaystyle{\leq \frac{C \epsilon + C_{\epsilon}}{\vert \sigma \vert^{p}} \vert\vert \phi \vert\vert}$, \,\,  $\sigma \in \Omega \,\, \quad \vert \sigma \vert \longrightarrow \infty$\\

\n Hence, \\

\n $\displaystyle{ \vert\vert T_{2} (T_{1} - \sigma I)^{-1}\phi \vert\vert < 1}$  \,\,  $\sigma \in \Omega \,\, \quad \vert \sigma \vert \longrightarrow \infty$\\

\n Then, by Newman identity:\\

\begin{equation}
\displaystyle{(T_{1} + T_{2}- \sigma I)^{-1} = (T_{1}- \sigma I)^{-1} \sum_{k=0}^{\infty}[-T_{2}(T_{1}- \sigma I)^{-1} ]^{k}}
\end{equation}

\n We get:\\

\n $\displaystyle{ \vert\vert (T_{1}  + T_{2} - \sigma I)^{-1}\vert\vert \leq C  \vert\vert (T_{1} - \sigma I)^{-1}\vert\vert  \leq \frac{C}{\vert \sigma \vert^{p}}}$ \,\,  $\sigma \in \Omega \,\, \quad \vert \sigma \vert \longrightarrow \infty$. \hfill { } $\blacksquare$\\

\begin{corollary}
From (4.9), if $\displaystyle{(T_{1} - \sigma I )^{-1} \in \mathfrak{C}_{p}}$ then $\displaystyle{(T_{1} + T_{2} - \sigma I )^{-1} \in \mathfrak{C}_{p}}$
\end{corollary}

\begin{theorem} ({\color{blue}[23]})\\

\n Let $T_{1}$ be a normal operator with compact resolvent, and let\\
$$\displaystyle{T_{1}\phi = \sum_{n=1}^{\infty} \sigma_{n} < \phi, e_{n} > e_{n}}$$
\n be its spectral decomposition. For $p > 0$ let \\

$$\displaystyle{T_{1}^{p}\phi = \sum_{n=1}^{\infty} \sigma_{n}^{p} < \phi, e_{n} > e_{n}, \quad D(T_{1}^{p}) = \{\phi :  \sum_{n=1}^{\infty} \vert \sigma_{n}\vert^{2p} \vert < \phi, e_{n} > \vert^{2} < \infty \}}$$

\n where $\displaystyle{\sigma^{p} = \vert \sigma \vert^{p}exp(ip.\text{arg}\,\sigma)}$, and $\text{arg}\,\sigma \in (-\pi , \pi]$.\\

\n (i) If $T_{2}$ is subordinate to $T_{1}^{p}$ \, $(0 < p < 1)$, then it is $p$-subordinate to $T_{1}$.\\

\n (ii) If  $T_{2}$ is $p$-subordinate to $T_{1}$  \, $(0 < p < 1)$ and $\notin \sigma(T_{1})$ ($\sigma(T_{1})$ is the spectrum of $T_{1}$), then $T_{2}$ is subordinate to $T_{1}^{q}$  for  any $q > p$.\\
 
\n This implies  that if $ T_{2}$ is $p$-subordinate \, $(0 \leq p < 1)$ to  a nomal operator $T_{1}$ with compact resolvent, then $T_{2}$ is compact relative to $T_{1}$.\\
\end{theorem}

\n {\bf {\color{red} Proof}}\\

\n (i) is the theorem 12.2 of the reference {\color{blue}[23]} and (ii) is  the theorem 12.3 of the reference {\color{blue}[23]}. \hfill { } $\blacksquare$\\

  \subsection{ {\color{blue} Some spectral properties of the Hamiltonian \\ $\displaystyle{\mathbb{H} = A^{*^{k}}A^{k} +\sum_{i+ j < 2k}a_{i,j}A^{*^{i}}A^{j}}$ on Bargmann space}}
 
 \begin{proposition}

\n Let $k \in \mathbb{N}^{*}$ and $\displaystyle{P_{m}(A^{*}, A) = \sum_{i + j \leq m}a_{i,j}A^{*^{i}}A^{j},\,\, a_{i,j} \in \mathbb{C}; m \leq 2k - 1}$  with minimal domain:\\

$\displaystyle{D_{min} := D_{min}(P(A^{*} , A)) = \{\varphi \in \mathbb{B} \,\, \mbox{such that there exist} \,\, p_{n} \,\, \mbox {in} \,\, \mathcal{P} \,\, \mbox{(polynomials set) }}$\\
\begin{equation}
\n \mbox{and} \,\, \displaystyle{\psi \in \mathbb{B} \,\, \mbox{with}\,\,  p_{n} \longrightarrow \varphi \,\, \mbox{and} \,\, P(A^{*} , A)p_{n} \underset {n \longrightarrow + \infty}{ \longrightarrow} \psi \}}
\end{equation}
\n then we have:\\
\begin{equation}
\displaystyle{\forall \,\, \epsilon > 0, \exists C_{\epsilon} > 0 ; \mid <  \sum_{i + j \leq 2k}a_{i,j}A^{*^{i}}A^{j},\varphi, \varphi >\mid \leq \epsilon < A^{*k}A^{k}\varphi, \varphi > + C_{\epsilon}\mid\mid \varphi \mid\mid^{2}},
\end{equation}
\n for all $\displaystyle{\varphi \in D(A^{2k})}$.\\

\n (ii) Let $k \in \mathbb{N}^{*}$ and $\displaystyle{H = S_{k} + P_{m}(A^{*}, A) ; m \leq 2k -1}$ then \\

\n $\displaystyle{ \exists C > 0 ; \Re e < H\phi, \phi > \geq -C\mid \mid \phi \mid\mid^{2} \quad \forall \phi \in D_{min}(H)}$.\\

\n (iii)  $\displaystyle{ \exists \beta_{0} \in \mathbb{R}; H^{min} + \beta_{0}I}$ is invertible from $D_{min}(H)$ to $\mathbb{B}$.\\
\end{proposition}

\n {\bf{\color{red} Proof}}\\

\n (i) This property is the corollary  is 3.33 of the fundamental lemma.\\

\n \n (ii) is an immediate consequence of (i).\\

\n (iii) Let $\beta > 0$, we consider $\displaystyle{H = S_{k} + P_{m}(A^{*}, A) ; m \leq 2k -1}$ then  it follows that $\displaystyle{S_{k} +  P_{m}(A^{*}, A) + \beta I = [I +  P_{m}(A^{*}, A)(S_{k} + \beta I)^{-1}](S_{k} + \beta I)}$.\\

\n Now as $(S_{k} + \beta I)$ is invertible then so just show that $\displaystyle{\mid\mid  P_{m}(A^{*}, A)(S_{k} + \beta I)^{-1}\mid\mid < 1}$: \\

\n Let $\psi \in \mathbb{B}$, then we have :\\

\n $\forall \epsilon > 0, \exists C_{\epsilon} > 0$;  $\displaystyle{\mid\mid P_{m}(A^{*}, A)(S_{k} + \beta I)^{-1}\psi\mid\mid \leq \epsilon \mid\mid S_{k}(S_{k} + \beta I)^{-1}\psi\mid\mid  + }$ \\

\n $\displaystyle{C_{\epsilon}\mid\mid(S_{k} + \beta I)^{-1}\psi\mid\mid}$.\\

\n $\leq \epsilon \mid\mid (S_{k} + \beta I - \beta I)(S_{k} + \beta I)^{-1}\psi\mid\mid  + C_{\epsilon}\mid\mid(S_{k} + \beta I)^{-1}\psi\mid\mid$.\\

\n $\leq \epsilon \mid\mid \psi\mid\mid  + (\epsilon\beta + C_{\epsilon})\mid\mid(S_{k} + \beta I)^{-1}\psi\mid\mid$.\\

\n As $\mid\mid(S_{k} + \beta I)^{-1}\mid\mid \leq \frac{1}{\beta}$ then we deduce that \\

\n  $\mid\mid P_{m}(A^{*}, A)((S_{k} + \beta I)^{-1}\psi \mid\mid \leq (2\epsilon + \frac{C_{\epsilon}}{\beta})\mid\mid \psi \mid\mid$.\\

\n  in particular  for $ \epsilon < \frac{1}{2}$ and  $ \beta > \frac{C_{\epsilon}}{1-2\epsilon}$ we get\\

 \n $\mid\mid P_{m}(A^{*}, A)(S_{k} + \beta I)^{-1}\mid\mid < 1$ \hfill { } $\blacksquare$\\
 
 \begin{proposition} 
 
 [$P_{m}(A^{*}, A)$ is {\color{red}strictly dominated} by $A^{*^{k}}A^{k}$]\\
 
\n  (i) $\displaystyle{ \forall \,\, \epsilon > 0 , \,\, \exists \,\, C_{\epsilon} > 0;  \vert\vert P_{m}(A^{*}, A)\phi \vert\vert^{2} \leq  \epsilon \vert\vert S_{k}\phi \vert\vert^{2} + C_{\epsilon} \vert\vert \phi \vert\vert^{2}}$. where $\displaystyle{m \leq 2k-1}$ and $\displaystyle{S_{k} = A^{*^{k}}A^{k}}$\\

 \n (ii) If $ m \leq 2k-1 $, $k = 1, 2, ...$   then  for every $\beta > 0$, $\displaystyle{ P_{m}(A^{*}, A)(S_{k}  + \beta I)^{-1}}$ is compact.\\
 
 \n (iii) For $ m \leq 2k-1 $, $k = 1, 2, ...$, the resolvent of $\displaystyle{H = S_{k}  +  P_{m}(A^{*}, A)}$ is nuclear.\\
 \end{proposition}
 
 \n {\bf{\color{red} Proof}}
 
 \n (i) \n Let $\displaystyle{\phi(z) = \sum_{n=0}^{\infty}a_{n}\frac{z^{n}}{\sqrt{n!}}}$ then we have :\\

\n {\color{red}$\bullet_{1}$} $\displaystyle{\vert\vert \phi \vert\vert^{2} = \sum_{n=0}^{\infty} \vert a_{n} \vert^{2}}$\\

\n {\color{red}$\bullet_{2}$} $\displaystyle{\vert\vert S_{k}\phi \vert\vert^{2} = \vert\vert A^{*^{k}}A^{k}\phi \vert\vert^{2} = \sum_{n=0}^{\infty} n^{2k} \vert a_{n} \vert^{2}}$\\

\n {\color{red}$\bullet_{3}$} $\displaystyle{\exists \,\, C > 0; \vert\vert P_{m}(A^{*}, A)\phi \vert\vert^{2} \leq C \sum_{n=0}^{\infty} n^{m} \vert a_{n} \vert^{2}}$ where $m = 2k-1$\\

\n Applying the property (ii) :\\

\n  $\displaystyle{\forall \epsilon > 0, \exists C_{\epsilon} > 0 ;  ab \leq \epsilon a^{p} + C_{\epsilon}b^{q} \quad \forall a \geq 0, \quad \forall b \geq 0.}$\\

\n of classical lemma to couple $(a , b) = (n^{m}, 1)$ to  obtain :\\

\n  $\displaystyle{\forall \epsilon > 0, \exists C_{\epsilon} > 0 ;  n^{m} \leq \epsilon n^{mp} + C_{\epsilon} .}$\\

\n i.e.\\

\n $\displaystyle{\forall \epsilon > 0, \exists C_{\epsilon} > 0 ;  n^{2k-1} \leq \epsilon n^{(2k -1)p} + C_{\epsilon} .}$\\
 
 \n If we choose $p$ ; such that $(2k -1)p = 2k$ it  follows that $p = \frac{2k -1}{2k}$ and $q = \frac{1}{2k}$ satisfy $\displaystyle{\frac{1}{p} + \frac{1}{q} = 1}$ and $\displaystyle{n^{2k-1} \leq \epsilon n^{2k} + C_{\epsilon}}$. In particular :\\
 
 \n $\displaystyle{n^{2k-1}\vert a_{n} \vert^{2} \leq \epsilon n^{2k} \vert a_{n} \vert^{2} + C_{\epsilon}\vert a_{n}\vert^{2}}$
  
  \n and\\
  
\n  $\displaystyle{\vert\vert P_{m}(A^{*}, A)\phi \vert\vert^{2} \leq  \epsilon \vert\vert S_{k}\phi \vert\vert^{2} + C_{\epsilon} \vert\vert \phi \vert\vert^{2}}$.\\

\n (ii) As $P_{m}(A^{*}, A)$ is {\color{red}strictly dominated} by $A^{*^{k}}A^{k}$ then from proposition 4.7, we deduce that  $P_{m}(A^{*}, A)$ is $A^{*^{k}}A^{k}$-compact and from proposition 4.6 it follows that for every $\beta > 0$ $P_{m}(A^{*}, A)$$(A^{*^{k}}A^{k} + \beta I)^{-1}$ is compact. \\

\n (iii) by Newman identity, we have:\\
\begin{equation}
\displaystyle{(H + \sigma I)^{-1} = (S_{k} + \sigma I)^{-1} \sum_{j=0}^{\infty}[- P_{m} (A^{*}, A)(S_{k} +\sigma I)^{-1} ]^{j}}
\end{equation}\\

\n Now, as  $\displaystyle{(S_{k} + \sigma I)^{-1} \in \mathfrak{C}_{p}}$ with $p > \frac{1}{k}$, it is nuclear if $k \geq 2$. It follows that  the resolvent of $\displaystyle{H = S_{k}  +  P_{m}(A^{*}, A)}$ is nuclear for $k \geq 2$ and if $k = 1$ the resolvent of $\displaystyle{H = S_{1}  +  P_{m}(A^{*}, A)}$ belongs  to $\displaystyle{\mathfrak{C}_{1 + \epsilon}, \, \forall \,\, \epsilon > 0}$ .\\

\begin{lemma}
\n Let $k \in \mathbb{N}$ then  $P_{k}(A^{*}, A)$ is $\frac{1}{2}$-subordinate to $S_{k}$. i.e. there exists a strictly positive constant $C$ such that \\
\begin{equation}
\displaystyle{\vert\vert P_{k}(A^{*}, A) \phi \vert\vert \leq C \vert\vert S_{k} \phi \vert\vert^{\frac{1}{2}}. \vert\vert \phi \vert\vert^{\frac{1}{2}} \,\,  \text{for every} \,\, \phi \in D(S_{k})}
\end{equation}
\end{lemma}

\n {\bf{\color{red} Proof}}\\

\n Let $\mathcal{N} = A^{*}A + I$ and $B_{k} = A^{*^{k}}A^{k}$, then we have \\

\n  $\displaystyle{D(\mathcal{N}^{\frac{k}{2}}) \subset D(P_{k}(A^{*}, A))}$ and $\displaystyle{\vert\vert \mathcal{N}^{k} \phi \vert\vert \sim \vert\vert B_{k} \phi \vert\vert}$  \\

\n As  $\displaystyle{D(\mathcal{N}^{\frac{k}{2}}) \subset D(P_{k}(A^{*}, A))}$, we deduce that  \\

\n $\displaystyle{\exists \, C > 0 ; \vert\vert P_{k}(A^{*}, A))\phi \vert\vert  \leq C \vert\vert \mathcal{N}^{\frac{k}{2}} \phi \vert\vert}$ for every $\phi \in D(\mathcal{N}^{\frac{k}{2}})$.\\

\n Now, as $\displaystyle{ \vert\vert \mathcal{N}^{\frac{k}{2}} \phi \vert\vert^{2} =  < \mathcal{N}^{\frac{k}{2}}\phi , \mathcal{N}^{\frac{k}{2}}\phi > =   < \mathcal{N}^{k}\phi ,\phi >  \leq \vert\vert\mathcal{N}^{k}\phi\vert\vert. \vert\vert \phi \vert\vert}$. It follows that \\

\n $\displaystyle{\vert\vert P_{k}(A^{*}, A) \phi \vert\vert \leq C \vert\vert S_{k} \phi \vert\vert^{\frac{1}{2}}. \vert\vert \phi \vert\vert^{\frac{1}{2}} \,\,  \text{for every} \,\, \phi \in D(S_{k})}$. \hfill { } $\blacksquare$\\

$\displaystyle{\mathbb{H} = A^{*^{k}}A^{k} +\sum_{i+ j < 2k}a_{i,j}A^{*^{i}}A^{j}}$

 \section{ {\color{red}On  regularized trace formula of  order one of the Hamiltonian $\displaystyle{\mathbb{H} = A^{*^{k}}A^{k} +\sum_{i+ j < 2k}a_{i,j}A^{*^{i}}A^{j}}$ on Bargmann space  }}

  \subsection{{\color{blue}Results of Sadovnichii and Podolski on regularized traces of abstract discrete operators}}
  
\n  As it is known, the trace of a finite-dimensional matrix is the sum of all the eigenvalues. But in an infinite dimensional space, in general, ordinary differential operators do not have a finite trace.\\

\n  In 1953, Gelfand and Levitan considered the Sturm-Liouville operator\\

\n $\left\{\begin{array}[c]{l}-y''(x) + q(x)y(x) = \sigma y(x)\\ 
\quad \\ 
y'(0) = 0, y'(\pi) = 0\\
\quad\\

q(x) \in \mathcal{C}^{1} [0, \pi],\quad \displaystyle{\int_{0}^{\pi}q(x)dx} = 0 \\
 \end{array}\right.\hfill { }  (*)$\\

and derived the formula\\

\n $\displaystyle{\sum_{n=1}^{\infty}(\sigma_{n} -\lambda_{n}) = \frac{1}{4}(q(0) + q(\pi))}$ $\hfill { }  (**)$\\

\n where $\sigma_{n}$ are the eigenvalues of the above operator and $\lambda_{n} = n^{2}$ are the eigenvalues of the same operator with $q (x) = 0$.\\

\n The proof of this regularized trace formula for the Sturm-Liouville operator can been found in {\bf{\color{blue}[14]}}\\

\n The same regularized trace formula for the same problem was obtained with different method by Dikii {\bf{\color{blue}[9] }}.\\

\n  For the scalar Sturm-Liouville problems, there is an enormous literature on estimates of large eigenvalues and regularized trace formulae which may often be computed explicitly in terms of the coefficients of operators and boundary conditions.\\

\n After these studies, several mathematicians were interested in developing regularized trace formulae for different differential operators. .\\

\n According Sadovnichii and  Podolskii, these formulae gave rise to a large and very important theory, which started from the investigation of specific operators and further embraced the analysis of regularized traces of discrete operators in general form.\\

\n Among the results of Sadovnichii and Podolskii {\color{blue}[31]} established for abstract operators, we can recall that following :\\

\n Let $A_{0}$ be a self-adjoint positive discrete operator of domain $D(A_{0})$ acting in a Hilbert space, we denote by $\{\lambda_{n}\}$ its eigenvalues arranged in ascending order, $\{\phi_{n}\}$ is an orthonormal basis formed by the eigenvectors of $A_{0}$ and $R_{0}(\sigma)$ is its resolvent . By $B$ we denote the perturbing operator, by $\{\sigma_{n}\}$ we denote the eigenvalues of the operator $A_{0} + B$ numbered in increasing order of their real parts, and $R(\sigma)$ stands for its resolvent.\\

\n Also, assume that $BA_{0}^{-\delta}$, $\delta > 0$ is a compact operator belonging to some finite-order Schatten von Neumann class, i.e. the set of compact operators whose singular numbers form a convergent series $\displaystyle{\sum_{n=1}^{\infty}s_{n}^{p}}$ for some $p > 0$ is traditionally denoted by $\mathcal{C}_{p}$.\\

\n For operators $A_{0}$ and $B$ in {\color{blue} [31]}, the following theorem is proved.\\

\begin{theorem} (Sadovnichii-Podol'skii {\bf{\color{blue} [31]}})\\

\n Consider operator $A_{0}$ and $B$ be such that $A_{0}^{-1} \in C_{1}$ and $D(A_{0}) \subset D(B)$, and  suppose that there exist $\delta \in [0, 1)$ such that the operator $BA_{0}^{-\delta}$ can be continued to a bounded operator. Further, suppose that there exists $\omega \in [0, 1)$ such that $ \omega + \delta < 1$ and  $A_{0}^{-(1-\delta -\omega)}$ is a trace class operator, i.e. in $\mathcal{C}_{1}$. Then, there exist a subsequence of natural numbers $\{n_{m}\}_{m=1}^{\infty}$ and a subsequence of contours $\Gamma_{m} \subset \mathbb{C}$, that  for $\omega \geq  \frac{\delta}{l}$ the following relation holds:\\
\begin{equation}
\lim\limits_{m \longrightarrow \infty} (\displaystyle{\sum_{j=1}^{n_{m}}(\sigma_{j} - \lambda_{j}) +\frac{1}{2\pi i}\int_{\Gamma_{m}}\sum_{k=1}^{l}\frac{(-1)^{k-1}}{k} Tr((BR_{0}(\sigma))^{k}d\sigma) = 0}
\end{equation}

\n In particular, for $l = 1$ we have \\
\begin{equation}
\lim\limits_{m \longrightarrow \infty} \displaystyle{\sum_{j=1}^{n_{m}}(\sigma_{j} - \lambda_{j} -  <B\phi_{j}, \phi_{j}>) = 0}
\end{equation}
\end{theorem}

\begin{remark}

\n  1) This theorem has been successfully applied to concrete ordinary differential operators as well as to partial differential operators, we can see some examples given in {\bf{\color{blue} [32}} {\bf{\color{blue} [33}}).\\

\n 2) We can found in {\bf{\color{blue} [34]}} or in {\bf{\color{blue} [35]}}, an excellent survey dedicated by Sadovnichii and Podolskii to the history of the state of the art in the theory of regularized traces of linear differential operators with discrete spectrum and a detailed list of publications related to the present aspect.\\

\n 3)  By applying above theorem of  Sadovnichii-Podolskii, we have given in { \color{blue} [21]} the number of corrections sufficient for the existence of finite formula of the trace of  concrete magic Gribov's operator:\\
\begin{equation}
H_{\lambda'',\lambda',\mu,\lambda} = \lambda{''}A^{*3}A^{3}+\lambda{'}A^{*2}A^{2} + \mu A^{*}A +i\lambda A^{*}(A + A^{*})A
\end{equation}
\n where  in Reggeon field theory, the real parameters $\lambda{''}$ is the magic coupling of Pomeron, $\mu$  is Pomeron intercept, $\lambda$ is the triple coupling of Pomeron and $i^{2} = -1$.
\end{remark}
\n We have obtained the following theorem: \\

\n \begin{theorem} {\color{red} ({\color{blue}[21]}  2016)}\\

\n  Let $\mathbb{B} $ be the Bargmann space, $ H = \lambda{''}G + H_{\mu,\lambda}$ acting on $\mathbb{B}$ where $ G = A^{*3}A^{3}$ and $ H_{\mu,\lambda} = \mu A^{*}A + i\lambda A^{*}( A + A^{*})A$ with $A$ and $A^{*}$ are the standard Bose annihilation and creation operators.\\

\n Then there exists an increasing sequence of radius $r_{m}$ such that $r_{m} \rightarrow \infty$ as $ m \rightarrow \infty$  and\\
\begin{equation}
\displaystyle{\lim\limits_{m \longrightarrow \infty} (\sum_{n=0}^{m}(\sigma_{n} - \lambda{''}\lambda_{n}) + \frac{1}{2i\pi}\int_{\gamma_{m}} Tr[\sum_{k=1}^{4}\frac{(-1)^{k-1}}{k}[H_{\mu,\lambda}(\lambda{''}G - \sigma I)^{-1}]^{k}]d\sigma) = 0}
\end{equation} 
\n  Where\\

 \n - $\sigma_{n}$ are the eigenvalues of the operator $ H = \lambda{''}G + H_{\mu,\lambda}$\\
 
\n  - $\lambda_{n} = n(n-1)(n-2)$ are the eigenvalues of the operator $ G $\\

 \n - $ (\lambda{''}G - \sigma I)^{-1}$ is the resolvent of the operator $\lambda{''}G$\\
 
\n  and\\
 
\n  - $\gamma_{m}$ is the circle of radius $r_{m}$ centered at zero in complex plane.\\
\end{theorem}
  
 \subsection{{\color{blue} The regularized trace formula for the Hamiltonian \\ $\displaystyle{\mathbb{H} = A^{*^{k}}A^{k} +\sum_{i+ j  < 2k}a_{i,j}A^{*^{i}}A^{j}}$ on Bargmann space}}
 
 \begin{proposition} 

(i) For $m \leq 2k - 3$ , th resolvent of $\displaystyle{ P_{k}(A^{*}, A)(S_{k} + \sigma I)^{-1}}$ is nuclear.\\

 (ii) \n Let $\displaystyle{\phi = \sum_{n=0}^{\infty} < \phi, e_{n} >e_{n}}$ belonging to the domain $\displaystyle{D(S_{k})}$ where  $\displaystyle{S_{k} = A^{*^{k}}A^{k}}$, $k = 1, 2, ...$ and $\displaystyle{e_{n} = \frac{z^{n}}{\sqrt{n!}}}$. Then \\

\n $\displaystyle{\exists M > 0; \vert\vert P_{m}(A^{*}, A) \phi \vert\vert \leq M \vert\vert (S_{k} + I)^{p}\phi \vert\vert}$ where $m = 2k - 1$ and $\displaystyle{p = \frac{m}{2k} = 1 - \frac{1}{2k}}$.\\
 
\end{proposition}

\n {\bf{\color{red}Proof}}\\

\n {\color{red}$\bullet_{1}$} For $k \in \mathbb{N}$, $S_{k}$ is a discrete self-adjoint non-negative in Bargmann space $\mathbb{B}$ operator with eigenvalues $\displaystyle{\lambda_{n,k} = n(n-1) ......... (n -(k-1))}$ and corresponding eigenfunctions $\displaystyle{e_{n}(z) = \frac{z^{n}}{\sqrt{n!}}, n \in \mathbb{N}}$.\\

\n  {\color{red}$\bullet_{2}$} $\displaystyle{\lim\limits_{n \longrightarrow +\infty}\frac{\lambda_{n,k}}{n^{k}} = 1}$  $\lambda_{n,k} \sim n^{k}$  ($n \geq k$).\\

\n  {\color{red}$\bullet_{3}$} Let $\displaystyle{\Delta \lambda_{n,k} = \lambda_{n+1, k} - \lambda_{n,k}}$ ($n \geq k$) then from lemma 3.37, we deduce that\\

\n  $\displaystyle{ \Delta \lambda_{n,k} = k \lambda_{n, k-1} \sim n^{k-1} \sim (n +1)^{k-1} = [(n+1)^{k}]^{1 - \frac{1}{k}} \sim \lambda_{n + 1, k}^{q} }$ where \\ $\displaystyle{q = 1 - \frac{1}{k} \in (0, 1)}$ if and only if {\color{red}$k >1$}.\\

\n  {\color{red}$\bullet_{4}$} For all non-negative integers $i, j, n$, we recall that:\\

\n  {\color{red}$\star_{1}$} $\displaystyle{A^{*^{i}} e_{n} = \sqrt{\lambda_{n+i, i}}\, e_{n {\color{red}+ i}} }$ ($n, i \in \mathbb{N}$).\\

\n  {\color{red}$\star_{2}$} $\displaystyle{A^{j} e_{n} = \sqrt{\lambda_{n, j}}\, e_{n {\color{red}- j}}}$ for ($0 \leq j \leq n$). and $ A^{j}e_{n} = 0$ if ($0 \leq n < j$).\\

\n  {\color{red}$\star_{3}$} $\displaystyle{A^{*^{i}}A^{j} e_{n} = \alpha_{n,i,j}e_{n+i-j}}$ if $i \in \mathbb{N}$ and ($0 \leq j \leq n$) and $\displaystyle{A^{*^{i}}A^{j} e_{n} = 0}$ if $0 \leq n < j$\\

\n where $\displaystyle{ \alpha_{n,i,j} = \sqrt{\lambda_{n+i-j, i}\, \lambda_{n,j}}}$, $ i \in  \mathbb{N}$ and $0 \leq j \leq n$.\\

\n  {\color{red}$\rhd_{1}$} $\displaystyle{ \alpha_{n,i,j} = \sqrt{\lambda_{n+i-j, i}\, \lambda_{n,j}} \sim \sqrt{(n + i +j)^{i}n^{j}} \sim n^{\frac{i + j}{2} }}$, $ i \in  \mathbb{N}$ and $0 \leq j \leq n$.\\

\n  {\color{red}$\rhd_{2}$} The elements of matrix of $\displaystyle{P_{m}(A^{*}, A)}$ are $\displaystyle{O(n^{\frac{m}{2}})}$ and  the elements of matrix of $\displaystyle{(S_{k} + \sigma I)^{-1}}$ are $\displaystyle{O(n^{-k})}$.\\

\n  {\color{red}$\rhd_{3}$} The serie $\displaystyle{\sum_{n=1}^{\infty} \frac{n^{\frac{m}{2}}}{n^{k}} }$ $\displaystyle{= \sum_{n=1}^{\infty} \frac{1}{n^{k -\frac{m}{2}}}}$ is convergent if \,\, $\displaystyle{k -\frac{m}{2} > 1}$ i.e. if \,\,{\color{red} $\displaystyle{k > \frac{m}{2} + 1}$}.  If  {\color{red}$m \leq 2k - 3$}, this last inequality is verified and it follows that :\\

\n $\displaystyle{ P_{m}(A^{*}, A) (S_{k} + \sigma I)^{-1}}$ is nuclear.\\

\n As $\displaystyle{\alpha_{n,i,j} \sim n^{\frac{i+j}{2}}}$ then $\displaystyle{\exists \,\, \beta_{i,j} > 0, \gamma_{i,j} > 0}; \,\, \gamma_{i,i} n^{\frac{i+j}{2}} \leq \alpha_{i,j} \leq  \beta_{i,j} n^{\frac{i+j}{2}}$\\

\n In particular, there exist the positive constants $ \beta_{i,j}$ such that \\

\begin{equation}
 \displaystyle{ \alpha_{n,i,j} \leq \beta_{i,j} n^{\frac{i+j}{2}}}, (j \leq n)
 \end{equation}
 \n It follows that\\
 \begin{equation}
 \displaystyle{A^{*^{i}}A^{j}\phi = \sum_{n=0}^{\infty} < \phi, e_{n} >e_{n}  =  \sum_{n=0}^{\infty} \alpha_{n,i,j}< \phi, e_{n} >e_{n {\color{red}+ i - j}} }
 \end{equation}
\n and\\
 \begin{equation}
 \displaystyle{\vert\vert A^{*^{i}}A^{j}\phi\vert\vert^{2} =  \sum_{n=0}^{\infty} \vert \alpha_{n,i,j}\vert^{2}\, \vert < \phi, e_{n} >\vert^{2} \leq \beta_{i,j}^{2}  \sum_{n=0}^{\infty} (n + 1)^{i+j} \vert < \phi, e_{n} > \vert^{2}}
 \end{equation}
 
 \n Now, as $\displaystyle{P_{m}(A^{*}, A) \phi = \sum_{i+j \leq m} c_{i,j}A^{*^{i}}A^{j} \phi}$ then \\
 
 \n  $\displaystyle{\vert\vert P_{m}(A^{*}, A) \phi \vert\vert \leq \sum_{i+j \leq m} \vert c_{i,j}\vert.\vert\vert A^{*^{i}}A^{j} \phi\vert\vert \leq \sum_{i+j \leq m}^{\infty}\vert c_{i,j}\vert \beta_{i,j} \sqrt{\sum_{n=j}^{\infty} (n +1)^{i+j} \vert < \phi , e_{n} > \vert^{2} }}$
 
 \n In particular\\
 \begin{equation}
 \displaystyle{\vert\vert P_{m}(A^{*}, A) \phi \vert\vert \leq M \sqrt{\sum_{n=0}^{\infty} (n +1)^{m} \vert < \phi , e_{n} > \vert^{2} }}\,; \,\, \displaystyle{M =  \sum_{i+j \leq m}^{\infty}\vert c_{i,j} \vert \beta_{i,j}}
\end{equation}

\n In the following we write this equation as follows:\\

\n  $\displaystyle{\vert\vert P_{m}(A^{*}, A) \phi \vert\vert \leq M \sqrt{\sum_{n=0}^{\infty} [(n +1)^{k}]^{2\frac{m}{2k}}\vert < \phi , e_{n} > \vert^{2} }}\,; \,\, \displaystyle{M =  \sum_{i+j \leq m}^{\infty}\vert c_{i,j} \vert \beta_{i,j}}$\\

 \n or \\ 

\n $\displaystyle{\vert\vert P_{m}(A^{*}, A) \phi \vert\vert \leq M \sqrt{\sum_{n=0}^{\infty} [(n +1)^{k}]^{2\delta}\vert < \phi , e_{n} > \vert^{2} }}\,; \,\, \displaystyle{M =  \sum_{i+j \leq m}^{\infty}\vert c_{i,j} \vert \beta_{i,j}}$ \\

\n  where {\color{red}$\displaystyle{ \delta = \frac{m}{2k}}$}\\
 
\n AS $\displaystyle{(n +1)^{k}] \sim \lambda_{n +1,k}}$ we deduce that \\

\n $\displaystyle{\sqrt{\sum_{n=0}^{\infty} [(n +1)^{k}]^{2\delta}\vert < \phi , e_{n} > \vert^{2} } = \sqrt{\sum_{n=0}^{\infty} \lambda_{n +1}^{2\delta}\vert < \phi , e_{n} > \vert^{2} } = \vert\vert (S_{k} + I)^{\delta}\phi\vert\vert}$\\

\n  where {\color{red}$\displaystyle{ \delta = \frac{m}{2k}}$} and  {\color{red}$\displaystyle{m = 2k - 3}$}\\

\n In particular we have\\
 \begin{equation}
  \displaystyle{\vert\vert P_{m}(A^{*}, A) \phi \vert\vert \leq M  \vert\vert (S_{k} + I)^{\delta}\phi \vert\vert}
\end{equation} \hfill { } $\blacksquare$\\

\begin{lemma} ({\color{red}checking the assymptions of Sadovnichii -Podolski's theorem})\\

\n (i) Let $\displaystyle{ S_{k} = A^{*^{k}}A^{k}}$ and its eigenvalues are $\lambda_{n,k} \sim n^{k}$, then we have $\displaystyle{ S_{k}^{-1} \in \mathfrak{C}_{1}}$  If $ k \geq 2$.\\

\n (ii) There exist $\delta = \frac{m}{2k} \in [0, 1)$ where $m = 2k - 3$ such that the operator $\displaystyle{P_{m}(A^{*},A)(S_{k} + I)^{-\delta}}$  can be continued  to a bounded operator.\\

\n (iii) For $\displaystyle{\delta = \frac{m}{2k} = 1 - \frac{3}{2k}}$ and $\displaystyle{\omega < \frac{1}{2k}}$, we have $\displaystyle{S_{k}^{-(1- \delta - \omega)} \in \mathfrak{C}_{1}}$ with $ \omega \in [0 , 1)$ and $\omega + \delta < 1$.\\
\end{lemma}
\n {\bf{\color{red} Proof}}\\

\n (i) $\displaystyle{\sum_{n=1}^{\infty} \frac{1}{\lambda_{n,k}} = \sum_{n=1}^{\infty} \frac{1}{n^{k} } < \infty}$ if {\color{red} $k > 1$}.\\

\n (ii)  From inequality (4.22) :\\

\n $\displaystyle{\exists \, \delta = \frac{m}{2k} = 1 - \frac{3}{2k} \,\, \text{we have}\,\, \vert\vert P_{m}(A^{*},A) \phi \vert\vert \leq M \vert\vert (S_{k} + I)^{\delta}\phi }$ with $m = 2k - 3$ and $k \geq 2$\\

\n we deduce that\\

\n $\displaystyle{\vert\vert P_{m}(A^{*},A)(S_{k} + I)^{-\delta} \phi \vert\vert \leq M \vert\vert \phi \vert\vert }$ with $m = 2k - 3$ and $k \geq 2$.\\

\n (iii) $\displaystyle{\sum_{n=1}^{\infty} \frac{1}{\lambda_{n,k}^{1 - \delta - \omega}} = \sum_{n=1}^{\infty} \frac{1}{n^{k(1 - \delta - \omega)}} < \infty}$\,\,  if\,\, $\displaystyle{ {\color{red}k(1 - \delta - \omega) > 1} \iff  k(1 - \frac{m}{2k} - \omega) > 1 }$\\ 
 
 \n $\displaystyle{\iff k - \frac{m}{2} - k \omega > 1 \iff k - \frac{2k - 3}{2} - k \omega > 1 \iff  k - k + \frac{3}{2} - k\omega  > 1 \iff }$\\
 
   \n $\displaystyle{{\color{red}\omega < \frac{1}{2k}}}$\\
   
   \n and it easy to observe that $\displaystyle{{\color{red}\delta + \omega} < 1 - \frac{3}{2k} + \frac{1}{2k} = 1 - \frac{1}{k} {\color{red}< 1}}$\hfill { } $\blacksquare$\\

\begin{remark}
\quad\\

\n (i) To show there is a subsequence of natural numbers $\displaystyle{\{n_{s}\}_{s=1}^{\infty}}$ and a subsequence of contours $\displaystyle{\Gamma_{s} \subset \mathbb{C}}$ and to apply  Sadovnichii -Podolski's theorem recalled above to our operator, we choose  $\displaystyle{l \geq \frac{\delta}{\omega} > \frac{m}{2k}2k = m = 2k - 3}$ ( ${\color{red} l \geq 2(k -1)}$) .\\

\n(ii)  For $k = 3$ as in {\color{blue}[21]}, we obtain $\displaystyle{\delta = \frac{m}{2k} = \frac{2k-3}{2k} = \frac{3}{3} = \frac{1}{2}}$, $\displaystyle{\omega < \frac{1}{6}}$ and  ${\color{red} l \geq 2(k -1) = 4}$.\\
\end{remark}
\n Of above lemma and remark, the following theorem holds:\\

\n \begin{theorem} ({\color{green}$\star\star\star\star\star$})\\

\n  Let $\mathbb{B}$ be the Bargmann space, $\displaystyle{ \mathbb{H} = S_{k} + P_{m}(A, A^{*})}$ acting on $\displaystyle{\mathbb{B}}$ where $\displaystyle{S_{k} = A^{*^{k}}A^{k}}$ and $\displaystyle{ P_{m}(A, A^{*}) = \sum_{i+j \leq m}a_{i,j}A^{*^{i}}A^{j}}$ with $A$ and $A^{*}$ are the standard Bose annihilation and creation operators and $a_{i,j} \in \mathbb{C}$ where $m \leq 2k - 3$.\\

\n Then there exists an increasing sequence of radius $r_{m}$ such that $r_{s} \rightarrow \infty$ as $ s \rightarrow \infty$  and\\
\begin{equation}
\displaystyle{\lim\limits_{s \longrightarrow \infty} (\sum_{n=0}^{s}(\sigma_{n} - \lambda_{n,k}) + \frac{1}{2i\pi}\int_{\gamma_{s}} Tr[\sum_{t=1}^{l}\frac{(-1)^{t-1}}{t}[ P_{m}(A, A^{*}) (S_{k} - \sigma I)^{-1}]^{t}]d\sigma) = 0}
\end{equation} 
\n  Where\\

 \n - $\sigma_{n}$ are the eigenvalues of the operator$ \mathbb{H} = S_{k} + P_{m}(A, A^{*})$\\
 
\n  - $\displaystyle{\lambda_{n,k} = n(n-1)(n-2) ......(n-(k-1))}$ are the eigenvalues of the operator $ \displaystyle{S_{k} = A^{*^{k}}A^{k}} $\\

 \n - $ (S_{k} - \sigma I)^{-1}$ is the resolvent of the operator $S_{k}$\\
 
\n  and\\
 
\n  - $\gamma_{s}$ is the circle of radius $r_{s}$ centered at zero in complex plane.\\
\end{theorem}
  
\begin{remark}
\quad\\

\n It is useful to resume all the techniques of the proof of Sadovnichii-Podolskii's \\

\n theorem to our operator as it was done in {\color{blue}[21]} with $k = 3$ for the magic operator\\

\n  of Gribov.\\
\end{remark}

\n {\bf BIBLIOGRAPHIE}\\

 \n {\bf {\color{blue}[1]}} Aftalion, A., Blanc,  X. and  Nier, F., Lowest Landau level functional and Bargmann spaces for Bose-Einstein condensates, J. Funct. Anal. 241 (2006), 661-702.\\
 
 \n {\bf {\color{blue}[2]}} Agarwal, R. P., Difference Equations and Inequalities, Marcel Dekker, 2000.\\

 \n {\bf {\color{blue}[3]}} Alpay, D., Colombo, F., Diki, K. and Sabadini, I., Reproducing kernel Hilbert spaces of poly-analytic functions of infinite order, arXiv:2112.14367v1 [math.CV] 29 Dec 2021.\\
 
 \n  {\bf {\color{blue}[4]}}  Bargmann, V., ,On a Hilbert space of analytic functions and an associated integral transform I, Comm. Pure Appl. Math. 14 (1961) 187-214. \\

\n {\bf {\color{blue}[5]}}  Berezanskii, Yu. M.: Expansion in eigenfunctions of selfadjoint operators. Providence, RI: Am. Math Soc., 1968.\\

\n {\bf {\color{blue}[6]}}  Courrege, Ph.\,et Renourd, P. : Oscillateur anharmonique processus de diffusion et mesures quasi-invariantes, Socit Mathmatiques de France, Astrique, 22-23 (1975).\\

\n {\bf {\color{blue}[7]}} Daubechies, I. : A time frequency localization operator: A geometric phase space approach, IEEE. Trans. Inform. theory. vol. 34, p. 605-612 (1988)\\

\n {\bf {\color{blue}[8]}} Daubechies, I.:  Ten Lectures on Wavelets, Rutgers University and $AT \& T$ Bell Laboratories(1992)\\

\n\n {\bf {\color{blue}[9]}} Dikii, L. A.  About a formula of Gelfand-Levitan, Uspekhi Mat. Nauk 8(2) (1953) 119-123.\\

\n {\bf {\color{blue}[10]}} Dikii, L. A.  New method of computing approximate eigenvalues of the Sturm-Liouville problem, Dokl. Akad. Nauk SSSR 116 (1957) 12-14.\\

 \n\n {\bf {\color{blue}[11]}}   Dunford, N. and Schwartz, J.T., Linear Operators, Vol. II. John Wiley and Sons, (1988)\\
 
 \n {\bf {\color{blue}[12]}}  Fischer C.F and  Usmani, R.,  Properties of some tridiagonal matrices and their application to boundary value problems, SIAM J. Numer. Anal. 6 (1969) 127-142.\\
 
 \n {\bf {\color{blue}[13]}} Folland, G.B., Harmonic Analysis in Phase Space,122 (Princeton Univer- sity Press, Princeton, NJ, 1989).\\
  
 \n \n {\bf {\color{blue}[14]}}  Gelfand, I. M. and Levitan, B. M., On a simple identity for the characteristic values of a differential operator of second order, Dokl. Akad. Nauk SSSR 88 (1953) 593-596.\\
 
\n  {\bf {\color{blue}[15]}}  G\'erard, P., G\'ermain, P. and Thomann, L.,  On the cubic Lowest Landau Level equation, Arch. Rat. Mech. Appl. 231 (2019) 1073-1128.\\

\n \n {\bf {\color{blue}[16]}} Gohberg, I. G. and  Krein, M.G.,  Introduction to the Theory of Linear Non Selfadjoint Operators, Transl. Math. Monogr., vol. 18 , Amer. Math. Soc., Providence, RI, 1969.\\

\n  {\bf {\color{blue}[17]}} Golub, G.,  Bounds for eigenvalues of tridiagonal symmetric matrices computed by the LR method, Mathematics of Computation 16 (1962), p. 428- 447.\\

\n \n {\bf {\color{blue}[18]}} Grochenig, K., Foundations of Time-Frequency Analysis, Birkhuser-Verlag, Basel, Berlin, Boston (2000)\\

\n  {\bf {\color{blue}[19]}}  Gribov, V. N.,Technique du diagramme de Regge, ZhETF. - 1967. - 53. - S. 654-672.\\

\n  {\bf {\color{blue}[20]}} Hall, B. C. :.  Holomorphic methods in analysis and mathematical physics, arXiv:quant-ph/9912054v2 14 Sep 2000.\\

\n {\bf {\color{blue}[21]}} Intissar A. Formule de trace r\'gularis\'ee de l'o\'erateur magique de Gribov sur l'espace de Bargmann, J. Math. Anal. Appl.- 2016. - 437, n 1. - P. 59-70.\\
 
\n \n {\bf {\color{blue}[22]}}  Kato, T., Perturbation Theory for Linear Operators. Second edition, Springer- Verlag, Berlin, (1976)\\

\n {\bf {\color{blue}[23]}} Krasnoselskii, M.A., Zabreyko, P. P. Pustylnik, E. I. and Sobolevski, P. E., Integral operators in spaces of summable functions, Nauka44, Moscow, 1966; English transl., Noordho, 1975.\\ 

\n  {\bf {\color{blue}[24]}} Lang, S., Complex Analysis, Graduate Texts in Mathematics, Spronger, Fourth edition, (1999) \\

\n {\bf {\color{blue}[25]}} Lewis, J.W., Inversion of tridiagonal matrices, Numer. Math. 38 (1982) 333-345.\\

\n  {\bf {\color{blue}[26]}} Mallik,R. K., Solutions of linear difference equations with variable coeffi- cients, J. Math. Anal. Appl. 222 (1998), pp. 79-91\\

\n \n {\bf {\color{blue}[27]}} Markus, A.S. and Matsev, V. I., Comparaison theorelms for spectra of linear operators , and spectral asymptotic, Trans. Moscow Math. Soc. (1984).\\

\n \n {\bf {\color{blue}[28]}} Merdas, B.,  Remarks on Schatteen-VonNeumann classes $\mathfrak{}_{p}$, Demonstratio Mathematica,Vol XXII, No 4, 1989\\

\n \n {\bf {\color{blue}[29]}} Reed, M. and Simon, B.,  Methods of modern mathematical physics, Vol. IV, Academic Press, 1978.\\
 
\n \n {\bf {\color{blue}[30]}} Reed, M. and Simon, B., Methods of Modern Mathematical Physics I. Func- tional Analysis, rev. and enl. edition, Academic Press, San Diego, 1980.\\

\n \n {\bf {\color{blue}[31]}} Sadovnichy, V. A. and   Podol'skii, V. E. Traces of operators with relativement compact perturbation, Mat. Assis. - 2002. - 193, n 2. - S. 129-152.\\

\n \n {\bf {\color{blue}[32]}} Sadovnichy, V. A ,  Konyagin, S.V and  Podol'skii, V. E. Trace r\'egularis\'ee d'un op\'erateur avec un r\'esolvant nucl\'eaire perturb\'e par un r\'esolvant born\'e, Dokl. COURU. - 2000. - 373, n 1. - S. 26-28.\\

\n \n {\bf {\color{blue}[33]}} Sadovnichiy, V. A. and Podol'skii,V. E. Trace r\'egularis\'ee d'une perturbation born\'ee d'un op\'erateur avec un r\'esolvant nucl\'eaire, Differ. \'eq. - 1999. - 34, n 4. - S. 556-564.\\

\n \n {\bf {\color{blue}[34]}} Sadovnichiy, V. A. and  Podol'skii, V. E.  Traces d'op\'erateurs avec une perturbation relativement nucl\'eaire , Dokl. - 2001. - 378, n 3. , S. 1-2.\\

\n {\bf {\color{blue}[35]}} . Sadovnichii, V. A. and  V.E. Podolskii, V. E.Regularized traces of discrete operators, Proc. Steklov Inst. Math. (Suppl. 2) (2006) 161-177.\\

\n  {\bf {\color{blue}[36]}} Saidi, A., Yahya Mahmoud, A. and Vall ould Moustapha, M.,  Bargmann Transform With Application To Time-Dependent Schrdinger Equation, International Journal of Scientific $\&$ Technology Research, Volume 9, Issue 02 February (2020)\\

\n {\bf {\color{blue}[37]}} Shigekawa, I., Eigenvalue problems for the Schrdinger Operator with the magnetic field on a compact Riemannian maniforld, J. Funct. Anal, 75 (1987), 92-127.\\

\n\n {\bf {\color{blue}[38]}} Simon, B.,Trace Ideals and Their Applications, Cambridge University Press, (1979).\\

\n {\bf {\color{blue}[39]}}  Teschl, G. , Jacobi Operators and Completely Integrable Nonlinear Lattices, Mathematical Surveys and Monographs Volume 72,\\ https://www.mat.univie.ac.at/ gerald/ftp/book- jac/jacop.pdf.\\

\n {\bf {\color{blue}[40]}} Usmani, R. Inversion of a tridiagonal Jacobi matrix, Linear Algebra Appl. 212/213 (1994) 413-414.\\

\n {\bf {\color{blue}[41]}}  Usmani, R. Inversion of Jacobi's tridiagonal matrix, Comput. Math. Appl. 27 (1994) 59-66.\\

\n {\bf {\color{blue}[42]}}  Walter, G.G.: Wavelet and the Orthogonal Systems with Applications, CRC Press, Boca Raton, Florida(1994)\\
 
\n\n {\bf {\color{blue}[43]}}  Weidmann, J. Linear Operators in Hilbert Spaces, Springer, New York, 1980.\\

\n {\bf {\color{blue}[44]}} Welstead, S.T.: Selfadjoint extensions of Jacobi matrices of limit-circle type. Math. Anal. and Appl. 89, 315-326 (1982).\\

\n {\bf {\color{blue}[45]}}  Yafaev, D. R. , Spectral analysis of Jacobi operators and asymptotic behavior of orthogonal polynomials, arXiv:2202.02087v1\\

\n {\bf {\color{blue}[46]}}  Yoshino, K.,   Eigenvalue problem of Toeplitz operators in Bargmann -Fock space, Operator Theory, Advances and Applications, Birkhuser, vol. 260, p. 276-290 (2017)\\

\n {\bf {\color{blue}[47]}} Young, R. M. Euler's constant, Math. Gazette 75, (1991), vol. 472, p. 187- 190\\

\n \n {\bf {\color{blue}[48]}} Zhu, K.,  Analysis on Fock spaces, Graduate Texts in Mathematics 263,  41, 1992.\\

\n {\bf {\color{blue}[49]}} \n  Zhu, K.,   Analysis on Fock Spaces, Springer, New York, 2012.\\

\n\n {\bf {\color{blue}[50]}}  Zhu, K.,  Towards a dictionary for the Bargmann transform, arXiv:1506.06326v1 [math.FA] 21 Jun 2015.\\
\end{document}